\documentclass[aps,twocolumn,prx,superscriptaddress,groupedaddress]{revtex4}

\usepackage{ragged2e}
\usepackage{amssymb}
\usepackage{amsmath}
\usepackage{color,graphicx}
\usepackage{caption}
\usepackage{times}
\usepackage{amssymb}
\usepackage{amsmath}
\usepackage{amsthm}
\usepackage{mathrsfs}
\usepackage{mathtools}
\usepackage{caption}
\usepackage{times}
\usepackage{float}
\usepackage[colorinlistoftodos]{todonotes}
\usepackage{marginnote}
\usepackage{ulem}
\usepackage{mathtools}
\usepackage{amsthm}
\usepackage{tabularx}
\usepackage{comment}
\usepackage{ragged2e}
\usepackage{array}

\DeclareUnicodeCharacter{2062}{}

\usepackage{hyperref}
\hypersetup{
 colorlinks=true,
 citecolor=red,
 linkcolor=blue,
 urlcolor=blue,
 pdfpagemode=UseNone,
 pdfstartview=FitH}

\usepackage{physics}

\newcommand{\orcidicon}[1]{\href{https://orcid.org/#1}
{\includegraphics[height=\fontcharht\font`\B]{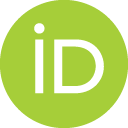}}}
\newcommand{\cH}{\mathcal{H}}
\newcommand{\cT}{\mathcal{T}}

\begin{document}

\author{O. Angeli\,\orcidicon{0009-0002-3618-8829}}
\email{oliviero.angeli@phd.units.it}
\affiliation{Department of Physics, University of Trieste, Strada Costiera 11, 34151 Trieste, Italy.}
\affiliation{INFN, Sezione di Trieste, Via Valerio 2, 34127 Trieste, Italy.}

\author{S. Donadi\,\orcidicon{0000-0001-6290-5065}}
\email{sandrodonadi@gmail.com}
\affiliation{INFN, Sezione di Trieste, Via Valerio 2, 34127 Trieste, Italy.}
\affiliation{Centre for Quantum Materials and Technologies, School of Mathematics and Physics, Queen’s University, Belfast BT7 1NN, United Kingdom.}

\author{G. Di Bartolomeo\,\orcidicon{0000-0002-1792-7043}}
\affiliation{Department of Physics, University of Trieste, Strada Costiera 11, 34151 Trieste, Italy.}
\affiliation{INFN, Sezione di Trieste, Via Valerio 2, 34127 Trieste, Italy.}

\author{J.L. Gaona-Reyes\,\orcidicon{0000-0003-1199-5501}}
\affiliation{Quantum Machines Unit, Okinawa Institute of Science and Technology
Graduate University, Onna, 904-0495, Okinawa, Japan.}
\affiliation{Department of Physics, University of Trieste, Strada Costiera 11, 34151 Trieste, Italy.}
\affiliation{INFN, Sezione di Trieste, Via Valerio 2, 34127 Trieste, Italy.}

\author{A. Vinante\,\orcidicon{0000-0002-9385-2127}}
\affiliation{Istituto di Fotonica e Nanotecnologie IFN--CNR, 38123 Povo, Trento, Italy}

\author{A. Bassi\,\orcidicon{0000-0001-7500-387X}}
\email{abassi@units.it}
\affiliation{Department of Physics, University of Trieste, Strada Costiera 11, 34151 Trieste, Italy.}
\affiliation{INFN, Sezione di Trieste, Via Valerio 2, 34127 Trieste, Italy.}

\title{Probing the Quantum Nature of Gravity through Classical Diffusion}
\begin{abstract}
The question of whether gravity is fundamentally quantum remains one of the deepest open problems in modern physics. A recently explored approach consists of testing gravity’s ability to entangle quantum systems, which requires preparing and controlling massive quantum states—a formidable experimental challenge. We propose an alternative strategy that circumvents the need for quantum state engineering. We show that, if gravity is classical, it must necessarily modify momentum statistics in a non-unitary way. We consider the corresponding linearized master equation for two harmonically trapped objects interacting gravitationally and establish a lower bound on the noise that any classical gravitational interaction must induce. We then outline an experimental protocol based on a high-precision torsion pendulum at millikelvin temperatures, showing that the predicted diffusion, if present, is in principle detectable with near-term technology. Our approach offers a novel route to testing the classical versus quantum nature of gravity without requiring macroscopic quantum superpositions or precise control of the system’s quantum state, thereby significantly reducing the experimental complexity.
\end{abstract}
\maketitle

\section{Introduction}
At the Chapel Hill conference in 1957, Richard Feynman raised the question whether gravity is quantum or not~\cite{dewitt2011role}; he favored the first option, yet leaving open the (bare, in his words) possibility that quantum mechanics might fail at some point. After more than 60 years of intense theoretical and experimental efforts, the question remains unanswered. 

Testing the quantum nature of gravity can be understood in two broad ways. The first, more ambitious way, which aligns with ambitious research programs such as string theory \cite{Schellekens2013} and loop quantum gravity \cite{Rovelli2008}, is to assess whether the graviton, the quantum of gravitational waves, exists. A direct test of the graviton \cite{Dyson2013,Rothman2006,Carney2024,Palessandro2024}, for example in scattering experiments, requires reaching the Planck scale, which is not accessible; this is the reason why recent attention is shifting towards indirect tests, possibly at low energies \cite{Parikh2020,Kanno2021,Tobar2024}, which however still necessitate formidable technological advances.  

The other, more phenomenological approach, which actually aligns with Feynman's 1957 discussion and is the subject of this work, concerns in assessing whether the gravitational field generated by a mass in spatial superposition is  the superposition of the gravitational fields produced by the mass at the two locations, or something else. In this regard, Feynman makes an analogy with electromagnetism: as the Coulomb field entering the Schr\"odinger equation already makes the electromagnetic field quantum, in the same way the Newtonian potential entering the Schr\"odinger equation makes the gravitational field quantum.  A direct consequence of the  Schr\"odinger equation with the Newtonian potential is that a mass superimposed at two different locations generates the superposition of the corresponding gravitational fields. 

There is an ongoing debate about the relation between these two ways of assessing the quantization of gravity~\cite{belenchia2018quantum,rydving2021gedanken,Carney2022,danielson2022gravitationally,Fragkos2022inf}, which goes beyond the scope of our work.  While detecting the graviton remains the holy grail, we subscribe Feynman's argument: assessing the validity of the superposition principle for the gravitational field, also in its Newtonian limit, would prove its quantum nature.

Before moving on, it is worthwhile noticing that the `bare' possibility briefly mentioned by Feynman, i.e. that the gravitational field remains classical because Quantum Mechanics fails at the macroscopic scale, over the years has become a field of research, in the form of gravity-induced wave function collapse models~\cite{karolyhazy1966gravitation,diosi1987universal,penrose2014gravitization,penrose1996gravity,gasbarri2017gravity}, and hybrid quantum-classical gravitational models~\cite{diosi1984gravitation,tilloy2018ghirardi,oppenheim2023postquantum,layton2023weak,donadi2022seven}. The two are intimately connected. The first assume that gravity, for some reason yet to be understood, causes the collapse of the quantum wave function; the second assume that gravity is fundamentally classical when coupling  to quantum matter, predicting as a byproduct the collapse of the wave function. Our analysis will also clarify this connection.

In recent years, it has been proposed to test the quantum nature of gravity, by assessing its entangling properties \cite{bose2017spin,marletto2017gravitationally,krisnanda2020observable,christodoulou2019possibility,Guff2022,Tilly2021,Schut2022,Li2023,VandeKamp2020,Marshman2020,Schut2023,Gunnik2023,Christodoulou2023,Großardt2020,Kent2021,Chevalier2020,Matsumara2020,Krisnanda2023,Miki2024,Higgins2024,Miki2021,Feng2022,Carney2021,hosten2022constraints,ma2022limits,streltsov2022significance,carney2022erratum,Marletto2021,Nguyen2020,Yant2023,Ghosal2024,Cui2023,Bose2022,Toroš2021,Zhang2024a,Zhang2024b}.  Specifically, the argument, which actually is still debated~\cite{martin2023gravity,Fragkos2022inf,Ma2022,Gollapudi2025,Husain2022,Hanif2024}, relies on the fact that if gravity is classical in the sense of  being a LOCC  (Local Operation and Classical Communication), it cannot entangle two initially separated  systems~\cite{horodecki2009quantum}; therefore if entanglement is detected, gravity must be quantum---or nonlocal, which is excluded by special relativity. With these premises, the proposal is to build two adjacent matter-wave interferometers, which create a spatial superposition for each  mass; the masses are large enough, and the interferometers are close enough, to generate an appreciable Newtonian potential which, if quantum in the sense specified before and for the given setting, turns an initially factorized state into an entangled one.  

Performing the proposed experiment  represents a daunting  challenge~\cite{rijavec2021decoherence,aspelmeyer2022zeh}, because it requires the control of two coupled large-mass interferometers; while reaching the Planck scale is not needed anymore, several technologies breakthroughs will still be necessary to reach the goal. 

Actually, creating a spatial superposition of the form `here + there', which the reason why interferometers are proposed, is not needed to exploit the LOCC argument: in principle, any quantum state does the job. This is the rationale behind a follow up work \cite{krisnanda2020observable}, which proposes to use two trapped particles, interacting gravitationally, and initialized in (highly) squeezed states; once again, if gravity is quantum,  the initially factorized state of the two masses becomes entangled over time. The experimental design simplifies, with the interferometers being replaced by harmonic traps.

Detecting entanglement is not easy, especially when dealing with large masses, large enough to generate an appreciable gravitational interaction. For this reason, a  recent analysis \cite{lami2024testing}, proposes to bypass entanglement detection with an ingenious argument. Given an initial quantum states for two (or more) gravitationally interacting systems, we know how it should evolve according to the Schr\"odinger equation with the Newtonian potential. One can then ask to which extent a LOCC can reproduce this unitary evolution: the analysis shows that, no matter how well this can be done,  a difference always remains, which is quantified in terms of a bound. If the bound is violated, then gravity cannot be a LOCC. 
The proposed protocol, by circumventing entanglement as a figure of merit, does not require highly delocalized quantum states; instead, it amounts to computing  the fidelity between the state---as given by the experiment---of two gravitationally interacting systems and the state that is expected by the presence of the unitary Newtonian interaction.

In all the above proposals, as well as in others \cite{Datta2021,Margalit2021,kryhin2025distinguishable,howl2021non}, the fundamental limiting factor is the necessity of controlling the quantum state of one or more masses which are large enough to generate an appreciable gravitational field: preparing and then controlling large-mass quantum states is notoriously very difficult \cite{fein2019quantum,Ma2021,Bild2023}. 

We propose a different and significantly easier strategy: we will show that if gravity is classical, then it must be random in order to couple consistently with quantum matter. This randomness causes diffusion in the motion of a probe interacting with the gravitational field,
even if the probe is in a {\it classical state}. Therefore, instead of having to deal with large-mass quantum states which, in order to be generated and controlled, ultimately limit the size of the system and the gravitational effect to be detected, our proposal requires only the monitoring of the center of mass motion of a classical probe, which can be conveniently large to generate a strong enough gravitational field. This procedure is simpler and already proved to be a very effective strategy in other contexts~\cite{Bahrami2014,Diosi2015,Vinante2017,bilardello2016bounds,schrinski2017collapse,altamura2024noninterferometric,donadi2021novel,donadi2021underground,arnquist2022search,carlesso2022present}.

The paper is organized as follows. In Section \ref{due}, we show with a Gedankenexperiment that if gravity is classical, it must collapse spatial superpositions of the wave function of the systems it couples to and induce  fluctuations in their motion. In Section~\ref{sec:math} we provide the mathematical proof underlying the Gedankenexperiment.  In Section \ref{tre}, we discuss how this diffusion, together with the requirement of recovering the Newtonian interaction for well-localized systems, leads to a specific master equation for two spatially confined, gravitationally interacting systems. In Section \ref{secIV}, we derive the constraints that the diffusive terms must satisfy to prevent the generation of entanglement, resulting in a lower bound on the strength of the diffusive effects. Finally, in Section \ref{secV}, we explore the feasibility of detecting these effects with current and near-future technologies.

\section{Classical gravity makes quantum systems' motion diffuse}\label{due}
Similarly to the above mentioned proposals, our working hypotheses are:
\begin{enumerate}
    \item Matter is quantum. 
    \item Gravity is classical.
    \item Gravity is local.
    \item Classical systems follow Newton's laws. 
\end{enumerate}

The first assumption should be straightforward. The second one implies that the gravitational potential has a well-defined value at each point in space and, therefore, cannot be in a superposition; as such, when it interacts with matter in  classical state, the state remains classical. The third assumption about locality is meant in the sense of no faster-than-light signaling between far-away parties. In the following, we consider the Newtonian limit, where gravity is actually instantaneous and thus nonlocal. This is not in contradiction with the no-signaling condition we are imposing. The non-locality in the Newtonian potential is only an artifact of the non-relativistic approximation and, as it will be clear from the following argument, it will play no role. The no-signaling condition we are referring to, is the possibility for two parties to communicate at a distance exploiting quantum non-locality; this is what is excluded. Assumption 4 states that when matter is in a state that, for all practical purposes, is classical, the gravitational force follows Newton's law (or Einstein’s equations in the relativistic case), up to small fluctuations. In other words, within experimental error, gravity behaves classically in the appropriate limit.

We now follow the spirit of Feynman's thought experiment at Chapel Hill, and take it to its natural conclusions under the above assumptions. 

\begin{figure*}[t!]
\centering
\includegraphics[width=\linewidth]{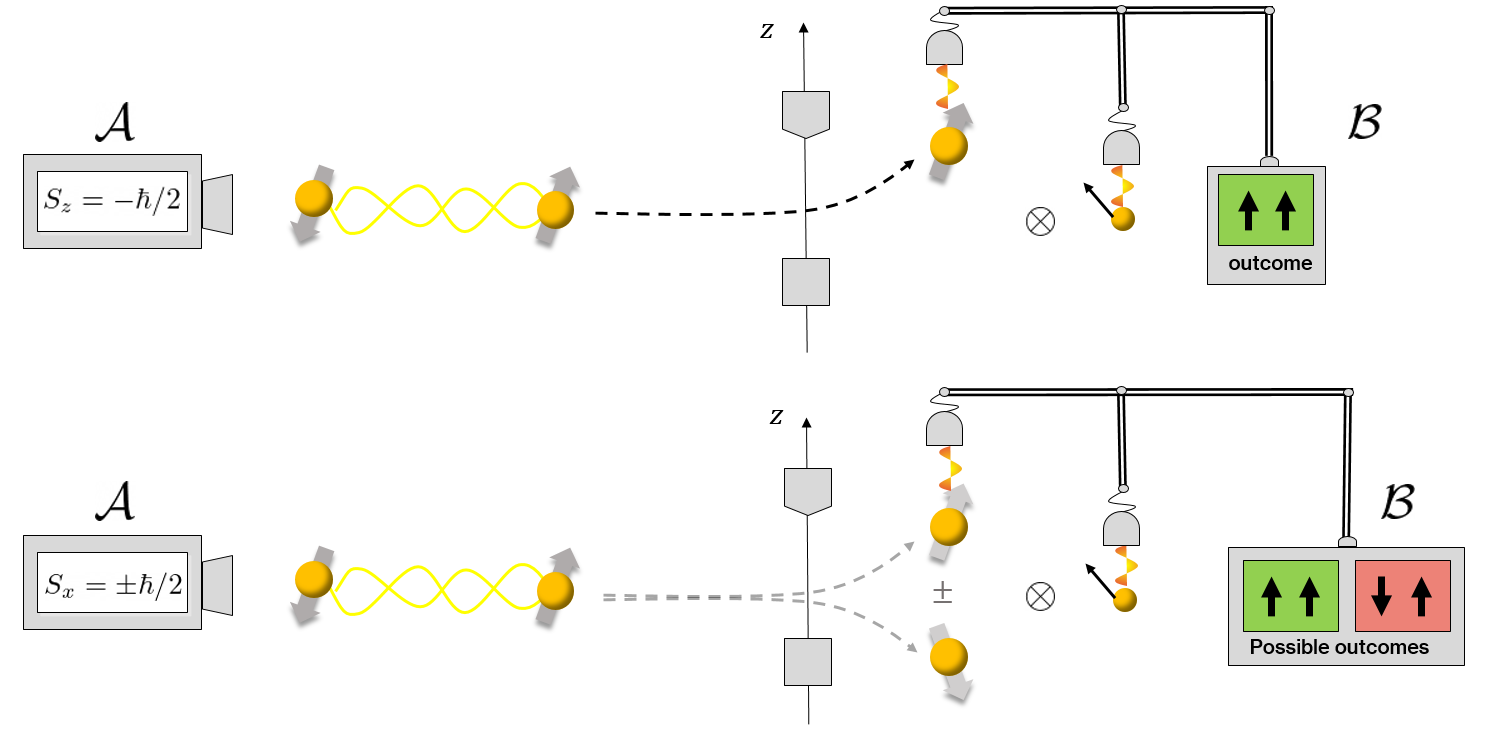}
\caption{\justifying\textbf{Gedankenexperiment.}
Pairs of spin 1/2 particles (represented by the yellow spheres with an arrow) are generated from a source in a spin-singlet state, with one particle traveling toward Alice and the other toward Bob. Alice performs her spin measurements first and is free to choose the direction of   measurement. On Bob's side, a fixed Stern-Gerlach apparatus with the magnetic field directed along the $z$ direction interacts with the incoming particles; beyond it,  a probe (represented by a small yellow sphere) detects the gravitational pull exerted by the outgoing particles.
{\it Top Panel -- } Alice measures the spin along the $z$-axis; Bob's particles have 50\% chance of going upward and a 50\% chance of going downward when passing through the Stern-Gerlach apparatus; accordingly, the probe moves  half of the times upward and the other half downward.
{\it Bottom panel -- } Alice measures the spin along the $x$-axis; Bob's particles always end up in spatial superposition of states, when exiting the Stern-Gerlach apparatus. If gravity is classical, the probe cannot be driven in a superposition state and, to avoid the possibility of faster-than-light signaling, it must again move upward half of the times and downward the other half. Moreover, if Bob's spin particle has not collapsed in space after the interaction with the probe, there would be a non-zero probability of an outcome which is impossible if Alice had measured along $z$ (e.g. the probe moves upward and the spin particle downward). This must be forbidden to avoid the possibility of having faster-than-light signaling.}
\label{fig1}
\end{figure*}

{\it The classical-quantum interaction must induce the collapse of the wave function. ---} Consider  Alice (A) and Bob (B), who are well apart from each other and share pairs of spin-$1/2$ particles in the entangled singlet state:
\begin{equation}
\ket{\Psi} = \frac{1}{\sqrt{2}}[\ket{+1^A_z}\ket{-1^B_z} - \ket{-1^A_z}\ket{+1^B_z} ]; 
\end{equation}
we omit the spatial part of the wave function for now, as it will become relevant later. Bob sets up a Stern-Gerlach apparatus through which his particles eventually pass, with the magnetic field oriented along the $z$ direction. Beyond the apparatus, a probe interacts gravitationally with the outgoing spin particles to test the gravitational field they generate (see Fig.~\ref{fig1}).

If Alice measures the spin of her particles along the $z$ direction, Bob's  particles will end up in one of the two  eigenstates of the spin operator $\hat{S}^B_z$ and, when crossing the Stern-Gerlach, they will either  be deflected upwards (if $S^A_z = -1$, modulo $\hbar/2$ which from now on we omit) or downwards (if $S^A_z = +1$); the two cases each occur with probability $1/2$, and each particle's wavefunction is supposed to be well localized in space to be considered classical for all practical purposes. Then, because of assumption 4 the probe, which  is  also assumed to be in a classical state, will be pulled upwards or downwards, each case occurring again with probability $1/2$.

If Alice measures the spin along the $x$ direction, then Bob's particles will end up in one of two eigenstates of $\hat{S}^B_x$ and, thus, in one of the two superposition states  $\left( \ket{+1^B_z} \pm \ket{-1^B_z}\right)/\sqrt{2}$ of spin  along the $z$ direction, which results, after passing through the Stern-Gerlach apparatus, in one of the two superposition states 
\begin{equation} \label{eq:thdgfs}
\ket{B^\pm}= \frac{1}{\sqrt{2}}\left[|+1^B_z\rangle|\text{up}^B_z\rangle \pm |-1^B_z\rangle|\text{down}^B_z\rangle\right]\,,
\end{equation}
where $|\text{up}^B_z\rangle$ ($|\text{down}^B_z\rangle$) describes the particle moving upwards (downwards). Now we have included the spatial part of the wave function, which before was omitted. The question is what can we say about the gravitational field generated by these delocalized states. To avoid superluminal signaling, the classical gravitational field has to be such that the probe reacts as in the previous case, otherwise Bob would be able to realize from a distance what Alice measured: the probe must be deflected either upwards or downwards, each with probability $1/2$. 

Suppose that, in a specific run of this second type of experiment, the probe is attracted upwards along \( z \). Then, the state of Bob's spin particle  
\textit{must} collapse to \( |+1^B_z\rangle|\text{up}^B_z\rangle \) because, if it did not, Bob could further measure its position and there would be a \( \frac{1}{2} \) probability of finding it in the down state; this would mark a difference with respect to the case where Alice measured the spin along the \( z \) direction, opening up again the way to superluminal signaling. Hence, when the probe moves guided by the gravitational field generated by Bob's delocalized particle, the state of the latter must collapse accordingly in space.  

Note that if the gravitational interaction is quantum-mechanical, there is no need to appeal to the collapse of the wave function, as the measurement of the gravitational field by the probe (or the measurement of the probe's position after the interaction with the spin particle) would naturally collapse the particle's state, since the probe and particle would be gravitationally entangled with each other. A classical interaction does not allow for the generation of entanglement; therefore, the collapse must be caused by something else, rather than by measuring the probe.

{\it The collapse must be random. ---} At this stage, it could be that the gravitational interaction and the collapse are deterministic in the sense that when Bob's spin particle, exiting the Stern-Gerlach apparatus, is in the state \( |B^+\rangle \), then the probe is always attracted, for example, upwards, and correspondingly the state collapses to \( |+1^B_z\rangle|\text{up}^B_z\rangle \). Similarly, when the state is \( |B^-\rangle \), the gravitational pull occurs downwards, and the state collapses to \( |-1^B_z\rangle|\text{down}^B_z\rangle \). However, even this situation would again cause superluminal signaling and therefore must be excluded, for the following reason.

The states \( |B^\pm\rangle \) are entangled states of the spin degree of freedom and the spatial degree of freedom; in our case, they refer to the same particle of Bob. However, in principle, the spin state could refer to another particle, which is far away from Bob's.  
Therefore, one can imagine a new Gedankenexperiment, where Alice and Bob now share pairs of particles in the entangled state:
\begin{equation}
   |\tilde B^+\rangle = \frac{1}{\sqrt{2}}\left(|+1^A_z\rangle|\text{up}^B_z\rangle + |-1^A_z\rangle|\text{down}^B_z\rangle\right),
\end{equation}
which is mathematically equivalent to \( |B^+\rangle \); now the spin state of Alice's particles is entangled with the position state of Bob's ones.  
As before, a massive probe on Bob's side samples the gravitational field generated by the quantum state of the particle flying towards it. In this case, based on our previous assumptions, the probe would be pulled upwards, and the entangled state would collapse accordingly. However, before Bob's particle reaches the probe, Alice can choose to apply a \( Z \) gate to the spin state of her particle and change the entangled state to
\begin{equation}
   |\tilde B^-\rangle = \frac{1}{\sqrt{2}}\left(|+1^A_z\rangle|\text{up}^B_z\rangle - |-1^A_z\rangle|\text{down}^B_z\rangle\right), 
\end{equation}
which is equivalent to $|B^-\rangle$: in this case the probe would be pulled downwards. Then again, a protocol for superluminal signaling could be established between Alice and Bob.

{\it The dynamics of quantum systems must be diffusive. ---} It has been shown~\cite{donadi2023collapse} that a stochastic dynamics, which is covariant under space translations and encodes the collapse of the wave function, must increase the variance of the momentum distribution, i.e. it is diffusive. The fundamental reason is the following. The collapse of the wave function is a {\it nonlinear} operation on $|\psi\rangle$, since it must suppress superpositions (in this case, in space). However, at the statistical level, when averaging over all possible ways the collapse occurs, the effect must be equivalent to a {\it linear} operation on the corresponding density matrix $\hat\rho$, otherwise---again---the possibility would open for superluminal signaling. This was shown long ago~\cite{gisin1989stochastic,polchinski1991weinberg} as a response to attempts to formulate nonlinear extensions of the quantum dynamics~\cite{weinberg1989testing,weinberg1989precision}. Since the collapse operator is linear on $\hat\rho$ and cannot be the identity, it  unavoidably changes  the state of the system, in particular by shifting its center in space: this shift, over time, amounts to a diffusion process. We thus arrive at the conclusion that the four assumptions above imply that classical and local gravity must come with a diffusion mechanism acting on matter. 

As a matter of fact, all models in the literature which assume that gravity is classical~\cite{kafri2014classical,tilloy2016sourcing,oppenheim2023postquantum,oppenheim2023gravitationally},  those where gravity plays a role in the emergence of classicality~\cite{karolyhazy1966gravitation,diosi1989models,penrose1996gravity} as well as more general arguments~\cite{galley2023any} are in agreement with our general conclusion, except for the Schr\"odinger-Newton equation~\cite{diosi1984gravitation,penrose2014gravitization}, which however allows for superluminal signalling \cite{bahrami2014schrodinger}.

For what follows, it is important to realize that the collapse must be present also when the system is in a \textit{classical} state. In this case, the state will not change dramatically, since it is already localized; yet, a tiny effect must be present in the form of a diffusion process. This unlocks the possibility of testing the diffusion process in classical experiments, a possibility that was precluded in previous proposals.

The reason for the presence of this gravity-induced diffusion also in classical systems is the following. Consider again the situation represented in Fig.~1: when Alice measures the spin of her particles along the \( x \)-direction, Bob's spin particles end up in one of the two states of Eq.~\eqref{eq:thdgfs}, each occurring with probability \( \frac{1}{2} \). As discussed before, the two states must collapse because of the classical gravitational interaction with the probe. Let us call \( \hat{\rho}_B \) the density matrix associated to the spin particles before the ``gravitational'' collapse occurs, and \( \hat{\rho}'_B \) the density matrix after the collapse; the difference between the two also encodes information about the change in momentum (diffusion) triggered by the collapse.

Suppose now that Alice measures the spin of her particles along the \( z \)-direction. To avoid superluminal signaling, Bob should not be able to tell the difference with respect to the previous case, meaning that also in this case the density matrix must initially be \( \hat{\rho}_B \) and, after having interacted classically with the probe, \( \hat{\rho}'_B \). Although the states are always localized in space, they must change (slightly) over time.

The next section converts the argument here presented into a mathematical proof.

\section{The mathematical proof}
\label{sec:math}
Consider Alice and Bob, far away. On Alice's side there is a system and on Bob's side two. In our case, the three systems are the two massive spin-particles and the gravitational probe, but for the sake of the proof we can keep them general. 

We will introduce very mild conditions which ought to be satisfied if gravity is classical. These are by no means sufficient to characterize what one would usually regard as a classical mediator~\cite{bose2017spin,kafri2014classical,tilloy2016sourcing,oppenheim2023postquantum}; nevertheless, they are sufficient to show that it must collapse quantum superpositions and simultaneously alter the momentum distribution, which is the goal of our analysis.

\subsection{Assumption 1: Matter is quantum}
We assume that matter is described quantum mechanically. Let $\cH_A$ be the Hilbert space for Alice's system, and $\cH_{B_1}$, $\cH_{B_2}$ those for Bob's; let $\cH_B = \cH_{B_1} \otimes \cH_{B_2}$. Alice and Bob can perform any type of local measurement on their systems, allowed by quantum theory; their statistics follows quantum mechanical rules.

Bob's two systems interact with each other gravitationally. We describe their dynamics at the level
of wave functions, taking place during a given time interval,  by a norm-preserving map
\begin{equation}
\mathcal V_\omega:\mathcal H_B \longrightarrow \mathcal H_B,
\end{equation}
where $\omega$ encodes potential randomness.
At this stage, as usually proposed in the literature,  $\mathcal{V}_\omega$ is not assumed to be necessarily linear  \cite{diosi1984gravitation,kafri2014classical,tilloy2016sourcing, tilloy2018ghirardi,oppenheim2023gravitationally,oppenheim2023postquantum,piccione2025hybrid}. As a matter of fact, as anticipated in the previous section, classical gravity implies the collapse of the wave function, therefore the dynamics must be non-linear.  
If stochasticity is present, the physical output state on $B$ is obtained by averaging over all realizations:
\begin{equation}\label{av_V}
    \hat\rho_{\text{\tiny I}} =|\psi\rangle \langle \psi| \longrightarrow \; \hat\rho_{\text{\tiny F}} = {\mathbb E} [|\mathcal{V}_\omega (\psi)\rangle\langle \mathcal{V}_\omega (\psi)|],
\end{equation}
where $\mathbb E$ denotes the stochastic average.
The central question is how to characterize the evolution~\eqref{av_V} according to our remaining assumptions.

\subsection{Assumption 2: Gravity is local}
We exploit locality in the  form of no-superluminal signaling among arbitrarily distant parties, which is a necessary ``relativistic'' request also for our non-relativistic setting (it corresponds to the micro-causality condition of relativistic quantum field theory). The following holds.

{\bf Theorem 1: no-signaling implies linearity.} Let $\cT(\cH)$ denote the space of trace-class operators on $\cH$. To avoid superluminal signaling, the operator $\mathcal{V}_\omega$ must  induce a  map on density matrices:
\begin{equation} \label{eq:avmap}
    \mathcal M: \cT(\mathcal{H}_{B}) \to \cT(\mathcal{H}_{B}),
\end{equation}
which is affine (linear by extension) and trace-preserving (TP); adding subsystem consistency yields complete positivity (CP).

{\it Proof. ---} This is known in the literature \cite{gisin1989stochastic,Simon2001}; because it shows the crucial role played by the no-signaling condition, we present the result  in a compact form. 

Fix $\hat{\rho}_{B}\in\cT(\cH_{B})$ and let us assume for simplicity assume finite-dimensional Hilbert spaces (or separable spaces with the appropriate technical conditions so that the Hughston–Jozsa–Wootters (HJW) theorem applies).
 By the HJW theorem \cite{hughston1993complete}, for any two ensembles of pure states $\{\!(p_i,|\psi_i\rangle)\!\}_i$ and $\{\!(q_j,|\phi_j\rangle)\!\}_j$ with the same average,
\begin{equation}
\hat{\rho}_{B}=\sum_i p_i\,|\psi_i\rangle\langle\psi_i|=\sum_j q_j\,|\phi_j\rangle\langle\phi_j|,
\end{equation}
there exists a purification $\ket{\Psi}_{AB} \in {\mathcal H}_A \otimes {\mathcal H}_B$ of $\hat{\rho}_{B}$ and local POVMs on $A$ that remotely prepare at $B$ either ensemble.
No-signaling requires that Alice's choice of POVM on $A$ cannot affect
the statistics observed at $B$ after the evolution has taken place.
Therefore, the corresponding output states must coincide:
\begin{equation}
\label{eq:affinity-states}
\sum_i p_i\, \mathbb{E}[\ket{\mathcal{V}_\omega(\psi_i)}\bra{\mathcal{V}_\omega(\psi_i)}]
=
\sum_j q_j\,\mathbb{E}[\ket{\mathcal{V}_\omega(\phi_j)}\bra{\mathcal{V}_\omega(\phi_j)}].
\end{equation}
Equation \eqref{eq:affinity-states} is ensemble–independence, which implies the existence of a map ${\mathcal M}$ as in Eq.~\eqref{eq:avmap}, defined as
$$
\mathcal M(\hat\rho_B)
:=
\sum_i p_i\,\mathbb{E}[\ket{\mathcal{V}_\omega(\psi_i)}\bra{\mathcal{V}_\omega(\psi_i)}],
$$
where $\{\!(p_i,|\psi_i\rangle)\!\}_i$ is any ensemble averaging to
$\hat\rho_{B}$. By construction, $\mathcal M$ coincides with the statistical action of
$\mathcal{V}_\omega$ on mixed states; furthermore it is affine on the convex set of density operators:
\begin{equation}
\mathcal M\big(\lambda\hat{\rho}+(1-\lambda)\hat{\sigma}\big)\ =\ \lambda\,\mathcal M(\hat{\rho})+(1-\lambda)\,\mathcal M(\hat{\sigma}),
\end{equation}
with $\lambda\in[0,1]$.
The affine map is extended uniquely to a linear map on all trace-class operators \cite{Heinosaari2012,Watrous2018}.

$\mathcal M$ must preserve the trace, otherwise Bob's probabilities of measurements would not sum to 1. To conclude complete positivity, we invoke {\it subsystem consistency}: the same reduced dynamics
$\mathcal M$ must be applicable to $B$ also when it is viewed as part of a larger system $AB$, with $A$ left untouched. Thus, for any Hilbert space $\mathcal H_A$ and any joint state
$\hat\rho_{AB}\ge 0$ on $\mathcal H_A\otimes\mathcal H_B$, given the identity map $\mathcal I_A$ on $\cT(\cH_A)$; the extended map $\mathcal I_A\otimes \mathcal M$ must send $\hat\rho_{AB}$ to a positive operator, i.e.
$(\mathcal I_A\otimes \mathcal M)(\hat\rho_{AB})\ \ge\ 0.$
By definition, this requirement is equivalent to complete positivity, hence $\mathcal M$ is CP.
 Therefore $\mathcal M$ is CPTP. \qed

As mentioned in the previous section, the Schr\"odinger Newton equation \cite{diosi1984gravitation}, which is often used as a benchmark for a classical theory of gravity \cite{aziz2025classical} violates the no-superluminal-signaling condition \cite{gisin1989stochastic,Simon2001} and therefore does not allow for a density matrix description in accordance with Theorem 1. As such, it falls outside the present formalism.
 
It is also worthwhile stressing that the possibility of signaling between Alice and Bob is not tied to the non-relativistic framework we are working with. As a matter of fact, no interaction is assumed to link the two parties: $A$ and $B$ are only assumed to be entangled, not to interact with each other (except for some time in the past when entanglement was created). It is quantum  nonlocality inherent to quantum measurements---which holds also for relativistic systems such as photons \cite{freedman1972experimental,aspect1982experimental,weihs1998violation,tittel1998violation,giustina2013bell,shalm2015strong} that can cause signaling if the map $\mathcal V_\omega$ (which acts locally only on Bob's side) does not satisfy the conditions of Theorem 1.

\begin{figure*}
    \centering
\includegraphics[width=0.9\linewidth]{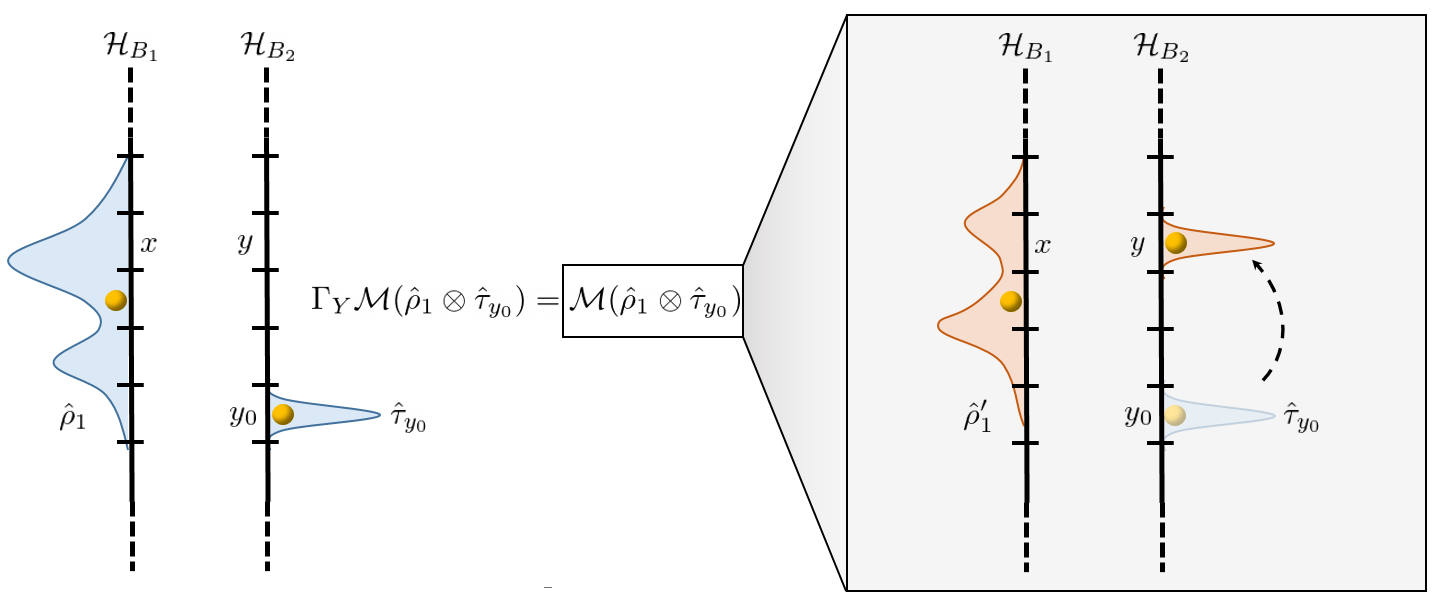}
    \caption{\justifying{\bf Classicality. } Schematic representation of the classicality condition~\eqref{eq:WeakC2}. 
    The Hilbert spaces of the particle and of the probe are represented by lines with intervals denoting  PVM decompositions $\{\Pi_x\}_{x\in X}$ and $\{\Pi_x\}_{y\in Y}$ which identify classical states; see main text. The probe $B_2$ is initialized in a state well localized in a specific sector $\Pi_{y_0}$, while the particle ${B_1}$ is in a generic state $\hat\rho_1$. According to the condition~\eqref{eq:WeakC2}, the state after the interaction $\mathcal{M}(\hat\rho_1\otimes\hat\tau_{y_0})$ is a fixed point of $\Gamma_Y$, the dephasing on $B_2$; this is another way of saying that the state of the probe $B_2$ remains well localized in a---potentially different---sector $\Pi_y$, after the interaction. The final state is not predetermined and occurs with probability $p(y) = \Tr[({\mathbb I}_{B_1} \otimes \Pi_y) {\mathcal M}(\hat\rho_1 \otimes \hat \tau_{y_{0}})]$ Meanwhile the reduced state on $\mathcal{H}_{B_1}$ evolves to $\hat \rho_1' =\Tr_{B_2}[\mathcal{M}(\hat\rho_1\otimes \hat\tau_{y_0})]$: without further constraints this evolution does not necessarily describe collapse.}
    \label{fig:classicality}
\end{figure*}
\subsection{Assumption 3: Gravity is classical}
The evolution $\mathcal V_\omega$---or, at the statistical level, the channel
$\mathcal M$---is supposed to implement a {\it classical} interaction. The literature often models a classical interaction with a LOCC \cite{kafri2014classical,di2021gravity,lami2024testing,bose2017spin}. Here we introduce a different characterization of a classical interaction, which we believe is better suited to the problem under investigation, because it is based not on the concept of local operations (followed by classical communication) performed on a system, but rather on how a classical mediator is expected to couple to quantum systems. As such, it neither includes nor excludes LOCC.

We assume that both $B_1$ and $B_2$ admit a preferred set of {\it classical
sectors}.  For $B_1$, we describe the classical sectors by a family of mutually
orthogonal projectors $\{\Pi_x\}_{x\in X}$ on $\mathcal H_{B_1}$, not necessarily
of rank one, satisfying the projection-valued measure (PVM) relations
\begin{equation}\label{eq:classical-PVM2}
\Pi_x\Pi_{x'}=\delta_{x,x'}\Pi_x,\qquad
\sum_{x\in X}\Pi_x=\mathbb{I}_{B_1}.
\end{equation}
We make a similar assumption for $B_2$ and denote by $\{\Pi_y\}_{y\in Y}$ the family of mutually
orthogonal projectors on $\mathcal H_{B_2}$. 
Here and in the following, by classical we mean that any state supported on an element of the sets $\{\Pi_{x}\}_{x\in X}$ or $\{\Pi_y\}_{y\in Y}$ is so well-localized to behave, for all practical purposes, as a classical state. Define the dephasing channel with respect to
$B_2$,
\begin{equation}\
\Gamma_Y(\hat{\rho}_{B}) :=  \sum_{y\in Y} (\mathbb{I}_{B_1}\!\otimes\! \Pi_y)\,\hat{\rho}_{B}\,(\mathbb{I}_{B_1}\!\otimes\! \Pi_y),
\label{eq:dephasing}
\end{equation}
which removes all coherences between distinct $y$--sectors.

Given the PVMs which define the classical sectors, we further define the set of localized states on $B_1$
\begin{equation}
\label{eq:sigma_x_def}
\mathsf S_x\!
:=\!\Big\{ \hat\sigma_x\!\in\!\mathcal T(\mathcal H_{B_1})\! :\! \hat\sigma_x\!\ge\! 0, \Tr_{B_1}[\hat\sigma_x]\!=\!1, \hat\sigma_x\!=\!\Pi_x\hat\sigma_x\Pi_x\Big\}.
\end{equation}
with an analogous definition holding for the PVM $\{\Pi_y\}_{y\in Y}$ in $B_2$.

We now impose the following {\it classicality condition} for the
interaction: whenever the probe $B_2$ is initially prepared in a definite
classical sector $y$, the output state must remain block--diagonal across
the $y$--sectors. Concretely, for every
$\hat\rho_1\in\mathcal T(\mathcal H_{B_1})$ and every normalized probe state $\hat{\tau}_y \in \mathsf{S}_y$ (and every $y\in Y$),
we require
\begin{equation}
\Gamma_Y\!\left(\mathcal M(\hat{\rho}_1 \!\otimes\! \hat\tau_y)\right)
=
\mathcal M(\hat{\rho}_1 \!\otimes\! \hat\tau_y).\tag{C.1}\label{eq:WeakC2}
\end{equation}
In other words, for such inputs the output has no off--diagonal blocks
$(\mathbb I_{B_1}\!\otimes\!\Pi_{y})\,\mathcal M(\cdot)\,(\mathbb I_{B_1}\!\otimes\!\Pi_{y'})$
with $y\neq y'$: the probe's initially classical state has not been turned into a superposition of different classical states.

Note that if $\Pi_y$ has rank higher than 1, the condition above enforces only block--diagonality across
different sectors $y\neq y'$, while allowing for nontrivial---in general coherent---dynamics within each sector. As such, the dynamics can retain some quantum features, however, we do not need to further constrain the dynamics in this respect, to reach the desired conclusion.  However, while some coherence might be present in the state $\mathcal{M}(\hat\rho_1\otimes\tau_{y})$, the block diagonal structure in $B_2$ implies that there is no coherence between distinct classical sectors of $B_2$. 
Thus, condition~\eqref{eq:WeakC2} predicts that the two systems cannot become entangled through coherences between distinct classical sectors.

Note also that that, so far, we are not committing to what happens to $B_2$ when initially it is in a superposition across distinct classical sectors. Condition~\eqref{eq:WeakC2} constrains
the output only for inputs that are classical on $B_2$, without committing on what happens for other choices of initial state.
Nevertheless, equation~\eqref{eq:WeakC2} forces a well precise structure on the map $\mathcal M$.

{\bf Theorem 2: Instrument structure.}
Consider a linear and CPTP map $\mathcal M:\mathcal T(\mathcal H_B)\to\mathcal T(\mathcal H_B)$
that satisfies the classicality condition~\eqref{eq:WeakC2} with respect to a family of mutually
orthogonal projectors $\{\Pi_y\}_{y\in Y}$ on $\mathcal H_{B_2}$. Assume that the probe is initially prepared in a normalized classical state
$\hat\tau_{y_0}\in \mathsf{S}_{y_{0}}$ supported on the sector $\Pi_{y_0}$.

Then there exists a collection of linear maps
\begin{equation}
\{\widetilde{\mathcal J}_{y}:\mathcal T(\mathcal H_{B_1})\to\mathcal T(\mathcal H_{B})\}_{y\in Y}
\end{equation}
such that, for all $\hat\rho_1\in\mathcal T(\mathcal H_{B_1})$,
\begin{align}
\label{eq:block_instrument_quantum_output}
\mathcal M(\hat\rho_1\otimes \hat\tau_{y_0})
& = \sum_{y\in Y}\widetilde{\mathcal J}_{y}(\hat\rho_1),
\nonumber \\
\widetilde{\mathcal J}_{y}(\hat\rho_1)
& = (\mathbb I_{B_1}\!\otimes\!\Pi_y)\,\widetilde{\mathcal J}_{y}(\hat\rho_1)\,(\mathbb I_{B_1}\!\otimes\!\Pi_y).
\end{align}
Each $\widetilde{\mathcal J}_y$ is completely positive and trace--non--increasing (CP--TNI), and
$\sum_y\widetilde{\mathcal J}_y$ is CPTP. In other words, $\{\widetilde{\mathcal J}_y\}_{y\in Y}$
forms a quantum instrument~ \cite{Wilde2017}. Moreover, defining the reduced instrument on $B_1$ by
\begin{equation}\label{eq:coarse_instrument_B1}
\mathcal J_y(\hat\rho_1):=\Tr_{B_2}\!\big[\widetilde{\mathcal J}_y(\hat\rho_1)\big],
\end{equation}
each $\mathcal J_y$ is also CP--TNI and $\Phi:=\sum_{y\in Y}\mathcal J_y$ is CPTP. 

\smallskip
{\it Proof. ---}
By assumption~\eqref{eq:WeakC2}, the operator
$\mathcal M(\hat\rho_1\otimes \hat\tau_{y_0})$ is a fixed point of $\Gamma_Y$, hence it is block--diagonal
across the $y$--sectors:
\begin{equation}
\mathcal M(\hat\rho_1\otimes \hat\tau_{y_0})=\sum_{y\in Y}(\mathbb I_{B_1}\!\otimes \Pi_y)\,\mathcal M(\hat\rho_1\otimes \hat\tau_{y_0})\,(\mathbb I_{B_1}\!\otimes \Pi_y)\,.
\end{equation}
Define
\begin{equation}
\widetilde{\mathcal J}_y(\hat\rho_1):=(\mathbb I_{B_1}\!\otimes \Pi_y)\,\mathcal M(\hat\rho_1\otimes \hat\tau_{y_0})\,(\mathbb I_{B_1}\!\otimes \Pi_y)\,,
\end{equation}
which yields~\eqref{eq:block_instrument_quantum_output}. Complete positivity of each $\widetilde{\mathcal J}_y$ follows from the fact that it is the composition of three completely positive maps: (i) the assignment map $\Lambda_{y_0}:\mathcal T(\mathcal H_{B_1})\to \mathcal T(\mathcal H_B)$, with $\Lambda_{y_0}(\hat\rho_1):=\hat\rho_1\otimes \hat\tau_{y_0}$, which is CP because $\hat\tau_{y_0}\ge 0$; (ii) the channel $\mathcal M$, which is CP by assumption; (iii) the projection
$\mathcal P_y(X):=(\mathbb I_{B_1}\!\otimes\!\Pi_y)\,X\,(\mathbb I_{B_1}\!\otimes\!\Pi_y)$, which is CP. Hence $\widetilde{\mathcal J}_y=\mathcal P_y\circ \mathcal M\circ \Lambda_{y_0}$ is completely positive.  Moreover, each branch $\widetilde{\mathcal J}_y$ is trace--non--increasing because it is positive and their traces sum to one. Finally, \eqref{eq:coarse_instrument_B1} is a composition of CP maps and the partial trace, hence each $\mathcal J_y$ is CP; trace--non--increasing and $\sum_y\mathcal J_y$ being CPTP follow by tracing the
corresponding statements for $\widetilde{\mathcal J}_y$. \qed

\smallskip
According to the Theorem~2, the probability for the probe to be
in the classical sector $y$ {\it after} the interaction, given that the system $B_1$ is initially prepared in the state
$\hat\rho_1$, is
\begin{equation}
p(y|\hat\rho_1):=\Tr_{B_1}\!\big[\mathcal J_y(\hat\rho_1)\big]
=
\Tr_{B}\!\big[\widetilde{\mathcal J}_y(\hat\rho_1)\big].
\end{equation}
Conditioned on the record $y$, the corresponding post--measurement state of $B_1$ is
\begin{equation}
\hat\varrho_{1|y} := \frac{\mathcal J_y(\hat\rho_1)}{\Tr_{B_1}[\mathcal J_y(\hat\rho_1)]},
\end{equation}
whenever $p(y|\hat\rho_1)>0$.
This conditional update does not yet imply that $\hat\rho$ collapses in any particular sector,
because we have not characterized the structure of the maps $\mathcal J_y$ (and/or
$\widetilde{\mathcal J}_y$) beyond complete positivity and trace conditions; this will be done in the
next sub-section.

The decomposition in Theorem~2 is {\it point-wise} in the initial probe preparation.
More explicitly, for an arbitrary classical probe state $\hat\tau_{y_0}$ supported on $\Pi_{y_0}$ one
should write
\begin{align}
\mathcal M(\hat\rho_1\otimes \hat\tau_{y_0})
& =
\sum_{y\in Y}\widetilde{\mathcal J}^{(\hat\tau_{y_0})}_{y}(\hat\rho_1),
\nonumber \\
\widetilde{\mathcal J}^{(\hat\tau_{y_0})}_{y}(\hat\rho_1)
& =
(\mathbb I_{B_1}\otimes \Pi_y)\,\widetilde{\mathcal J}^{(\hat\tau_{y_0})}_{y}(\hat\rho_1)\,(\mathbb I_{B_1}\otimes \Pi_y),
\end{align}
and correspondingly the induced coarse instrument on $B_1$ is
\begin{equation}
\mathcal J^{(\hat\tau_{y_0})}_y(\cdot):=\Tr_{B_2}\!\big[\widetilde{\mathcal J}^{(\hat\tau_{y_0})}_y(\cdot)\big], \qquad \Phi^{(\hat\tau_{y_0})}:=\sum_{y\in Y}\mathcal J^{(\hat\tau_{y_0})}_y.
\end{equation}
In what follows we fix one reference preparation $\hat\tau_{y_0}$ to lighten notation; allowing for
different initial classical preparations does not change the conclusions, but only relabels the
associated families of maps.

As mentioned earlier, if the projectors $\Pi_y$ have rank $>1$, condition~\eqref{eq:WeakC2} enforces only
block--diagonality across different $y$--sectors, but still allows nontrivial
dynamics within each sector. Consequently, the probe may retain quantum degrees of freedom ``inside'' each classical sector, and the joint state within a
given $y$--block need not factorise.  One may rule out residual quantum behavior within each sector by
adding a stronger classicality postulate (e.g.\ a rank--$1$ refinement), but we do not need to consider this here. In this regard, we note that a similar behaviour is present in the Tilloy--Di{\'o}si measurement--and--feedback model of classical gravity \cite{tilloy2016sourcing} The model does predict entanglement generation on small enough lengthscales \cite{Trillo2025,Angeli2025} however this stems not from the gravitational mediator, which is assumed to be classical, but from the fact that its “measurement” step involves a Gaussian spatial smearing and is therefore not a strictly local operation with respect to the bipartition.

\subsection{Assumption 4: Newtonian (macroscopic) limit}
At macroscopic scales, gravity is accurately described by Newtonian mechanics, and a model of a
classical gravitational interaction should respect this large-scale behavior. In particular, macroscopic objects should approximately follow Newtonian trajectories. Therefore we put forward few minimal assumptions incorporating the interaction, which will allow us to reach the conclusion concerning collapse and momentum  statistics.

Assume the conditions of Theorem~2. Suppose, as in the thought experiment, that Bob reads the probe record $y\in Y$ and then performs the projective
measurement $\{\Pi_x\}_{x\in X}$ on $B_1$.
We assume that the probe provides an
{\it $\varepsilon$--faithful readout} of the sector $x$ in the following sense. Assume that there exist an injective map $g:X\to Y$ and a parameter
$\varepsilon\in[0,1)$ such that for every $x\in X$ with $y_x:=g(x)$
\begin{equation}
\tag{F.1}\label{eq:faithfulness-eps-narrative}
\sum_{y\neq y_x}\Tr_{B_1}\!\big[\mathcal J_y(\hat\sigma_x)\big]\ \le\ \varepsilon\quad \forall\ \hat\sigma_x\in\mathsf S_x\,.
\end{equation}
The inequality says that, for any input state entirely supported in sector $x$, the probe record equals
the ``correct'' value $y_x$ with probability at least $1-\varepsilon$. Moreover, we assume:
\begin{equation}
\tag{F.2}\label{eq:faithfulness-eps-narrative2}
\Tr_{B_1}\!\big[(\mathbb I_{B_1}-\Pi_x)\,\mathcal J_{y_x}(\hat\sigma_x)\big]\ \le\ \varepsilon \qquad \forall\,\hat\sigma_x\in\mathsf S_x.
\end{equation}
This second inequality says that, conditioned on the correct record $y_x$, the system is found outside
sector $x$ with probability at most $\varepsilon$ under Bob's subsequent test
$\{\Pi_x, \mathbb{I}_{B_1}-\Pi_x\}$.

Note that, to keep notation simple, we are implicitly working in  particle $B_1$'s frame, where its position does not change. Moving to a different frame would imply that its classical configurations change due to the gravitational interaction, which is notationally heavy to keep track of.

In probabilistic terms, the conditions~\eqref{eq:faithfulness-eps-narrative}
and~\eqref{eq:faithfulness-eps-narrative2} imply, for every sector-supported input
$\hat\sigma_x\in\mathsf S_x$,
\begin{equation}\label{eq:faithfulness-eps-conditional}
p(y_x | \hat\sigma_x )\ \ge\ 1-\varepsilon,
\qquad
p(x | y_x,\hat\sigma_x)\ \ge\ 1-\frac{\varepsilon}{1-\varepsilon}.
\end{equation}

\begin{figure}[t!]
    \centering
\includegraphics[width=\linewidth]{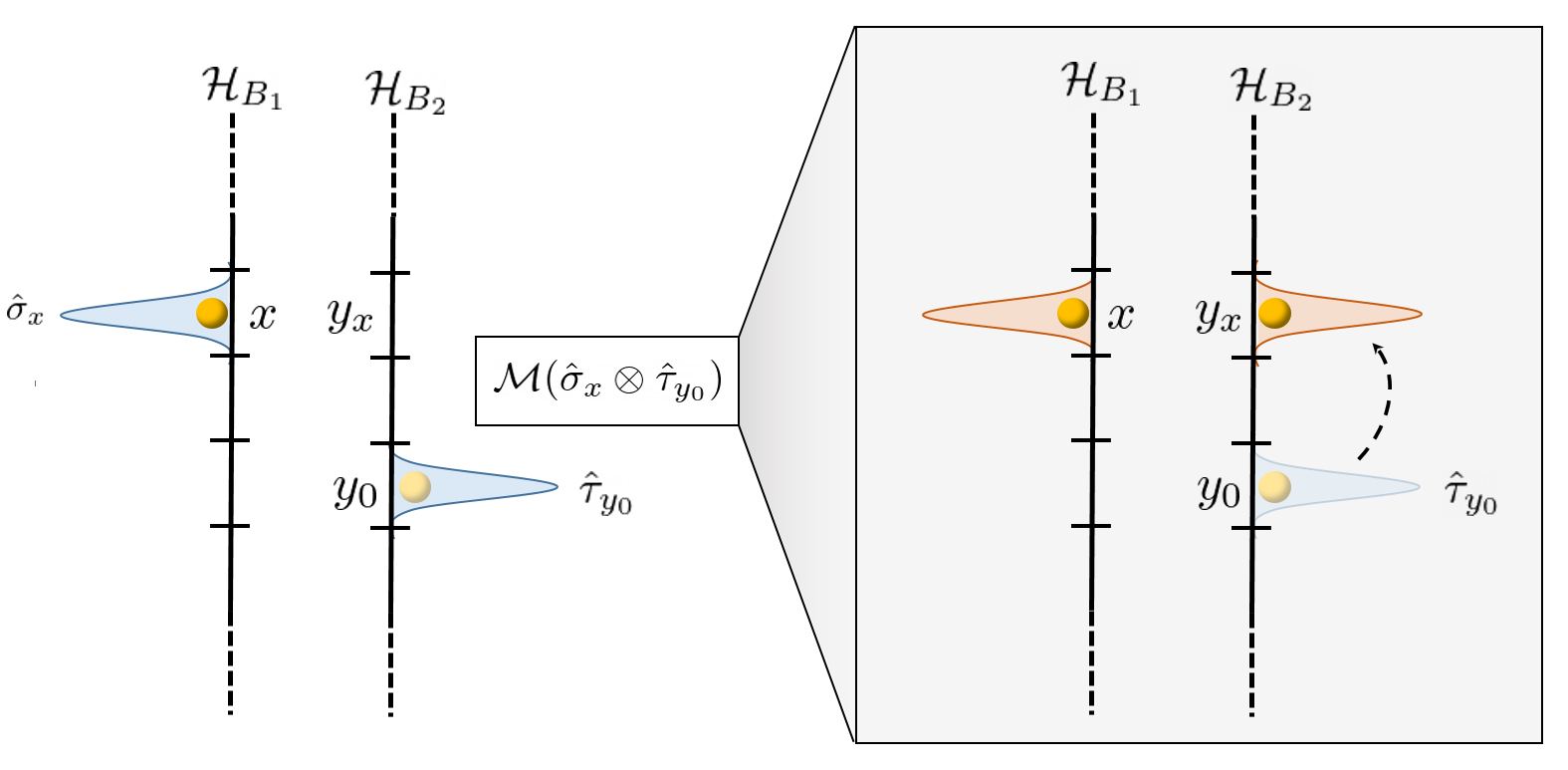}
    \caption{\justifying{\bf Faithfulness. } Schematic representation of the faithfulness conditions. The probe and the particle are prepared in well localized states with respect to their PVMs.  Then, in the frame of particle $B_1$, the probe $B_2$ moves to the sector $y_x=g(x)$, corresponding to the expected classical trajectory (condition~\eqref{eq:faithfulness-eps-narrative}) with a probability $p(y_x | \hat\sigma_x )\ \ge\ 1-\varepsilon$. In such a frame, furthermore, particle $B_1$ almost surely remains localized in its original sector (condition~\eqref{eq:faithfulness-eps-narrative2}) with a probability $p(x | y_x,\hat\sigma_x)\ \ge\ 1-\varepsilon/(1-\varepsilon)$.}
    \label{fig:placeholder}
\end{figure}

In the Gedankenexperiment of the previous Section, we used these conditions when we assumed that, if the spin-particle is deflected upwards (downwards) by the Stern-Gerlach device, then the probe is attracted upwards (downwards)---this is assumption~\eqref{eq:faithfulness-eps-narrative}. The function $y_x = g(x)$ simply assigns to an initial preparation the corresponding final position of the probe according the expected Newtonian trajectory. Implicitly we also assumed that the spin-particle does not flip position for whatever reason---this is assumption~\eqref{eq:faithfulness-eps-narrative2}.

In the following, we will assume $\varepsilon =0$, which we will call exact faithfulness, to keep the results easy to follow. The more general case $\varepsilon \neq 0$ is considered in Appendix~\ref{app:varepsilon}, without altering the general conclusions. We first prove that conditions~\eqref{eq:WeakC2} and~\eqref{eq:faithfulness-eps-narrative} imply that the map $\mathcal M$ cannot be unitary \cite{diosi1995quantum, Diosi2000,oppenheim2023postquantum,oppenheim2023gravitationally,diosi2023hybrid}. The intuitive reason is the following. A unitary evolution is linear and preserves purity: in particular pure states in a superposition, must be mapped to superpositions of pure states. This, however, combined with the assumption that classical states become correlated (i.e. the presence of an interaction, assumptions \eqref{eq:faithfulness-eps-narrative} and \eqref{eq:faithfulness-eps-narrative2})
inevitably leads to the generation of coherences between different classical sectors. 

{\bf Theorem 3:  $\mathcal M$ is not unitary.}
Assume~\eqref{eq:WeakC2} and~\eqref{eq:faithfulness-eps-narrative} with $\varepsilon = 0$, and suppose there exist at least two distinct sectors
$x\neq x'$. Then $\mathcal M$ is not unitary.

\begin{proof}
Suppose by contradiction that $\mathcal M$ is unitary, i.e.
\begin{equation}
\mathcal M(\cdot)=U(\cdot)U^\dagger
\end{equation}
for some unitary $U$ on $\mathcal H_{B_1}\otimes\mathcal H_{B_2}$. Since $g$ is injective, $y_x:=g(x)$ and $y_{x'}:=g(x')$ are distinct and we can choose unit vectors
\begin{equation}
\ket{\phi_x}\bra{\phi_x}\in \mathsf S_x,
\qquad
|\phi_{x'}\rangle\bra{\phi_{x'}}\in \mathsf S_{x'}.
\end{equation}
Let $\hat\tau_{y_0}=\sum_i p_i\,|\eta_i\rangle\!\langle\eta_i|$
be a spectral decomposition of the fixed probe preparation, with
$|\eta_i\rangle\bra{\eta_i}\in\mathsf S_{y_0}$. By exact ($\varepsilon = 0$) faithfulness \eqref{eq:faithfulness-eps-narrative}, for every normalized
$\hat\sigma_x\in\mathsf S_x$ one has
\begin{equation}
\Tr_{B_1}\!\big[\mathcal J_{y_x}(\hat\sigma_x)\big]=1.
\end{equation}
Applying this to $\hat\sigma_x=|\phi_x\rangle\!\langle\phi_x|$ yields
\begin{align}\label{N1_uinit}
1
& =
\Tr_{B}\!\Big[(\mathbb I_{B_1}\otimes \Pi_{y_x})\,
U\big(|\phi_x\rangle\!\langle\phi_x|\otimes \hat\tau_{y_0}\big)
U^\dagger\Big]
\nonumber \\
&=
\sum_i p_i\,\big\|(\mathbb I_{B_1}\otimes \Pi_{y_x})\,U(|\phi_x\rangle\otimes|\eta_i\rangle)\big\|^2.
\end{align}
Since each norm in \eqref{N1_uinit} is bounded above by $1$ and $\sum_{i}p_i =1$, every term multiplying $p_i>0$ must itself equal $1$. Thus,
for all $i$ such that $p_i>0$,
\begin{equation}
U(|\phi_x\rangle\otimes|\eta_i\rangle)
\in
\mathcal H_{B_1}\otimes \left.\mathcal{H}_{B_2}\right|_{\Pi_{y_x}};
\end{equation}
The same argument applied to $x'$ gives
\begin{equation}
U(|\phi_{x'}\rangle\otimes|\eta_i\rangle)
\in
\mathcal H_{B_1}\otimes \left.\mathcal{H}_{B_2}\right|_{\Pi_{y_{x'}}}.
\end{equation}

Fix one $i$ with $p_i>0$ and define
\begin{equation}
|\psi\rangle:=\frac{|\phi_x\rangle+|\phi_{x'}\rangle}{\sqrt2}.
\end{equation}
Then, by linearity of $U$,
\begin{equation}
U(|\psi\rangle\otimes|\eta_i\rangle)
=
\frac{
U(|\phi_x\rangle\otimes|\eta_i\rangle)
+
U(|\phi_{x'}\rangle\otimes|\eta_i\rangle)
}{\sqrt2}.
\end{equation}
The two addends are nonzero and belong to the orthogonal probe sectors
$\Pi_{y_x}$ and $\Pi_{y_{x'}}$, respectively. Therefore the pure output state
\begin{equation}
U\big(|\psi\rangle\!\langle\psi|\otimes|\eta_i\rangle\!\langle\eta_i|\big)U^\dagger
\end{equation}
has nonvanishing off-diagonal blocks between the distinct probe sectors $y_x$ and $y_{x'}$. However, $|\eta_i\rangle\!\langle\eta_i|$ is supported on the classical sector $\Pi_{y_0}$, so the
classicality condition~\eqref{eq:WeakC2} applies and requires
\begin{equation}
\Gamma_Y\!\left(\mathcal M\big(|\psi\rangle\!\langle\psi|\otimes|\eta_i\rangle\!\langle\eta_i|\big)\right)=\mathcal M\big(|\psi\rangle\!\langle\psi|\otimes|\eta_i\rangle\!\langle\eta_i|\big).
\end{equation}
This is a contradiction, hence $\mathcal M$ cannot be unitary.
\end{proof}

Theorem~3 restricts the set of possible maps $\mathcal{M}$, but does not specify its action. With the following theorem, we further characterize the map $\mathcal M$: we show that it must collapse initial superpositions of $B_1$, in agreement with the argument of the previous Section.

{\bf Theorem 4: $\mathcal M$ collapses.}
Assume Theorem~2, so that  $\mathcal M(\hat\rho_1\otimes \hat\tau_{y_0})=\sum_{y\in Y}\widetilde{\mathcal J}_y(\hat\rho_1)$, and that the faithfulness conditions
\eqref{eq:faithfulness-eps-narrative}--\eqref{eq:faithfulness-eps-narrative2}
hold exactly for an injective map $g:X\to Y$; then:

\smallskip
\noindent
(i) For each $x\in X$, the image of the branch $\widetilde{\mathcal J}_{y_x}$, with $y_x:=g(x)$, has support in the joint sector
$\Pi_x\otimes \Pi_{y_x}$ and depends only on the $x$--block of the input:
\begin{equation}\label{eq:lifted_sector_support_thm4}
\widetilde{\mathcal J}_{y_x}(\hat\rho_1)
=
(\Pi_x\otimes \Pi_{y_x})\,
\widetilde{\mathcal J}_{y_x}(\hat\rho_1)\,
(\Pi_x\otimes \Pi_{y_x})
=
\widetilde{\mathcal J}_{y_x}(\Pi_x\hat\rho_1\Pi_x),
\end{equation}
$\forall\,\hat\rho_1\in\mathcal T(\mathcal H_{B_1})$. Equivalently, there exists a CPTP map
\begin{equation}\label{eq:lifted_Rx_def}
\widetilde{\mathcal R}_x:\mathcal T(\mathcal H_{B_1}|_{\Pi_x})
\longrightarrow
\mathcal T(\mathcal H_B|_{\Pi_x\otimes\Pi_{y_x}})
\end{equation}
such that
\begin{equation}\label{eq:lifted_factorization_thm4}
\widetilde{\mathcal J}_{y_x}(\hat\rho_1)
=
\widetilde{\mathcal R}_x\!\big(\Pi_x\hat\rho_1\Pi_x\big).
\end{equation}

\smallskip
\noindent
(ii) For every input state $\hat\rho_1$ on $B_1$,
\begin{equation}\label{eq:lifted_projective_stats}
p(y_x|\hat\rho_1)
:=
\Tr_{B}\!\big[\widetilde{\mathcal J}_{y_x}(\hat\rho_1)\big]
=
\Tr_{B_1}[\Pi_x\hat\rho_1],
\end{equation}

\smallskip
\noindent
(iii) The unconditional output of $\mathcal M$ for the fixed probe preparation is
\begin{equation}\label{eq:lifted_unconditional_output}
\mathcal M(\hat\rho_1\otimes \hat\tau_{y_0})
=
\sum_{x\in X}\widetilde{\mathcal R}_x\!\big(\Pi_x\hat\rho_1\Pi_x\big),
\end{equation}
hence it is block--diagonal in the joint decomposition
$\{\Pi_x\otimes \Pi_{y_x}\}_{x\in X}$.

\smallskip
Before proving the theorem we comment on its meaning. (i) implies that given a generic state on $B_1$ the output has lost any coherence between sectors: any superposition that might have been present, as in the Gedankenexperiment, has collapsed due to the gravitational interaction with the probe. Moreover, (ii) guarantees that the probe readout statistics coincide with the projective measurement
$\{\Pi_x\}_{x\in X}$.
Therefore, whenever $p(y_x|\hat\rho_1)>0$, the corresponding conditional post--measurement state on
$B$ is
\begin{equation}\label{eq:lifted_conditional_state}
\hat\varrho_{B|y_x}
:=
\frac{\widetilde{\mathcal J}_{y_x}(\hat\rho_1)}{\Tr_B[\widetilde{\mathcal J}_{y_x}(\hat\rho_1)]}
=
\frac{\widetilde{\mathcal R}_x(\Pi_x\hat\rho_1\Pi_x)}{\Tr_{B_1}[\Pi_x\hat\rho_1]},
\end{equation}
and satisfies
\begin{equation}\label{eq:lifted_joint_support}
\hat\varrho_{B|y_x}
=
(\Pi_x\otimes\Pi_{y_x})\,\hat\varrho_{B|y_x}\,(\Pi_x\otimes\Pi_{y_x}).
\end{equation}
Thus the branch associated with $y_x$ collapses exactly into the joint sector
$x$ on $B_1$ and $y_x$ on $B_2$. Correspondingly (iii) characterizes the full map $\mathcal{M}$ as the non-selective result of these measurements.

\smallskip
{\it Proof. ---}
We now prove point (i). Define the effects
\begin{equation} \label{eq:effects} 
\widetilde E_y:=\widetilde{\mathcal J}_y^\dagger(\mathbb I_B)\ge 0.
\end{equation} 
Since $\sum_y\widetilde{\mathcal J}_y$ is trace-preserving, one has $
\sum_{y\in Y}\widetilde E_y=\mathbb I_{B_1}$. Fix $x\in X$ and  let $\hat\sigma_x\in\mathsf S_x$; in terms of the effects $\tilde E_y$, the exact faithfulness condition~\eqref{eq:faithfulness-eps-narrative} reads:
\begin{align}
\Tr_{B_1}[\widetilde E_{y_x}\hat\sigma_x]&=1,
\nonumber\\
\Tr_{B_1}[\widetilde E_{y}\hat\sigma_x]&=0,
\qquad (y\neq y_x).
\end{align}
Consider all $y\neq y_x$, then the relation
\(
\Tr_{B_1}[\widetilde E_y\hat\sigma_x]=0
\)
for all normalized $\hat\sigma_x\in\mathsf S_x$ implies
\begin{equation}\label{TildeE_pi_x}
\Pi_x\widetilde E_y\Pi_x=0.
\end{equation}
Moreover, since $\tilde E_y\ge 0$ it admits a square root $\tilde E_y = \tilde E_y^{1/2}\tilde E_y^{1/2}$ and therefore
\begin{equation}
\Pi_x\widetilde E_y\Pi_x
=
\Pi_x\widetilde E_y^{1/2}\widetilde E_y^{1/2}\Pi_x
=
(\widetilde E_y^{1/2}\Pi_x)^\dagger(\widetilde E_y^{1/2}\Pi_x),
\end{equation}
which implies
\begin{equation}\label{E00}
\widetilde E_y\Pi_x=0,
\;\;(y\neq y_x).   
\end{equation}
Furthermore, using the completeness of the effects $\sum_y\widetilde E_y=\mathbb I_{B_1}$, we get
\begin{equation}\label{EPi}
\widetilde E_{y_x}\Pi_x=\Pi_x. 
\end{equation}

Using Eq. (\ref{E00}) and Eq. (\ref{EPi}) and the completeness of the projectors $\Pi_x$ we have 
\begin{equation}\label{eq:lifted_effect_identity}
\widetilde{E}_{y_{x}}=\widetilde{E}_{y_{x}}\sum_{x'}\Pi_{x'}=\widetilde{E}_{y_{x}}\Pi_{x}=\Pi_{x}\qquad \forall\,x\in X
\end{equation}

Thus, the effect corresponding to a particular outcome $y_x$ is precisely the projector $\Pi_x$. Since $\widetilde{\mathcal{J}}_{y_x}$ is CP it admits a Kraus representation
\begin{equation}
\widetilde{\mathcal J}_{y_x}(\cdot)
=
\sum_\alpha K_{x,\alpha}(\cdot)K_{x,\alpha}^\dagger,
\qquad
K_{x,\alpha}:\mathcal H_{B_1}\to \mathcal H_B.
\end{equation}
Since we have proven that $\widetilde E_{y_x} = \Pi_x$, we have
\begin{equation}\label{kraus_Pi}
\sum_\alpha K_{x,\alpha}^\dagger K_{x,\alpha}=\Pi_x.
\end{equation}
Projecting Eq.~\eqref{kraus_Pi} onto  $\mathbb{I}_{B_1}-\Pi_x$, we obtain
\begin{equation}
    \sum_{\alpha}\big[K_{x,\alpha}(\mathbb{I}_{B_1}-\Pi_x)\big]^\dagger\big[K_{x,\alpha}(\mathbb{I}_{B_1}-\Pi_x)\big] = 0,
\end{equation}
which, in turn, implies
\begin{equation}\label{eq:K_right_support_lifted}
K_{x,\alpha}(\mathbb I_{B_1}-\Pi_x)=0,
\qquad\text{i.e.}\qquad
K_{x,\alpha}=K_{x,\alpha}\Pi_x.
\end{equation}
This characterization of the right action of the Kraus operators specifies the support of each $\widetilde{\mathcal J}_{y_x}$. We now move onto the characterization of the output. By Theorem~2 each branch $\widetilde{\mathcal J}_{y_x}$ has image in the probe sector
$\Pi_{y_x}$, so we may choose the Kraus operators so that
\begin{equation}\label{eq:K_probe_support_lifted}
K_{x,\alpha}
=
(\mathbb I_{B_1}\otimes \Pi_{y_x})K_{x,\alpha}.
\end{equation}

Let now $\hat{\sigma}_x = \ket{\phi}\bra{\phi}\in\mathsf S_x$, with $\ket{\phi}$ a unit vector, and apply exact
\eqref{eq:faithfulness-eps-narrative2} :
\begin{align}
0
&=\Tr_{B_1}\!\big[(\mathbb I_{B_1}-\Pi_x)\mathcal J_{y_x}(\hat\sigma_x)\big]
\nonumber\\
&=\Tr_{B}\!\big[((\mathbb I_{B_1}-\Pi_x)\otimes\mathbb I_{B_2})
\,\widetilde{\mathcal J}_{y_x}(\hat\sigma_x)\big]
\nonumber\\
&=\sum_\alpha \big\|
((\mathbb I_{B_1}-\Pi_x)\otimes\mathbb I_{B_2})K_{x,\alpha}|\phi\rangle
\big\|^2.
\end{align}
Hence $
((\mathbb I_{B_1}-\Pi_x)\otimes\mathbb I_{B_2})K_{x,\alpha}=0$ which, combined with \eqref{eq:K_right_support_lifted} and \eqref{eq:K_probe_support_lifted} gives
\begin{equation}
K_{x,\alpha}
=
(\Pi_x\otimes\Pi_{y_x})K_{x,\alpha}\Pi_x,
\qquad \forall\,\alpha.
\end{equation}
Therefore
\begin{equation}
\widetilde{\mathcal J}_{y_x}(\hat\rho_1)=(\Pi_x\otimes \Pi_{y_x})\,
\widetilde{\mathcal J}_{y_x}(\hat\rho_1)\,
(\Pi_x\otimes \Pi_{y_x})=\widetilde{\mathcal J}_{y_x}(\Pi_x\hat\rho_1\Pi_x),
\end{equation}
which proves the first claim of (i). We can now define
\begin{equation}
\widetilde{\mathcal R}_x(\hat\sigma)
:=\widetilde{\mathcal J}_{y_x}(\hat\sigma),
\qquad
\hat\sigma\in \mathcal T(\mathcal H_{B_1}|_{\Pi_x}).
\end{equation}
Since $\widetilde{\mathcal J}_{y_x}$ is CP,
also $\widetilde{\mathcal R}_x$ is CP. Moreover, for every
$\hat\sigma=\Pi_x\hat\sigma\Pi_x$,
\begin{align}
\Tr_B[\widetilde{\mathcal R}_x(\hat\sigma)]
& =
\Tr_B[\widetilde{\mathcal J}_{y_x}(\hat\sigma)]
=
\Tr_{B_1}[\widetilde E_{y_x}\hat\sigma]
\nonumber \\
& =
\Tr_{B_1}[\Pi_x\hat\sigma]
=
\Tr_{B_1}[\hat\sigma],
\end{align}
where we used \eqref{eq:lifted_effect_identity}. Hence $\widetilde{\mathcal R}_x$ is CPTP on its domain, completing the proof of (i).

Point (ii) follows immediately from the relation between the map $\widetilde{\mathcal{J}}_{y_x}$ and the effect $\widetilde{E}_y$ in \eqref{eq:effects}, and the relation $\widetilde{E}_{y_x} = \Pi_x$ [cf. \eqref{eq:lifted_effect_identity}].

As for point (iii), recall that the map $\mathcal{M}$ is defined as
\begin{equation}\label{M_iv}
    \mathcal{M}(\hat\rho_1\otimes\tau_{y_0}) = \sum_{y\in Y}\widetilde{\mathcal{J}}_y(\hat\rho_1).
\end{equation}

Notice that if $y\notin g(X)$ then $y\neq y_x$ for every $x$. Then Eq.~\eqref{E00} summed over all $x\in X$ implies $\widetilde E_y=0.$
Since $\widetilde{\mathcal J}_y$ is CP, this further implies
 $\tilde{\mathcal{J}}_y(\cdot) =0$. Indeed, for every positive trace--class operator
$\hat\tau\ge 0$,
\begin{equation}
\Tr_B[\widetilde{\mathcal J}_y(\hat\tau)]=\Tr_{B_1}[\widetilde E_y\hat\tau]
=0,
\end{equation}
but $\widetilde{\mathcal J}_y(\hat\tau)\ge 0\ \forall \hat\tau \ge0$, thus
$\widetilde{\mathcal J}_y=0$.
Finally, inserting this in~\eqref{M_iv}
\begin{align}
    \mathcal{M}(\hat\rho_1\otimes\tau_{y_0}) &= \sum_{y\in Y}\widetilde{\mathcal{J}}_y(\hat\rho_1)\nonumber\\
    &= \sum_{x\in X}\widetilde{\mathcal{J}}_{y_x}(\hat\rho_1)\nonumber\\
    &= \sum_{x\in X}\widetilde{\mathcal R}_x\!\big(\Pi_x\hat\rho_1\Pi_x\big).
\end{align}
The block--diagonality
of $\mathcal{M}$ is immediate because each summand
$\widetilde{\mathcal R}_x(\Pi_x\hat\rho_1\Pi_x)$ is supported in the single joint sector
$\Pi_x\otimes\Pi_{y_x}$. \qed

A similar result, whose proof follows closely that of Theorem~4, holds for $\mathcal{J}_y(\cdot)$ and the reduced map $\Phi$.

{\bf Corollary 1: reduced collapse on $B_1$.}
Assume the hypotheses of Theorem~4, and let
\begin{equation}
\mathcal J_y(\hat\rho_1):=\Tr_{B_2}\!\big[\widetilde{\mathcal J}_y(\hat\rho_1)\big],
\qquad
\Phi:=\sum_{y\in Y}\mathcal J_y .
\end{equation}
Then:

\smallskip
\noindent
(i) For each $x\in X$ and for all $\hat\rho_1\in\mathcal T(\mathcal H_{B_1})$, the branch $\mathcal J_{y_x}$, with $y_x:=g(x)$, has support in the $x$--sector:
\begin{equation}\label{eq:cor_sector_support_reduced}
\mathcal J_{y_x}(\hat\rho_1)=\Pi_x\,\mathcal J_{y_x}(\hat\rho_1)\,\Pi_x=\mathcal{J}_{y{_x}}(\Pi_x\hat\rho_1\Pi_x).
\end{equation}
Equivalently, there exists a CPTP map $\mathcal R_x$ on the sector algebra
$\mathcal T(\mathcal H_{B_1}|_{\Pi_x})$ such that
\begin{equation}\label{eq:cor_reduced_factorization}
\mathcal J_{y_x}(\hat\rho_1)=\mathcal R_x\!\big(\Pi_x\hat\rho_1\Pi_x\big).
\end{equation}

\smallskip
\noindent
(ii) For every input state $\hat\rho_1$ on $B_1$,
\begin{equation}\label{eq:cor_projective_stats_reduced}
p(y_x|\hat\rho_1)
:=
\Tr_{B_1}\!\big[\mathcal J_{y_x}(\hat\rho_1)\big]
=
\Tr_{B_1}[\Pi_x\hat\rho_1],
\end{equation}
i.e.\ the probe readout statistics coincide with the projective measurement
$\{\Pi_x\}_{x\in X}$.

\smallskip
\noindent
(iii) The reduced map $\Phi$ satisfies
\begin{equation}\label{eq:cor_unconditional_reduced}
\Phi(\hat\rho_1)
=
\sum_{x\in X}\mathcal R_x\!\big(\Pi_x\hat\rho_1\Pi_x\big),
\end{equation}
hence $\Phi(\hat\rho_1)$ is block--diagonal in the $\{\Pi_x\}$ decomposition, i.e. $
\Gamma_X\!\big(\Phi(\hat\rho_1)\big) = \Phi(\hat\rho_1)$ $\forall\,\hat\rho_1$.

\smallskip
In the exact-faithfulness decomposition~\eqref{eq:cor_reduced_factorization},
the map $\mathcal R_x$ describes the residual dynamics {\it within} the classical sector $x$. If the mediator is genuinely classical, one expects that, once the sector $x$ has been read out, the mediator should not be able to coherently control or manipulate quantum coherences inside it. Then, modulo the free evolution, $\mathcal R_x= \mathbb{I}_x$, or at most a map acting on sector populations, without implementing coherent intra-sector quantum processing. This is the case of collapse models (ignoring the tails of the gaussian smearing) \cite{Ghirardi1986,Pearle1989,Ghirardi1990} and of classical gravity as a measurement-and-feedback mechanism \cite{tilloy2016sourcing} where the operators are smeared out with suitable functions.

As a matter of fact, the dynamics on $B_1$  is mathematically
equivalent to the following two-step scheme. First, one performs a random collapse with collapse
operators $\{\Pi_x\}_{x\in X}$, obtaining outcome $x$ with Born probability
\begin{equation}
p_x=\Tr_{B_1}(\Pi_x\hat\rho_1),
\end{equation}
and conditional (selective) post-collapse state
\begin{equation}
\hat\rho_1 \longmapsto \frac{\Pi_x\hat\rho_1\Pi_x}{\Tr_{B_1}(\Pi_x\hat\rho_1)} \qquad (p_x>0).
\end{equation}
Second, conditioned on the same outcome $x$, one applies a sector-internal CPTP map $\mathcal R_x$.
Averaging over outcomes reproduces the unconditional map
$\hat\rho_1\mapsto\sum_{x\in X}\mathcal R_x(\Pi_x\hat\rho_1\Pi_x)$. In the special case $\mathcal R_x=\mathbb{I}_x$
for all $x$, one recovers the standard non-selective projective collapse. Sharp collapses generated by projectors $\Pi_x$ induce discontinuities at sector boundaries,
leading to unphysical high-momentum tails. This artifact can be avoided by replacing the sharp
projectors with suitably smeared localization operators by choosing $\mathcal R_x$ accordingly.

We have proven that the map $\mathcal{M}$ collapses the states onto the sectors of the privileged PVM $\{\Pi_x\}_{x\in X}$, but we have yet to specify to what observable these projectors correspond to. The answer is dictated by the dynamics in the semiclassical limit, codified in the faithfulness conditions~\eqref{eq:faithfulness-eps-narrative}-\eqref{eq:faithfulness-eps-narrative2}. In this regime we know that the gravitational interaction couples the {\it positions} of the masses, and the trajectories, described by the function $g$ should be Newtonian classical trajectories. Thus the sectors $\Pi_x$  ought to be coarse grained {\it positions}. Having established this, now we turn to the change in momentum; to this end, we have to consider explicitly that the Hilbert space is $L^2(\mathbb R)$. We will first show that the total momentum cannot be conserved, in spite of the fact that the global $B = B_1B_2$ system is closed.

To simplify the analysis, we fix an interval coarse--graining of width $\ell>0$:
\begin{equation}\label{eq:interval_projectors_main}
I_n:=[n\ell,(n+1)\ell),\qquad 
\Pi_n:=\int_{I_n}\!|x\rangle\!\langle x|\,dx,
\qquad n\in\mathbb Z .
\end{equation}
Let $\Gamma_\ell$ be the associated dephasing channel,
\begin{equation}\label{eq:Gamma_ell_def}
\Gamma_\ell(\hat\rho_1):=\sum_{n\in\mathbb Z}\Pi_n\,\hat\rho_1\,\Pi_n .
\end{equation}

{\bf Theorem 5: Total momentum statistics change.}
Assume $\mathcal H_{B_1}=\mathcal H_{B_2}=L^2(\mathbb R)$, and assume the hypotheses of Theorem~4 for the interval coarse--graining
\eqref{eq:interval_projectors_main}. Fix a classical probe preparation $\hat{\tau}_{y_0} \in \mathsf{S}_{y_0}$ and denote the total momentum operator on $\mathcal H_{B_1}\otimes\mathcal H_{B_2}$ by
\begin{equation}
\hat P_{\mathrm{tot}}:=\hat p_1+\hat p_2.
\end{equation}
Then there exists a density operator $\hat\rho_1\in\mathcal T(\mathcal H_{B_1})$ such that the
probability distribution of $\hat P_{\mathrm{tot}}$ in the output state
$\mathcal M(\hat\rho_1\otimes\hat\tau_{y_0})$
differs from that in the input $
\hat\rho_1\otimes\hat\tau_{y_0}$.
\medskip
\\
{\it Proof. ---}
For a trace--class operator $\hat\varrho$ on
$\mathcal H_{B_1}\otimes\mathcal H_{B_2}$, define the total--momentum characteristic function
\begin{equation}\label{eq:Xi_tot_def_rewritten}
\Xi_{\hat\varrho}(d)
:=
\Tr_{B}\!\left[\hat\varrho\,e^{-\,\frac{i}{\hbar}d(\hat p_1+\hat p_2)}\right],
\qquad d\in\mathbb R.
\end{equation}
This function determines uniquely the probability distribution of
$\hat P_{\mathrm{tot}}$. By Theorem~4 one has
\begin{equation}\label{eq:M_depends_on_blocks_only_rewritten}\mathcal M(\hat\rho_1\otimes\hat\tau_{y_0})
=\sum_{n\in\mathbb Z}\widetilde{\mathcal R}_n\!\big(\Pi_n\hat\rho_1\Pi_n\big),
\end{equation}
hence the output depends only on the block--diagonal part $
\Gamma_\ell(\hat\rho_1)$. Therefore, if two inputs $\hat\rho_1,\hat\sigma_1$ satisfy $
\Gamma_\ell(\hat\rho_1)=\Gamma_\ell(\hat\sigma_1),$
then
\begin{equation}\label{eq:same_output_same_blocks_rewritten}
\mathcal M(\hat\rho_1\otimes\hat\tau_{y_0})=\mathcal M(\hat\sigma_1\otimes\hat\tau_{y_0}).
\end{equation}
In particular, the corresponding output characteristic functions coincide identically. We now construct two input states with the same block--diagonal part but different
initial total--momentum statistics. Choose two adjacent intervals, say $I_n=[n\ell,(n+1)\ell)$ and
$I_{n+1}=[(n+1)\ell,(n+2)\ell)$.
Pick normalized vectors
\begin{equation}
|\phi_n\rangle\in \left.\mathcal{H}_{B_1}\right|_{\Pi_{n}},
\qquad
|\phi_{n+1}\left.\rangle\in\mathcal{H}_{B_1}\right|_{\Pi_{n+1}},
\end{equation}
with compact supports contained in small neighborhoods of the common boundary
$x=(n+1)\ell$, chosen so that for some $d$ with $0<d<\ell$ one has
\begin{equation}\label{eq:cross_term_nonzero_rewritten}
\langle \phi_{n+1}|e^{-\,\frac{i}{\hbar}d\hat p_1}|\phi_n\rangle\neq 0,
\end{equation}
while
\begin{equation}\label{eq:cross_term_zero_rewritten}
\langle \phi_n|e^{-\,\frac{i}{\hbar}d\hat p_1}|\phi_{n+1}\rangle=0.
\end{equation}
Condition \eqref{eq:cross_term_zero_rewritten} holds because
$e^{-\,\frac{i}{\hbar}d\hat p_1}$ translates wave functions to the right by $d$,
so for $0<d<\ell$ a vector supported in $I_{n+1}$ remains disjoint from $I_n$ after translation.

Define
\begin{equation}
|\psi_\pm\rangle:=\frac{|\phi_n\rangle\pm|\phi_{n+1}\rangle}{\sqrt2},
\qquad
\hat\rho_\pm:=|\psi_\pm\rangle\!\langle\psi_\pm|.
\end{equation}
Both states have the same block--diagonal parts:
\begin{equation}\label{eq:blockdiag_same_pm_rewritten}
\Gamma_\ell(\hat\rho_+)=\Gamma_\ell(\hat\rho_-)
=\frac12\,|\phi_n\rangle\!\langle\phi_n|+\frac12\,|\phi_{n+1}\rangle\!\langle\phi_{n+1}|.
\end{equation}
Hence $\mathcal M(\hat\rho_+\otimes\hat\tau_{y_0})=\mathcal M(\hat\rho_-\otimes\hat\tau_{y_0}),$
so the two outputs have exactly the same total--momentum statistics.

On the other hand, for a product input $\hat\varrho=\hat\rho_1\otimes\hat\tau_{y_0}$, the characteristic
function factorizes:
\begin{equation}\label{eq:product_factorization_tot_rewritten}
\Xi_{\hat\rho_1\otimes\hat\tau_{y_0}}(d)
=\chi^{(1)}_{\hat\rho_1}(d)\,
\chi^{(2)}_{\hat\tau_{y_0}}(d),
\end{equation}
where
\begin{align}\label{chi1}
\chi^{(1)}_{\hat\rho_1}(d)
&:=\Tr_{B_1}\!\left[\hat\rho_1\,e^{-\,\frac{i}{\hbar}d\hat p_1}\right],\\
\chi^{(2)}_{\hat\tau_{y_0}}(d)
&:=\Tr_{B_2}\!\left[\hat\tau_{y_0}\,e^{-\,\frac{i}{\hbar}d\hat p_2}\right].
\end{align}
A direct expansion gives
\begin{align}
\chi^{(1)}_{\hat\rho_+}(d)\!-\!\chi^{(1)}_{\hat\rho_-}(d)
&=\langle \phi_n|e^{-\,\frac{i}{\hbar}d\hat p_1}|\phi_{n+1}\rangle
\!+\!\langle \phi_{n+1}|e^{-\,\frac{i}{\hbar}d\hat p_1}|\phi_n\rangle
\nonumber\\
&=\langle \phi_{n+1}|e^{-\,\frac{i}{\hbar}d\hat p_1}|\phi_n\rangle
\neq 0,
\end{align}
where in the second line we used
\eqref{eq:cross_term_zero_rewritten} and \eqref{eq:cross_term_nonzero_rewritten}. Moreover, $\chi^{(2)}_{\hat\tau_{y_0}}(0)=\Tr_{B_2}[\hat\tau_{y_0}]=1$,
so by continuity there exists a choice of $d\in(0,\ell)$ for which $\chi^{(2)}_{\hat\tau_{y_0}}(d)\neq 0$.
Therefore
\begin{equation}
\Xi_{\hat\rho_+\otimes\hat\tau_{y_0}}(d)
\neq \Xi_{\hat\rho_-\otimes\hat\tau_{y_0}}(d).
\end{equation}
Thus the two input states have different total--momentum statistics, whereas their outputs coincide and hence have the same total--momentum statistics. If $\mathcal M$ were such that it preserved the total--momentum distribution for both inputs
$\hat\rho_+\otimes\hat\tau_{y_0}$ and $\hat\rho_-\otimes\hat\tau_{y_0}$, we should have different total--momentum statistics after applying the map, which is not possible since $
\mathcal M(\hat\rho_+\otimes\hat\tau_{y_0})
=\mathcal M(\hat\rho_-\otimes\hat\tau_{y_0})$. Therefore, for at least one choice
\begin{equation} \label{rhochoices}
\hat\rho_1\in\{\hat\rho_+,\hat\rho_-\},
\end{equation}
the probability distribution of $\hat P_{\mathrm{tot}}$ changes under
$\mathcal M$. \qed

Theorem~5 concerns the full statistics of the total momentum on $B_1B_2$. We now record, for completeness, an immediate consequence of Corollary~1 at the level of the reduced dynamics on $B_1$.

\medskip
\noindent
{\bf Corollary 2: momentum change for the reduced dynamics.}
Assume the interval coarse--graining
\eqref{eq:interval_projectors_main}; assume the hypotheses of Corollary~1, so that the reduced map $\Phi$ on $B_1$ satisfies
\begin{equation}
\Phi(\hat\rho_1)=\sum_{n\in\mathbb Z}\mathcal R_n\!\big(\Pi_n\hat\rho_1\Pi_n\big),
\qquad
\Gamma_\ell\big(\Phi(\hat\rho_1)\big)=\Phi(\hat\rho_1),
\end{equation}
with $\Gamma_\ell$ defined in \eqref{eq:Gamma_ell_def}. 
Then, consider $\hat\rho_1\in\mathcal T(\mathcal H_{B_1})$ such that  there exists a $d\in\mathbb R,\ |d|>\ell$ for which $\chi_{\hat\rho_1}(d)\neq 0$; then $\chi_{\Phi\rho_1}(d)\neq \chi_{\rho_{1}}(d)$ and therefore the momentum distribution of $B_1$ changes under $\Phi$.

\smallskip
{\it Proof. ---}
Firstly notice that the momentum characteristic function can be written as:
\begin{equation}\label{eq:chi_def}
\chi_{\hat\rho_1}(d)=\int_{\mathbb R}\langle x|\hat\rho_1|x+d\rangle\,dx,
\qquad d\in\mathbb R.
\end{equation}
Since $\Phi(\hat\rho_1)$ is block--diagonal in the interval decomposition, its position kernel
vanishes across different cells: $\langle x|\Phi(\hat\rho_1)|x'\rangle=0$
whenever $x$ and $x'$ belong to different intervals $I_n$.
If $|d|>\ell$, then for every $x$ the points $x$ and $x+d$ cannot belong to the same interval
$I_n$, hence
\begin{equation}\label{eq:nonvanishing_condition}
\langle x|\Phi(\hat\rho_1)|x+d\rangle=0,
\qquad\forall\,x\in\mathbb R .
\end{equation}
Therefore, if \eqref{eq:nonvanishing_condition} holds, then $\chi_{\Phi(\hat\rho_1)}$ and $\chi_{\hat\rho_1}$ differ on the set of $|d|>\ell$ for which $\chi_{\rho_1}(d)\neq 0$. Since the momentum distribution is the Fourier transform of the characteristic function, the two momentum distributions cannot coincide. \qed

Note that the interval decomposition introduced above should be understood as an auxiliary coarse--graining, not as a fundamental preferred origin in space. More precisely, and with reference to the partition of $\mathbb R$, for every translated partition
\begin{align}
I_n^{(\delta)} &:=[\delta+n\ell,\delta+n+1)\ell),
\nonumber \\
\Pi_n^{(\delta)}&:=\int_{I_n^{(\delta)}} |x\rangle\!\langle x|\,dx,
\end{align}
$\delta\in[0,\ell)$, together with a correspondingly translated classical probe preparation $\tau_{y_0}^{(\delta)}$, the conclusions of Theorems~2--5 and Corollaries~1--2 hold {\it pointwise} with the replacements $\Pi_n\mapsto \Pi_n^{(\delta)}$, $\tau_{y_0}\mapsto \tau_{y_0}^{(\delta)}$, and therefore with a family of maps $\mathcal{M}^{(\delta)},\ \Phi^{(\delta)}$ rather than a single offset--independent channel. In this sense, the underlying joint map $\mathcal M$ may be kept fixed, while the classical sector structure and the corresponding probe preparation are translated together.

On physical grounds, one may further require that $\mathcal M$ be covariant under common spatial translations and Galilean boosts. Writing
\begin{equation}
T_B(a):=T_{B_1}(a)\otimes T_{B_2}(a),\qquad 
T_{B_j}(a):=e^{-\,\frac{i}{\hbar}a\hat p_j},
\end{equation}
translational covariance means 
\begin{equation}\label{Tcov}
\mathcal M\!\big(T_B(a)\,\hat X\,T_B(a)^\dagger\big)=T_B(a)\,\mathcal M(\hat X)\,T_B(a)^\dagger\,,
\end{equation}
$\forall\,a\in\mathbb R,\ \forall\,\hat X\in\mathcal T(\mathcal H_B)$. Likewise, the standard Galilean boost operator on the two-particle Hilbert space is
\begin{equation}\label{Vcov}
V_B(v):=V_{B_1}(v)\otimes V_{B_2}(v),\  \  V_{B_j}(v)\coloneqq e^{\frac{i}{\hbar}m_j v\hat x_j}\,.
\end{equation}
Since the map $\mathcal M$ describes the evolution over a time interval $t$, boost covariance is expressed as
\begin{align}\label{Bcov}
\mathcal M\!\big(V_B(v)&\,\hat X\,V_B(v)^\dagger\big)
=\nonumber\\
&=T_B(vt)V_B(v)\,\mathcal M(\hat X)\,V_B(v)^\dagger T_B(vt)^\dagger,
\end{align}
$\forall\,v\in\mathbb R$. Notice that the translation operators on the RHS of Eq.~(\ref{Bcov}) account for the spatial shift by $vt$ generated by the boost in the time interval from 0 to $t$. These covariance requirements are independent additional assumptions on the dynamics, and are compatible with the collapse conclusions obtained for each translated coarse--graining.

We now consider what happens to the center--of--mass motion; thus let $M= m_1+m_2$ and
\begin{equation}\label{comCoord}
\begin{cases}
\hat X_C:=(m_1\hat x_1+m_2\hat x_2)/M,
\quad
\hat r:=\hat x_1-\hat x_2,\\\\
\hat P_C:=\hat p_1+\hat p_2,
\qquad
\hat p_r:=(m_2\hat p_1-m_1\hat p_2)/M.
\end{cases}
\end{equation}

Let us rewrite the Hilbert space in center--of--mass and relative coordinates
\begin{equation}
\mathcal H_{B_1}\otimes\mathcal H_{B_2}\simeq \mathcal H_{C}\otimes\mathcal H_{r},
\end{equation}
where we denoted unitary equivalence with $\simeq$. The transformed state is $\hat \rho^C = \mathcal{U}_C\hat\rho = U_{Cr}\hat\rho U_{Cr}^\dagger$. The state in these new coordinates evolves according to the map
\begin{equation}
    \mathcal{M}_{Cr} = \mathcal{U}_{Cr}\circ\mathcal{M}\circ\mathcal{U}_{Cr}^{-1}\,.
\end{equation}
We will assume that the map $\mathcal M$ effectively decouples the center--of--mass ($C$) and relative ($r$) motions~\cite{Note1}, thus
\begin{equation}\label{Cr_Factorization}
    \mathcal{M}_{Cr} = \mathcal{M}_C\otimes\mathcal{M}_r\,,
\end{equation}
with both $\mathcal M_C$ and $\mathcal M_r$ CPTP.  Note the covariance requests of Eqs.~\eqref{Tcov} and~\eqref{Bcov} are nothing but the request of Galilei covariance for $\mathcal{M}_C$.

From the previous results we know that, under the classicality and faithfulness assumptions, the map $\mathcal M$ suppresses certain off--diagonal coherences in the original particle decomposition $B_1|B_2$. We also know that the map $\mathcal{M}_C$ cannot be a simple free evolution since, as we proved in Theorem 5, the statistics of the total momentum $\hat P_\text{tot}$, i.e. $\hat P_C$, must change under the evolution $\mathcal{M}$, at least for some states. Since the free evolution is the only unitary Galilei covariant evolution \cite{FondaGhirardi1970}, which is excluded by Theorem 5, it follows that $\mathcal{M}_C$ must encode an irreversible evolution of the center of mass.

At the present level of generality,  coherence suppression can distribute between center--of--mass and relative coordinates is many ways: in principle, it could be attributed mostly to the relative channel. 
 While this is mathematically possible, a model of this kind would look highly artificial; none of the models in the literature behaves this way~\cite{Note2}. 

Motivated by this and by Theorem 5, we then assume that the map $\mathcal{M}_C$ is capable of discriminating different center-of-mass configurations.  To formalize this, we assume that $\mathcal H_C$ admits a finite family of mutually orthogonal projectors
$\{\Pi^C_\alpha\}_{\alpha\in A}$, and that $\mathcal H_r$ admits a nonzero projector $\Pi^r$. The projectors $\Pi^C_\alpha$ label distinct center--of--mass sectors, while $\Pi^r$ selects one fixed relative sector. We then define the corresponding family of projectors on the original two--particle Hilbert space by
\begin{equation}\label{eq:Cr_branch_projectors}
P_\alpha:=U_{Cr}^\dagger(\Pi^C_\alpha\otimes \Pi^r)U_{Cr},
\end{equation}
with $\alpha\in A$. The branches $\{P_\alpha\}_{\alpha\in A}$ therefore describe different placements of the whole two--particle system while keeping the same internal relative configuration. Define
\begin{equation}\label{eq:Cr_sector_algebra}
P:=\sum_{\alpha\in A}P_\alpha,
\quad
\mathcal T_P:=\{X\in\mathcal T(\mathcal H_{B_1}\otimes\mathcal H_{B_2}) : X=PXP\}.
\end{equation}

Requiring the map $\mathcal M$ to distinguish these branches can be formalized by assuming the existence of CP--TNI maps
\begin{equation}
\{\mathcal J_\alpha:\mathcal T_P\to \mathcal T(\mathcal H_{B_1}\otimes\mathcal H_{B_2})\}_{\alpha\in A}
\end{equation}
such that
\begin{equation}\label{eq:Cr_inst_sum}
\mathcal M(X)=\sum_{\alpha\in A}\mathcal J_\alpha(X),
\qquad \forall\,X\in \mathcal T_P .
\end{equation}

Define the set of normalized states supported in branch $P_\alpha$ by
\begin{equation}\label{eq:Cr_branch_states}
\mathsf S_\alpha^{Cr}
\!\!:=\!\!
\Big\{\!
\hat\sigma\!\in\!\mathcal T(\mathcal H_{B_1}\otimes\mathcal H_{B_2})\! :\!
\hat\sigma\!\ge\! 0,\ \Tr[\hat\sigma]=1,\ \hat\sigma=P_\alpha \hat\sigma P_\alpha
\Big\}.
\end{equation}
We impose the following $\varepsilon$--faithful readout condition: assume that there exists
$\varepsilon\in[0,1)$ such that, for every $\alpha\in A$ and every
$\hat\sigma_\alpha\in \mathsf S_\alpha^{Cr}$,
\begin{equation}\label{eq:Cr_eps_faithfulness1}
\sum_{\beta\neq \alpha}\Tr[\mathcal J_\beta(\hat\sigma_\alpha)]\le \varepsilon,
\end{equation}
meaning --since there is no probe ``reading" the center-of-mass position-- that if the input is supported in the center-of-mass sector $\alpha$, then the total probability assigned to the other classical branches is at most $\varepsilon$. Assuming $\varepsilon = 0$, it is possible to prove, using the same derivation as in Theorem~4, that the off--diagonal coherences between distinct branches are suppressed:
\begin{equation}\label{eq:Cr_offdiag_killed1}
\mathcal M(P_\alpha X P_\beta)=0,
\qquad
\forall\,X\in\mathcal T_P,\ \alpha\neq\beta,
\end{equation}
i.e. the off-diagonal coherences between different center-of-mass branches are suppressed. To connect this statement with the center--of--mass channel, let $\hat\tau_r$ be any normalized state supported in $\Pi^r$, and choose
\begin{equation}
X:=U_{Cr}^\dagger\big((\Pi^C_\alpha Y \Pi^C_\beta)\otimes \hat\tau_r\big)U_{Cr},
\end{equation}
with $\alpha\neq\beta$ and $Y\in\mathcal T(\mathcal H_C)$. Then $X=P_\alpha X P_\beta$, hence \eqref{eq:Cr_offdiag_killed1} gives $\mathcal M(X)=0$. Using the factorization \eqref{Cr_Factorization}, we obtain
\begin{equation}
\mathcal M_C(\Pi^C_\alpha Y \Pi^C_\beta)\otimes \mathcal M_r(\hat\tau_r)=0.
\end{equation}
Since $\hat\tau_r$ is a density operator and $\mathcal M_r$ is CPTP, one has $\mathcal M_r(\hat\tau_r)\neq 0$. Therefore
\begin{equation}\label{eq:C_offdiag_killed}
\mathcal M_C(\Pi^C_\alpha Y \Pi^C_\beta)=0,
\qquad
\forall\,Y\in\mathcal T(\mathcal H_C),\ \alpha\neq\beta.
\end{equation}

{\bf Theorem 6: Center of mass diffusion.} Assume that $\mathcal{M}_C(\Pi^C_\alpha X\Pi^C_\beta) =0$ for two different center of mass sectors. Further assume that $\mathcal{M}_C$ is a Galilei covariant channel. Then
\begin{equation}\label{diff_COM}
    \text{Var}_\mathcal{M_C(\hat\rho)}(\hat P_C) =\text{Var}_{\hat\rho}(\hat P_C) + \Delta, 
\end{equation}
where $\Delta >0$.
\\
\\
{\it Proof.} 
Since $\mathcal{M}_C$ is a Galilei covariant channel, let us first pin down its general structure. Let $W_C(p,q)= e^{\frac{i}{\hbar}(p\hat X_C + q\hat P_C)}$ be the Weyl operators associated with the center of mass degrees of freedom. Note that a translation by $x$ is obtained by conjugation by $W_C(0,-x)$ and, correspondingly, a boost by $+v$ by conjugation of $W_C(Mv,0)$. Let us define the auxiliary map $\mathcal{N}_t(X) = U^\dagger_t\mathcal{M}_C(X)U_t$ where $U_t = e^{-\frac{i}{\hbar}\hat P_C^2t/2M}$. From Eqs.~\eqref{Tcov} and~\eqref{Vcov} it follows that $\mathcal{M}_C$ is Galilei covariant {\it iff} $\mathcal{N}_t$ is Weyl covariant, that is
\begin{equation}
    \mathcal{N}_t(W_C(p,q)XW_C^\dagger(p,q)) =W_C(p,q)\mathcal{N}_t(X)W_C^\dagger(p,q)\,. 
\end{equation}
The general structure of Weyl covariant channels is known~\cite{Haapasalo2019,holevo2005additivity} and they are random Weyl displacements, that is, there exists a measure $\mu_t(p,q)dpdq$ such that
\begin{equation}
    \mathcal{N}_t(X) = \int \!\!dqdp\mu_t(p,q) W_C(p,q)XW_C^\dagger(p,q)\,.
\end{equation}

Therefore, mapping back to $\mathcal{M}_C$ we obtain that it must have the form
\begin{equation}
    \mathcal{M}_C(\hat \rho) = U_t\left(\int dpdq \mu_t(p,q)W_C(p,q)\hat\rho W_C^\dagger(p,q) \right)U_t^\dagger,
\end{equation}
The map introduces a shift in the average momentum:
\begin{align}\label{avPCM}
\langle\hat P_C\rangle_{\mathcal{M}_C(\hat\rho)}&=\Tr[\hat{P}_C\mathcal{M}_C(\hat\rho)]\nonumber\\
&=\int dp\,dq\, \mu_t(p,q)\Tr[ W_C^\dagger(p,q) \hat P_CW_C(p,q)\hat\rho]    \nonumber\\
&=\Tr[\hat P_C \hat\rho]+\int dp\,dq\, \mu_t(p,q)p    \nonumber\\
&=\langle\hat P_C\rangle_{\hat{\rho}}+\bar p_t,
\end{align}
where we introduced 
\begin{equation} \label{barp}
    \bar p_t= \int dp\,dq\, \mu_t(p,q)p:= \int dp\nu_t(p)p,
\end{equation}
while the variance of the total momentum of $\mathcal{M}_C(\hat\rho)$ is:
\begin{align}
    &\text{Var}_{\mathcal{M}_C(\hat\rho)}(\hat P_C) =\Tr[(\hat P_C-\langle\hat P_C\rangle_{\mathcal{M}_C(\hat\rho)})^2\mathcal{M}_C(\hat\rho)]\nonumber\\
    &\overset{(1)}{=}\int dpdq\mu_t(p,q)\times \nonumber\\
    &\qquad\Tr[(\hat P_C-\langle\hat P_C\rangle_{\hat\rho}-\bar p_t)^2UW_C(p,q)\hat \rho W_C^\dagger(p,q)U^\dagger]\nonumber\\
    &\overset{(2)}{=}\int dpdq\mu_t(p,q)\Tr[(\hat P_C-\langle\hat P_C\rangle_{\hat\rho} + p-\bar p_t)^2\hat \rho]\nonumber\\
    &=\Tr[(\hat P_C-\langle\hat P_C\rangle)^2\hat \rho] +\int dpdq\mu_t(p,q)(p-\bar p_t)^2\nonumber\\
    &= \text{Var}_{\hat\rho}(\hat P_{C}) +\text{Var}_{\nu_t}(p)\,,
\end{align}
where in $(1)$ we used the relation Eq.~\eqref{avPCM} and in (2) that $\comm{U}{f(\hat P_C)}=0$. Thus $\Delta = \text{Var}_{\nu_t}(p)$. Let us now assume by contradiction that $\Delta = \text{Var}_{\nu_t}(p) = 0$ so that $\nu_t(p) = \delta(p-\bar p_t)$.
Under a Galilei covariant channel the characteristic function of the momentum changes as
\begin{align} \label{chafunc}
&\chi_{\mathcal{M}_C(\hat\rho)}(d) =\nonumber\\
&=\int dpdq\mu_t(p,q)\Tr[UW_C(p,q)\hat \rho W_C^\dagger(p,q)U^\dagger e^{-\frac{i}{\hbar}d \hat P_C}]\nonumber\\
&=\int dpdq\mu_t(p,q)\Tr[\hat \rho W_C^\dagger(p,q) e^{-\frac{i}{\hbar}d\hat P_C}W_C(p,q)]\nonumber\\
&=\int dpdq\mu_t(p,q)\Tr[\hat \rho e^{-\frac{i}{\hbar}d(\hat P_C+ p)}]\nonumber\\
&=\Tr[\hat \rho e^{-\frac{i}{\hbar}d\hat P_C}]\int dpdq\mu_t(p,q)e^{-\frac{i}{\hbar}d p}\nonumber\\
&=\chi_{\hat \rho}(d) \int dp\nu_t(p)e^{- \frac{i}{\hbar}d p}\nonumber\\
&=\chi_{\hat \rho}(d)e^{- \frac{i}{\hbar}d \bar p_t},
\end{align}
 where in the last line we used the hypothesis $\nu_t(p) = \delta(p-\bar p_t)$. Thus the characteristic function can only vary by a phase under our assumption. It follows that, given any two states $\rho_{1,2}$
\begin{align}
|\chi_{\mathcal{M}_C(\hat\rho_1)}(d)-\chi_{\mathcal{M}_C(\hat\rho_2)}(d)| = |\chi_{\hat\rho_1}(d)-\chi_{\hat\rho_2}(d)|.
\end{align}
Let us then consider a state $\ket{\psi}$ belonging to some sector $\Pi_\alpha^C$. Then construct the states $\ket{\Psi_{\pm}} = \frac{1}{\sqrt{2}}(\ket{\psi_\alpha^C} \pm e^{\frac{i}{\hbar}d^*\hat P_C}\ket{\psi_\alpha^C})$ and $\hat\rho_{1,2} = \hat \rho_{\pm} = \ket{\Psi_{\pm}}\bra{\Psi_{\pm}}$ with $d^*> \ell$. Then, for $d=d^*$
\begin{equation}\label{diffChiMain}
    |\chi_{\rho_+}(d^*)-\chi_{\rho_-}(d^*)| = 1.
\end{equation}
However since $\ket{\psi_\alpha^C}$ and $e^{\frac{i}{\hbar}d \hat P_C}\ket{\psi_\alpha^C}$ are supported on different sectors $\forall\ d>\ell$  we have that
\begin{align}
&|\chi_{\mathcal{M}_C(\hat\rho_+)}(d)-\chi_{\mathcal{M}_C(\hat\rho_-)}(d)| = \nonumber\\
&|\Tr[\mathcal{M}_C\qty(e^{\frac{i}{\hbar}d^*\hat P_C}\ket{\psi_\alpha^C}\bra{\psi_\alpha^C}+\ket{\psi_\alpha^C}\bra{\psi_\alpha^C}e^{-\frac{i}{\hbar}d^*\hat P_C})e^{\frac{i}{\hbar}d\hat P_c}]\nonumber\\
&=0.
\end{align}
This is a contradiction with~\eqref{diffChiMain}. \qed

One might be tempted to interpret this diffusion as a consequence of the Heisenberg uncertainty principle: localization in position implies spreading in momentum. This, however, is not correct. As it should be clear from Eq.~\eqref{diff_COM}, the diffusion is state-independent, and is closely related to Galilei covariance. In particular, translational covariance implies that the collapse cannot  occur always in the same point in space. Hence, for any wave packet of finite width, the ``collapse center'' is distributed over different positions, thus shifting the wave function, therefore  leading to a diffusive motion even when the wave packet is already localized (see \cite{bassi2005collapse} for a quantitative discussion of this mechanism in a simple model).
\\
\\
{\bf Summary.} The no signaling condition implies that, at the statistical level, the dynamics describing the interaction (whether classical or quantum) between two quantum systems must be a linear and CPTP map $\mathcal M$ (Theorem~1). As mentioned earlier, this is well established in the literature.

We have characterized the classicality of gravity by the classicality condition~\eqref{eq:WeakC2} and faithfulness condition~\eqref{eq:faithfulness-eps-narrative}.  In the exact faithfulness regime ($\varepsilon = 0$), these two conditions suffice to prove that $\mathcal M$ cannot be unitary (Theorem~3). Together with condition~\eqref{eq:faithfulness-eps-narrative2}, they imply that $\mathcal M$ must collapse an initial superposition (Theorem~4) and that the total momentum cannot be conserved (Theorem~5). This represents a distinctive feature of classical gravity that allows---in principle--- to discriminate it from its quantum counterpart. Adding Galilei covariance of the center of mass motion further characterizes the change of momentum as an increase of its spread (Theorem 6).

These results do not come as a surprise: all proposals in the literature that model gravity as classical have these features. The novelty here is showing that they must be so, under few, general assumptions.

Note that, up to experimental errors, conditions~\eqref{eq:faithfulness-eps-narrative} and~\eqref{eq:faithfulness-eps-narrative2} are experimentally verified for macroscopic systems. Therefore, provided the no-signaling condition holds (and that matter is quantum), if no deviation is registered in the momentum statistics the condition~\eqref{eq:WeakC2} cannot be correct. In this sense, our result is a robust, model independent test of the classicality of gravity.

It is worthwhile noting that the proposals for gravitationally induced entanglement (GIE) detection \cite{bose2017spin,marletto2017gravitationally} rely on similar assumptions: matter is described quantum mechanically, superluminal signaling is not allowed, gravity behaves as predicted by Newton for classical states. The first assumption is necessary to  formulate the entanglement question~\cite{Note3}, the second avoids the possibility for classical nonlocal models to mimic entanglement, the third is needed to make quantitative predictions.
As such,  our framework represents an equally strong, but much less challenging to realize, test of the quantum nature of gravity.

In the next Sections, we will turn it into an experimental proposal. 

\section{Dynamics of two masses interacting via classical gravity}\label{tre}
\begin{figure}[h!]
 \centering
  \includegraphics[width=\linewidth]{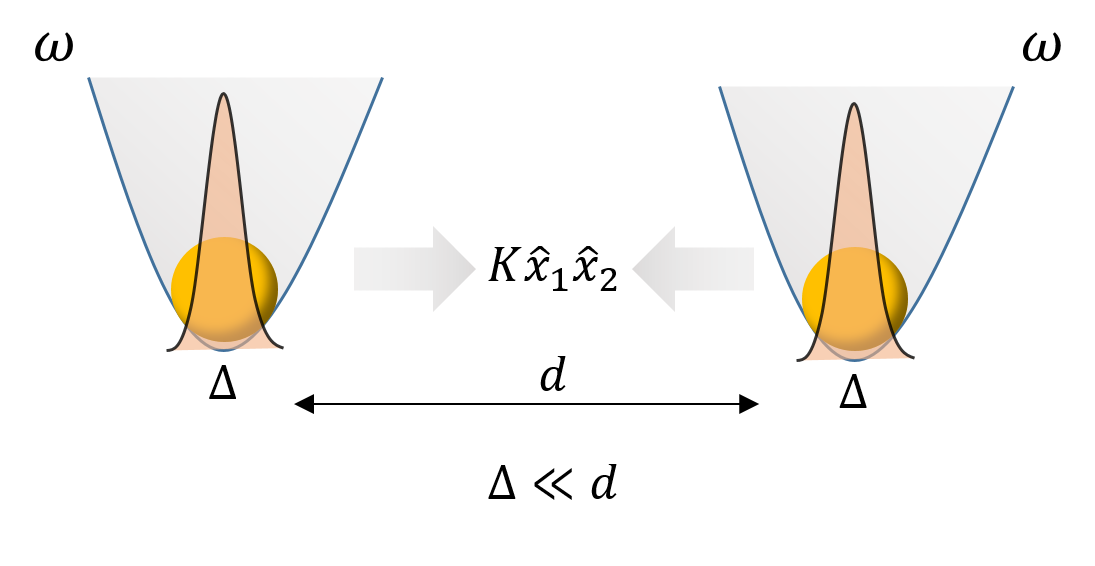}
  \caption{\justifying\textbf{Setup of the experiment.}
 We consider two particles confined in harmonic traps with a renormalized frequency $\Omega$, separated by a distance $d$ much larger than the spatial extent of their wave functions, denoted by $\Delta$. The dynamics of these two systems is described by Eq. (\ref{ME_main}).}
 \label{fig2}
\end{figure}

The channel $\mathcal M$ characterized in the previous section is rather general, as it describes the evolution of any two quantum systems interacting through what we characterized as a classical mediator, showing that the resulting dynamics must be diffusive. We now specialize the analysis to two harmonically trapped massive objects along the axis joining them, as illustrated in Fig.~\ref{fig2}.

Assuming that the dynamics is continuous over time, we describe it by a family of CPTP maps $\{\mathcal M_t^{G}\}_{t\ge 0}$, where we made explicit the dependence on Newton's constant $G$. In the absence of gravity, Assumption~1 implies that the evolution reduces to a unitary channel
\begin{equation}
\mathcal M_t^{G}\big|_{G=0}=\mathcal U_t,
\qquad
\mathcal U_t[\hat{\rho}]=e^{-i\hat H_0 t/\hbar}\,\hat{\rho}\,e^{i\hat H_0 t/\hbar},
\end{equation}
where the Hamiltonian $\hat H_0$ includes the free dynamics as well as all non-gravitational interactions. This allows us to define the interaction-picture map with respect to $\mathcal U_t$,
\begin{equation}
    \widetilde{\mathcal M}_t^{G}\coloneqq \mathcal U_t^{-1}\circ \mathcal M_t^{G},
\end{equation}
with a corresponding interaction picture state $   \widetilde{\hat{\rho}}(t)=\mathcal U_t^{-1}[\hat{\rho}(t)]$. Notice that  by construction we have $\widetilde{\mathcal{M}}_t^{G}\Big|_{G=0} = \mathcal{I}$, where $\mathcal{I}[\hat{\rho}]=\hat{\rho}$ is the identity superoperator.

Assuming that in our setting gravity is weak, we Taylor expand the dynamics to arrive at the well-defined form of a time-local master equation.

{\bf Theorem 7: Time-local master equation.} Assume that the map ${\mathcal{M}}^{G}_t$ defined here above admits perturbative expansion in terms of $G$, i.e. 
\begin{equation} \label{eq:perme}
\widetilde{\mathcal{M}}^{G}_t = \mathcal{I} + G \Lambda_t +O(G^2),
\end{equation}
with $G\norm{\Lambda_t}<1 \ \forall\ t\in [0,T]$. Neglecting terms of order $G^2$ or higher, the evolution can be written in terms of a time local generator of the form
\begin{align}\label{th7Claim}
    \frac{\dd }{\dd t}\hat\rho(t) &=-\frac{i}{\hbar}\comm{\hat H_0+\hat H^{G}(t)}{\hat\rho(t)} \nonumber\\
    &+ \sum_{\alpha,\beta}\lambda^{G}_{\alpha\beta}(t)\bigg(\hat f_{\alpha}\hat{\rho}(t)\hat f_{\beta}^\dagger -\frac{1}{2}\acomm{\hat f_{\beta}^\dagger \hat f_{\alpha}}{\hat\rho(t)}\bigg),
\end{align}
where $\hat H^{G}$ is a Hamiltonian contribution of order $G$  and the coefficients $\lambda^{G}_{\alpha\beta}$, also of order $G$, form a time-dependent Hermitian matrix.
\medskip
\begin{proof}
Given~\eqref{eq:perme} and
since $G\|\Lambda_t\|<1$, the map $\widetilde{\mathcal M}_t^{G}$ is invertible for each
$t\in[0,T]$, and its inverse can be defined through the Neumann series as
\begin{align}
    \big(\widetilde{\mathcal M}_t^{G}\big)^{-1}
    &=\big(\mathcal{I}+G\Lambda_t+O(G^2)\big)^{-1}\nonumber \\
    &=\sum_{n=0}^\infty (-1)^n\big(G\Lambda_t+O(G^2)\big)^n.
\end{align}
In particular, to first order in $G$,
\begin{equation}
    \big(\widetilde{\mathcal M}_t^{G}\big)^{-1}
    =\mathcal{I}-G\Lambda_t+O(G^2).
\end{equation}
Differentiating the interaction-picture state $\widetilde{\hat{\rho}}(t)=\widetilde{\mathcal M}_t^{G}\,\widetilde{\hat{\rho}}(0)$
we obtain
\begin{align}
    \frac{\dd }{\dd t} \widetilde{\hat{\rho}}(t)
    &=\dot{\widetilde{\mathcal M}}_t^{G}\,\widetilde{\hat{\rho}}(0) \nonumber\\ 
    &=\dot{\widetilde{\mathcal M}}_t^{G}
    \big(\widetilde{\mathcal M}_t^{G}\big)^{-1}\widetilde{\hat{\rho}}(t)
    \coloneqq \widetilde{\mathcal L}_t\,\widetilde{\hat{\rho}}(t).
\end{align}
Hence the interaction-picture dynamics is time local.
Returning to the Schr\"odinger picture
\begin{align}
\frac{\dd }{\dd t}\hat{\rho}(t)&=\dot{\mathcal{U}_t}\widetilde{\hat{\rho}}(t)+\mathcal{U}_t\widetilde{\mathcal{L}}_t \widetilde{\hat{\rho}}(t)=\dot{\mathcal{U}}_t\mathcal{U}^{-1}_t\hat\rho(t) +\mathcal{U}_t\widetilde{\mathcal{L}}_t\mathcal{U}^{-1}_t\hat\rho(t)\nonumber \\
&=-\frac{i}{\hbar}\comm{\hat H_0}{\hat\rho_t} +\mathcal{L}_t\hat\rho(t),
\end{align}
where we used $\dot{\mathcal{U}}_t\mathcal{U}^{-1} = -\frac{i}{\hbar}\comm{\hat H_0}{\cdot}$ and defined the extra contributions to the generator coming from the $G$ expansion as $\mathcal{L}_t = \mathcal{U}_t\circ \tilde{\mathcal{L}}_t\circ\mathcal{U}^{-1}_t$. Now $\mathcal L_t$ must preserve trace and Hermiticity: such generator can be written in the standard form~\cite{gorini1976completely}
\begin{align}
    \mathcal L_t \rho(t)
    &=-\frac{i}{\hbar}[\hat H^{G}(t),\hat \rho(t)] \nonumber\\
    &\quad+\sum_{\alpha,\beta}\lambda^{G}_{\alpha\beta}(t)
    \left(
    \hat f_\alpha \rho(t) \hat f_\beta^\dagger
    -\frac12\{\hat f_\beta^\dagger \hat f_\alpha,\hat \rho(t)\}
    \right),
\end{align}
for some operator basis $\{\hat f_\alpha\}$ and a Hermitian matrix
$\lambda^{G}_{\alpha\beta}(t)$. Adding the Hamiltonian contribution $\hat H_0$ we find the expression Eq.~\eqref{th7Claim}.
\end{proof}

To encode diffusion, this dynamics must deviate from a unitary evolution and can, in general, be non-Markovian. However, this non-Markovianity has a natural timescale: that required by the interaction to connect the two systems. For systems moving at non-relativistic speeds, which is the case we are considering here, this timescale is negligible, and thus the gravitational interaction and the corresponding diffusive effects can be considered instantaneous. This allows us to work in the Markovian limit, where the GKLS theorem guarantees that the matrix $\lambda^G_{\alpha \beta}(t) = \lambda^G_{\alpha \beta}\ge 0$ for the generator to describe a completely positive dynamics. Thus, using the spectral decomposition of $\lambda^G$ we have a
Lindblad master equation
~\cite{lindblad1976generators,gorini1976completely}:
\begin{align}\label{MAIN_EQ:GKLS}
    \frac{\dd }{\dd t}\hat\rho (t) = &-\frac{i}{\hbar}\comm{\hat H_0 + \hat H^{G}}{\hat \rho(t)} +\nonumber\\
    &\sum_{k}\lambda_{k}^{G}\Big[f_k(\hat c)\hat \rho(t) f_k^\dagger(\hat c)-\frac{1}{2}\acomm{\hat f_k(\hat c)^\dagger f_k(\hat c)}{\hat \rho(t)}\Big]\,.
\end{align}
where $f_k$ are functions of the positions and momenta of the two particles, $\hat c^T = (\hat x_1,\hat x_2,\hat p_1,\hat p_2)$. 

Now, we take into account the regimes of the experimental setup we want to consider. We assume that the two systems are separated by a distance much larger than the spatial extent of their wave functions and that their positions and momenta deviate only slightly from their equilibrium values. Moreover, such deviations are much smaller than the average distance between the two systems. We will elaborate further on the experimental realization in Section~\ref{secV}.

Within this regime, we can Taylor expand the Hamiltonian to quadratic order $\hat H^G \approx \hat H^G_q = \frac{1}{2} \sum_{ij}\hat{c}_i (H_q^G)^{ij}\hat{c}_j$ --we neglect linear terms as they will not play a role-- and the the functions $f_k(\hat c)$ to linear order in the quadratures: $f_k(\hat c)\approx \sum_{i=1}^4h_{ki}\hat c_i$ where $h_{ki} \in \mathbb{C}$. Then, the master equation \eqref{MAIN_EQ:GKLS} reduces to: 
\begin{align}\label{ME_main_1}
    \frac{\dd }{\dd t}\hat\rho(t) =& -\frac{i}{\hbar}\comm{\hat H_0  +\frac{1}{2}\sum_{ij=1}^4\hat c_i(H_q^G)^{ij}\hat c_j}{\hat \rho(t)} +\nonumber\\
    &+\sum_{ij=1}^4\gamma_{ij}\Big[\hat c_i\hat\rho(t)\hat c_j -\frac{1}{2}\acomm{\hat c_j\hat c_i}{\hat \rho(t)}\Big],
\end{align}
where we introduced:
\begin{equation}
\gamma_{ij}:=\sum_{k}\sqrt{\lambda_k^{G}}h_{ki}h^*_{kj}\sqrt{\lambda_k^{G}},
\end{equation} 
which are the elements of a manifestly Hermitian, positive semidefinite matrix. 
We have yet to tailor the dynamics to reproduce Newtonian gravity when the state of the two particles is effectively classical, as required by assumption 4. In the formalism of the previous section, we have yet to impose that the generator determines a map which respects the $\varepsilon$-faithfulness conditions~\eqref{eq:faithfulness-eps-narrative}-\eqref{eq:faithfulness-eps-narrative2} adapted to the Newtonian potential $\hat V_N$. The Newtonian potential, expanded to second order reads
\begin{equation}
\hat V_N\approx-\frac{K}{2}(\hat x_1-\hat x_2)^2 \,,\quad K =\frac{ 2Gm_1m_2}{d^3}\,,
\end{equation}
where we neglected an irrelevant constant and the linear terms have been removed by a change of coordinates $\hat{x}_i$, as described in Appendix~\ref{App:HamiltonianNewton}. With a slight abuse of notation, we retain the same symbols for the transformed coordinates. In the linearized limit of Eq.~\eqref{ME_main_1}, compatibility with Newtonian predictions requires (see Appendix~\ref{APP_NewtLimit}) that $\gamma_{ij}\in \mathbb{R}$, which means the equation is purely diffusive and, furthermore $\hat H^G = \hat V_N \approx -\frac{K}{2}(\hat x_1-\hat x_2)^2$.
Overall therefore the master equation reads
\begin{equation}\label{ME_main}
    \frac{\dd}{\dd t}  \hat \rho(t) = -\frac{i}{\hbar}\comm{\hat H_0 +K\hat x_1 \hat x_2}{\hat \rho (t)}-\frac{1}{2}\sum_{i j=1}^4\gamma_{ij}\comm{\hat c_i}{\comm{\hat c_j}{\hat \rho(t)}},
\end{equation}
where for he sake of readability we absorbed all non-coupling Hamiltonian terms into $\hat H_0$ and exploited that $\gamma = \gamma^T$ to rewrite the dissipator conveniently

Note that most hybrid classical-quantum models of gravity proposed in the literature~\cite{tilloy2016sourcing,tilloy2018ghirardi,kafri2014classical,oppenheim2023postquantum}, with the exception of those which violate Galilei covariance~\cite{bahrami2014role,di2023linear}, reduce to Eq.~\eqref{ME_main} in the appropriate limits~\cite{layton2023weak} and experimental regimes, albeit with different values for the entries of  $\gamma$ (c.f.~\cite{gaona2021gravitational}).  Appendix~\ref{APP_D} contain further discussion on the relation between Eq.~\eqref{ME_main} and those known in the literature.

\section{Separability bound}\label{secIV}
The master equation~\eqref{ME_main} describes two systems undergoing a general diffusive dynamics that reproduces (linearized) Newtonian gravity on average. 
We still need to codify into the master equation the classicality condition~\eqref{eq:WeakC2}, namely classical states ought to remain classical. A direct consequence is that, for separated systems prepared in well-localized states, the absence of coherence generation also precludes the generation of entanglement. On this basis, we will regard the ground states of the two harmonic oscillators in the setup of Fig.~\ref{fig2} as classical states, and therefore require that they should not become entangled under the dynamics described by Eq.~\eqref{ME_main}. This in turn will provide  a nontrivial constraint that the master equation must satisfy.

Loosely speaking, the bilinear interaction term \( K\hat{x}_{1}\hat{x}_{2} \) in the Hamiltonian induces entanglement between the particles, while the diffusive Lindblad terms work against its generation: only if the latter are sufficiently strong entanglement will not be generated.  
To be quantitative, we use the criterion of positivity under partial transposition (PPT)~\cite{peres1996separability,horodecki2009quantum} to compute the necessary conditions that Eq.~\eqref{ME_main} must obey to ensure that the separability of states is preserved.

For convenience, we temporarily work with the dimensionless quadratures  $\bar x_j=\sqrt{{m\Omega_j}/{\hbar}}~\hat x_j$ and $\bar p_j= \hat p_j/\sqrt{{m\hbar\Omega_j}}$; the master equation then reads
\begin{equation}\label{MAIN_EQ:ME_dimLess}
    \frac{\dd }{\dd t}\hat{\rho}(t)\!=\!-i[\bar{H}_0 + \frac{K \bar{x}_1\bar{x}_2}{m\sqrt{\Omega_1\Omega_2}},\hat{\rho}(t)]- \frac{1}{2}\sum_{ij=1}^4\bar{\gamma}_{ij}\comm{\bar{c}_i}{\comm{ \bar{c}_j}{\hat \rho(t)}}    \, ,
\end{equation}
where $\bar{H}_0=\sum_{i=1,2}\frac{\Omega_i}{2}(\bar p_i^2 +\bar x_i^2)$ and the total Hamiltonian is of the form $\frac{1}{2}\bar c^T H \bar c$  where $ H$ is a symmetric real matrix which reads
\begin{equation}\label{Hambar}
 H = \begin{bmatrix}
    \Omega_1 & K/m\sqrt{\Omega_1\Omega_2}&0&0\\
    K/m\sqrt{\Omega_1\Omega_2}& \Omega_2 & 0 &0\\
    0 & 0 & \Omega_1 & 0\\
    0&0&0&\Omega
\end{bmatrix}.\,
\end{equation}
In the Lindbladian, the coefficients rendering the variables dimensionless have been absorbed into $\bar\gamma=S\gamma S$ with $S=\sqrt{\hbar}\operatorname{diag}\left(
\sqrt{\frac{1}{m\Omega_1}},
\sqrt{\frac{1}{m\Omega_2}},
\sqrt{m\Omega_1},
\sqrt{m\Omega_2}
\right)$. 
The canonical commutation relations between the modes $\bar c_i$ can be written compactly by introducing the symplectic matrix $J$:
\begin{equation}
    \comm{\bar c_i}{\bar c_j} = i J_{ij},\quad J= \begin{bmatrix}
    0&\mathbb{I}_2\\
    -\mathbb{I}_2&0
    \end{bmatrix}\,.
\end{equation}
Furthermore, letting $\{\cdot,\cdot\}$ denote the anticommutator and defining the covariance matrix as $V_{ij}(t) = \langle\{(\bar c_i-\langle\bar c_i\rangle)(\bar c_j-\langle\bar c_j\rangle\}\rangle/2$; the crucial consequence of the non commutativity of the operators $\bar c_i$s, i.e. Heisenberg's uncertainty principle, can be expressed as the matrix inequality~\cite{ferraro2005gaussian}:
\begin{equation}\label{MAIN_EQ:UP}
    V(t) + \frac{i}{2}J \ge 0\,,    
\end{equation} 
which has to be satisfied by any quantum dynamics. 

Partial transposition has the physical interpretation of a partial time reversal, that is, for continuous variables, it can be recast geometrically as a partial reflection in phase space, which is codified in the matrix $\Lambda = \text{diag}[1,1,1,-1]$. The PPT condition amounts to the observation that separable quantum states must be mapped into valid, i.e. positive, quantum states by this reflection. This is equivalent to the request that the partially transposed state satisfies Heisenberg's uncertainty principle~\eqref{MAIN_EQ:UP}~\cite{simon2000peres}. Introducing  the partial reflection in phase space with respect to the second party, the PPT condition can thus be written concisely as a matrix inequality: 
\begin{equation}\label{MAIN_EQ:PPT}
 \Lambda V(t)\Lambda+\frac{i}{2} J \ge 0 \iff V(t)+\frac{i}{2}\Lambda J\Lambda \ge 0 \,,
 \end{equation} 
which must hold for any separable state with covariance $V(t)$. 

The covariance matrix under the evolution~\eqref{MAIN_EQ:ME_dimLess} evolves according to 
\begin{equation}\label{MAIN_EQ:CovEvo}
    \frac{dV(t)}{dt} = J H V(t)+V(t) (J H)^T - J \bar\gamma J,
\end{equation}
and it explicitly depends on the matrix $\bar\gamma$. Therefore the validity of the PPT condition over time~\eqref{MAIN_EQ:PPT} implicitly sets restrictions on the elements of $\bar\gamma$, which govern the diffusion.

We now require that the ground states of the two oscillators do not immediately entangle after a time $\varepsilon>0$, as would otherwise happen in absence of diffusion, due to the Newtonian interaction. Taking as initial covariance matrix $V_0 = \mathbb{I}_4$/2, by Taylor expanding the PPT condition we obtain:
\begin{equation}\label{MAIN_EQ:PPTshort}
    0 \le \frac{1}{2} z^\dagger\Big(\mathbb{I}_4 +i\Lambda J\Lambda\Big)z +\varepsilon z^\dagger\left.\dv{V(t)}{t}\right|_{t=0} z, \qquad\forall z\in \mathbb{C}^4.
\end{equation}
The first term on the right hand side is non-negative by construction and independent of both $\bar\gamma$ and $K$; however, by choosing $z_0=(a,-b,ia,ib)$, with $a,b \in \mathbb{C}$, it vanishes identically. Therefore, requiring that the separability of the ground state is preserved implies the positivity of the time derivative of the covariance, $\frac{dV}{dt}|_{t=0}$, on the subset of vectors $z_0$.
Inserting the above choice for $z_0$ in the inequality~\eqref{MAIN_EQ:PPTshort}, with the additional assumption $|a| = |b|$, one obtains a set of conditions which amount to (we recall that \(\bar\gamma\) is real and symmetric, so also its off-diagonal entries are real):
\begin{align}
    \frac{2K\sin\alpha}{m\sqrt{\Omega_1\Omega_2}} \le &\Tr[\bar\gamma] +2(\bar\gamma_{12}-\bar\gamma_{34})\cos\alpha\nonumber\\
    &-2(\bar\gamma_{14}+\bar\gamma_{23})\sin\alpha\,,
\end{align}
where $\alpha \in [0,2\pi)$ is simply the phase difference between $a$ and $b$.
The strongest of these conditions is set by maximizing the l.h.s. of the above equation, obtaining:
\begin{align}
    \text{Tr}[\bar\gamma] - 2(\bar\gamma_{14}+\bar\gamma_{23}) \ge \frac{2K}{m\sqrt{\Omega_1\Omega_2}} \,.
\end{align}
Finally, by exploiting the positivity of the $\bar\gamma$ matrix, which implies $|\bar\gamma_{ij}|\le \frac{\bar\gamma_{ii}+\bar\gamma_{jj}}{2}$, one arrives at the weaker, but more useful, bound:
\begin{equation}\label{MET_EQ:bound}
    \text{Tr}[\bar\gamma] \ge \frac{K}{m\sqrt{\Omega_1\Omega_2}} = \frac{2Gm}{d^3\sqrt{\Omega_1\Omega_2}}  \,,
\end{equation}
which, when written in terms of the dimensional matrix elements of $\gamma$, reads:
\begin{equation}\label{MAIN_EQ:BoundOmegas}
\Omega_2\gamma_{11}+\Omega_1\gamma_{22} +m^2 \Omega_1\Omega_2(\Omega_1\gamma_{33} +\Omega_2\gamma_{44})\ge \frac{2Gm^2}{\hbar d^3}\sqrt{\Omega_1\Omega_2}\,.
\end{equation}
Eq.~\eqref{MAIN_EQ:BoundOmegas} has an explicit dependence on the frequencies of the oscillators, that is, on the specific setup that has been chosen. This is expected: depending on the configuration of the two masses, the minimum noise required to keep them unentangled changes. Since gravity is supposed to be universal and therefore independent of the choice of experimental setup, the gravitational diffusion coefficients \( \gamma_{ij} \) are independent of the frequencies and depend only on the masses and distances involved—a reasonable expectation in the non-relativistic limit~\cite{Note4}. This implies that gravity should respect the bound~\eqref{MAIN_EQ:BoundOmegas} for any values of \( \Omega_1, \Omega_2 \). Noting that for equal masses, under the above assumption, \( \gamma_{11}=\gamma_{22} \) and \( \gamma_{33}=\gamma_{44} \), by varying the frequencies we arrive at the simple bound:
\begin{equation}\label{MAIN_EQ:bound}
     \gamma_{11} +m^2\omega^2\gamma_{33}\ge \frac{Gm^2}{\hbar d^3}\,,
\end{equation}
valid for any frequency $\omega$.

The physical meaning of equation \eqref{MAIN_EQ:bound} is straightforward: in order to prevent entanglement generation, the total decoherence rate, i.e. the sum of the diagonal entries of $\gamma$, has to be large enough to counteract the entangling Newtonian interaction, whose strength is given by $K$ and, hence, $G$. Similar bounds have been derived within the context of non-entangling Lindblad dynamics \cite{kafri2013noise}. If {\it no} extra diffusion compatible with the bound in Eq. (\ref{MAIN_EQ:bound}) is detected, one can conclude that gravity cannot be classical. On the other hand, since the above condition is only necessary for separability,  the detection of extra diffusion does not imply the classicality of gravity. Still, it would mark a striking departure from the strictly unitary evolution predicted by the Newtonian potential. 

A noisy behavior of quantum matter is also expected within quantum gravity scenarios due to  spacetime becoming ``fuzzy" at Planck lengths~\cite{karolyhazy1966gravitation,amelino1999gravity,amelino2001phenomenological,petruzziello2021quantum,verlinde2021observational,zurek2022vacuum,arzano2023fundamental,sharmila2025signatures}. These  effects scale inversely with powers of  the Planck energy (depending on the model) and, thus, are orders of magnitude weaker than the effects expected by classical gravity scenarios here considered, which are those satisfying the bound~\eqref{MAIN_EQ:bound}.

The bound~\eqref{MAIN_EQ:bound} is obtained for a purely diffusive and Markovian gravitational dynamics. In Appendix~\ref{APP:DissNonM} we show that this result is robust against dissipative extensions of the master equation~\eqref{ME_main} and transient non-Markovian behaviors. The question now moves to finding an efficient strategy to measure the presence or lack of the diffusion quantified by Eq.~\eqref{MAIN_EQ:bound}.

\section{Experimental signatures of classical gravity}\label{secV}
\begin{figure}[h!] \includegraphics[width=0.8\linewidth]{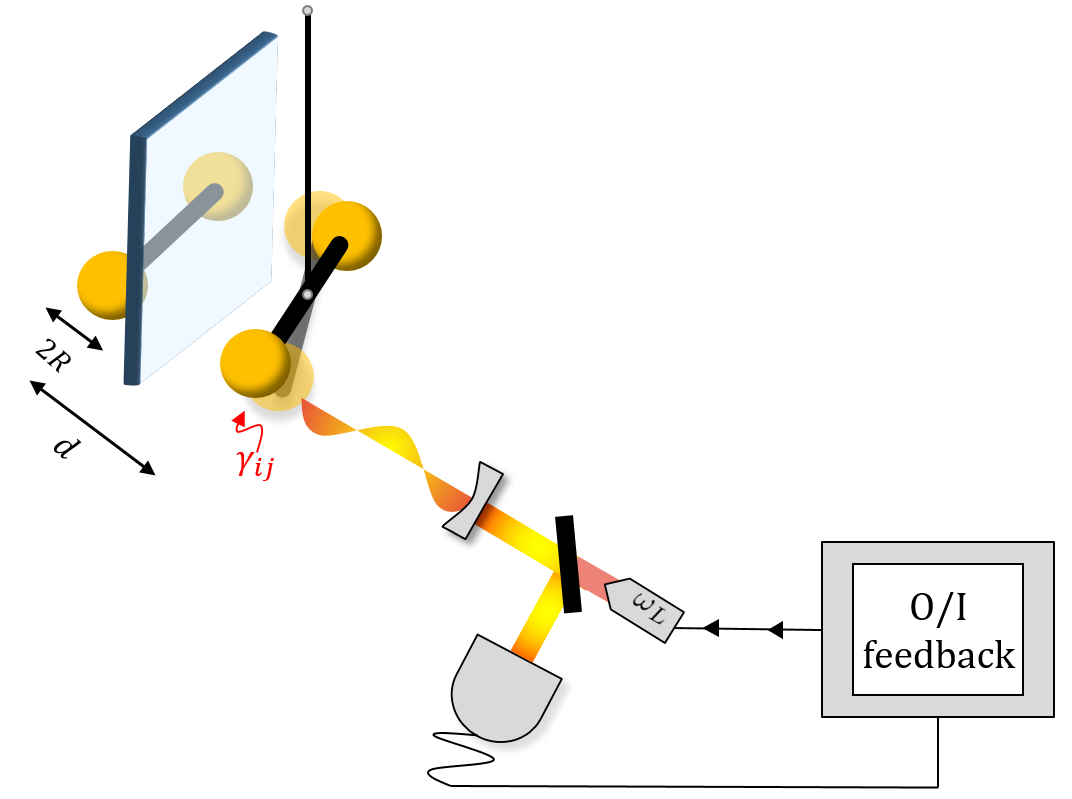} 
  \caption{\justifying
  \textbf{Implementation based on a torsion pendulum.} Two identical pairs of masses are arranged symmetrically, facing each other. The left pair serves as the source of  gravitational noise and can be considered fixed, while the right pair is part of a torsion pendulum with effective frequency \( \Omega \). A superconducting shield between the two pairs suppresses any electromagnetic interaction. The masses have radius \( R \), density \( \rho \), and equilibrium separation \( d \).  The relative displacement of the pendulum masses is measured by an apparatus with near-quantum-limited noise, enabling preparation of the pendulum close to its ground state via feedback cooling. The system is then allowed to evolve without measurement (i.e., ``in the dark"), and the mechanical amplitude is probed after a fixed time. This measurement is repeated multiple times over a total integration time \( t \), yielding the phonon heating rate \( \Gamma \). Provided that the thermal noise is sufficiently low, the heating rate produced by the gravitational noise will eventually become detectable. 
}  \label{fig3}  
\end{figure}

Diffusion leads to an increase in the variance of the position of the two oscillators; we now investigate its detectability. While the gravitational noise affects both systems, detecting it requires monitoring only the position of one of them. Therefore, we focus on a setup as shown in Fig.~\ref{fig3}, in which one mass is fixed, acting solely as a source of gravitational attraction and, correspondingly, as a source of gravitational noise for the other oscillating mass. 

Diffusion can be captured efficiently in the frequency domain by measuring the two-frequency correlation function of the Fourier transform of the position $\tilde x(\omega)$ of the oscillating particle. The correlation function is directly related to the density noise spectrum (DNS)~\cite{paternostro2006reconstructing}, which is defined as $S_{x_1x_1}(\omega)\delta(\omega+\nu) = \langle\tilde x_1(\omega) \tilde x_1(\nu) +\tilde x_1(\nu) \tilde x_1(\omega)\rangle/2$ and collects all  noisy properties of the motion of the particle's position. 
The calculation of the DNS is more conveniently performed in the Heisenberg picture, since the master equation~\eqref{ME_main} is  statistically equivalent to a set of Heisenberg-Langevin equations for the position and momentum operators of the two systems~\cite{Note5}. Having one mass fixed at an average distance distance $d$ from the other one, the equations of motion  for the latter are:
\begin{equation}  \label{MAIN_EQ:Lang}
\begin{split}
\frac{d\hat x}{dt} &= \hat p/m + \hbar w_3(t),\\
\frac{d\hat p}{dt} &= -m\Omega^2 \hat x -\eta \hat p +\hat \xi(t) - \hbar w_1(t)\,,
\end{split}
\end{equation}
with $\Omega^2=\omega_0^2 - K/m$, where $\omega_0$ is the proper frequency of the oscillator and where $w_i(t)$ are white noises, whose correlations encode the gravitational diffusion coefficients: $\mathbb{E}[w_i(t)w_j(t')]=\gamma_{ij}\delta(t-t')$.
In addition to the gravity related noise, we have also included in our analysis a thermal noise $\hat \xi(t)$ with temperature $T$ and correlation~\cite{paternostro2006reconstructing}
\begin{align} \label{eq:thermal}
\mathbb{E}[\langle\hat\xi_i(t)\hat\xi_j(t')\rangle]&=\delta_{ij}\frac{\hbar\eta_j m_j}{2\pi}\times\nonumber\\
&\int \dd \omega \, {\omega} e^{-i \omega (t-t')}\left[1+\coth\left(\frac{\hbar \omega}{2 k_\text{B} T} \right) \right],
\end{align}
and damping rate $\eta$ ($k_\text{B}$ stands for Boltzmann constant), which encode the effects of the other sources of decoherence. Solving the equations~\eqref{MAIN_EQ:Lang} in the frequency space, a straightforward calculation leads to

\begin{widetext}
\begin{equation}\label{main_eq:Spectrum}
    S_{xx}(\omega) = \frac{\hbar^2}{\Big|m(\Omega^2-\omega^2 - i \eta \omega)\Big|^2}\Big[\gamma_{11} + m^2\omega^2\gamma_{33} + \frac{\eta m \omega}{\hbar}\Big(1+\coth(\frac{\hbar \omega}{2k_\text{B} T})\Big) + m^2\eta^2\gamma_{33} - 2m\eta \gamma_{13}\Big]\,.
\end{equation}
\end{widetext}

Considering setups with high quality factor, we can safely neglect the last two terms in Eq. (\ref{main_eq:Spectrum}). Then a necessary condition for this setup to detect the gravity induced diffusion is that the thermal fluctuations (third term in Eq. (\ref{main_eq:Spectrum})) are lower than the noisy terms related to gravity (first two terms in Eq. (\ref{main_eq:Spectrum})). This condition, on resonance $\omega= \Omega$, keeping into account the bound in Eq. (\ref{MAIN_EQ:bound}) and neglecting the vacuum contribution of the environment, is fulfilled when:
\begin{equation}\label{final_cond}
\frac{\eta m}{\hbar}\Omega \coth\Big(\frac{\hbar\Omega}{2k_\text{B}T}\Big)
\leq 
\frac{Gm^2}{\hbar d^3}.
\end{equation} 

We now discuss the experimental requirements needed to satisfy the above inequality.
First, consider the simplest case of two identical spherical masses with radius $R$ and density $\rho$, placed at a distance $d=2 R \beta$, where $\beta > 1$. In the limiting case $\beta=1$ the two masses  touch each other. By writing the mass $m=4\pi/3 \rho R^3$ and taking the classical limit $k_b T \gg \hbar \Omega$ for the occupation number, Eq. (\ref{final_cond}) can be cast in the following form, which contains only frequencies:
\begin{equation}  \label{simplified_cond}
 \frac{12}{\pi}\beta^3 \Omega \Gamma \leq \omega_G^2,
\end{equation}
where we have introduced the phonon heating rate $\Gamma = \eta n_T$, where $  n_T = k_\text{B} T/\hbar \Omega  $ is the mean thermal occupation number, and we have defined the frequency $\omega_G = \sqrt{ G \rho}$, which characterizes the gravitational interaction between masses with density $\rho$. Note that Eq. (\ref{simplified_cond}) is independent of the radius $R$ and is therefore scale invariant. 

Since the gravitational interaction is easily overwhelmed by electromagnetic forces at the microscale, this clearly favors macroscopic experiments, for which the short distance condition $\beta \approx 1 $ can be approached, allowing an electromagnetic shield to be placed between the masses. The maximum density of solid-state materials given by Osmium $\rho = 2.26\times 10^4$ Kg/m$^3$, implies that $\omega_G ~=~ 1.1$~mHz at most, setting a quite tight condition on $\Gamma$. For example, for $\Omega/2\pi  = 0.1 $ mHz, achievable, for instance, by a torsion pendulum, and setting $\beta=1$ and Osmium as a material, we find $\Gamma \leq 0.6$ mHz to satisfy Eq. (\ref{simplified_cond}). This heating rate translates into a ratio $Q/T = 2 \times 10^{14}$ K$^{-1}$ where $Q=\Omega/\eta$ is the mechanical quality factor: even at $T=10$ mK, a $Q$ factor as large as $2 \times 10^{12}$ would be required.
We therefore arrive at the general conclusion that an experiment designed to detect the gravitational noise must be macroscopic and have a characteristic frequency $\Omega$ and a heating rate $\Gamma$ lower than $1$ mHz. This fully justifies the classical limit for $n_T$. 

At this point, we note that condition (\ref{simplified_cond}) is in principle conservative. If thermal noise is the only relevant source of noise in addition to the gravitational one, it can be accurately characterized, for example, by a set of measurements as a function of temperature \cite{vinante2020}, and subtracted. One can therefore estimate the gravitational noise as a fraction of thermal noise, which depends on the experiment accuracy and on the total measurement time. 

To this end, let us define a concrete measurement protocol. The very low phonon heating rate suggests a strategy based on measuring the reheating rate of the oscillator after preparation in a very low energy initial state by feedback cooling. In the best scenario, a quantum-limited detector allows one to cool the oscillator to the ground state. We discuss later the experimental feasibility. Importantly, reheating can be performed in the absence of any measurement, so we can neglect the back-action noise from the measurement apparatus. We assume that the reheating is performed in a time much shorter than the resonator relaxation time $\eta^{-1}$ and that the procedure is repeated and averaged many times. Under these conditions, it can be shown~\cite{vinantepontin2019} that the uncertainty in the determination of $\Gamma$ is $\sqrt{N \Gamma/ t}$ where $t$ is the total integration time and $N$ is the noise of the measurement apparatus expressed in units of $\hbar$. $N=1$ corresponds to a quantum-limited apparatus. 

The problem is now to distinguish the non-thermal contribution to $\Gamma$ coming from gravity which, according to Eq. (\ref{simplified_cond}), is given at the very least by:
\begin{equation}  \label{simplified_cond2}
 \Gamma_G = \frac{\pi \omega_G^2}{12 \beta^3 \Omega},
\end{equation}
from a thermal heating rate background:
\begin{equation}
\Gamma_{th}=\eta n_T = \frac{k_\text{B} T}{\hbar Q}.
\end{equation}
If we can characterize thermal noise with a relative uncertainty $r$, we can distinguish a gravitational noise which is $r$ times smaller ($\Gamma_G = r \Gamma_{th}$) within an integration time $t=N(1+1/r)/\Gamma_G \approx N/(r \Gamma_G)$. For example, for $r=0.01$ and $N=1$ we obtain an integration time $\sim  2$ days. This, in turn,  relaxes the requirement on $Q/T$ by a factor $r$. Given that robust  claims typically require to satisfy  the $5 \sigma$ rule,  the corresponding integration time increases by a factor 25.

Instead of a reheating protocol, one could use a different strategy to estimate the heating rate, based on continuous monitoring of the oscillator at thermal equilibrium and spectral estimation of the force noise. This strategy has been demonstrated, for example, in recent tests of spontaneous collapse models \cite{Vinante2016,Vinante2017,vinante2020}, which also implemented a subtraction of thermal noise. As discussed in Ref. \cite{vinantepontin2019}, reheating after cooling and stationary monitoring are, in principle, equivalent strategies. The stationary protocol is more illuminating in the crucial aspect that our proposal, in contrast to previous ones \cite{bose2017spin,marletto2017gravitationally,krisnanda2020observable,lami2024testing}, does not require the preparation and control of quantum states of motion. As shown in the following, in practice a near quantum-limited detector is sufficient: this is a much less demanding requirement.

Let us now discuss the feasibility of the proposed measurement. A resonance frequency range below millihertz can be readily achieved by torsion pendulums, suggesting as measurement scheme a symmetrical torsion balance as depicted in Fig. \ref{fig3}. Here, the first pair of masses can be thought as fixed, as its role is only to produce the classical gravity noise, while the second pair is a proper torsion pendulum; a summary of the parameters required by the proposed implementation is shown in Table \ref{table}. Regarding its feasibility, the biggest challenge is the extremely low level of dissipation and thermal noise required. The longest damping time demonstrated so far in torsion pendulums at $f \sim 0.1$ mHz is $\tau \sim 10^8$ s, with extrapolation up to $3\times 10^9$ \cite{Braginsky1972}, corresponding to a mechanical quality factor $Q \sim  10^6$. A similar $Q$ was extrapolated for a torsion pendulum at $1$ mHz in the zero pressure limit  \cite{cavalleri2009}. On the other hand, under the optimistic assumption previously discussed, namely that we could distinguish gravity-induced noise as a fraction $r=0.01$ of thermal noise at $T=10$ mK, we would still need $Q=2 \times 10^{10}$. This means that current experiments are at least 4 orders of magnitude away from the requirements. Although it is difficult to assess the effort and time needed to fill this gap, we will argue in the following that the required numbers can be in principle achieved.
\begin{table}[h!]
\vspace{0.4cm}
 \begin{tabular}{c |c  }
 \hline
 \hline
  Parameters & Value \\ 
 \hline
 $\Omega/2\pi$   & $10^{-4}$ Hz \\

 $\rho$   & $2.26 \times 10^4$ Kg/m$^3$ \\

 $R$   & $3$ cm \\

 $\beta$   & $\approx 1$ \\

 $T$   & $0.01$ K \\

 $Q$   & $2 \times 10^{10}$ \\

 $N$   & $\approx 1$\\

 $r$  &  $0.01$ \\
 \hline
 \hline
\end{tabular}
\caption{\justifying\textbf{List of parameters for a torsion pendulum implementation}. The frequency \( \Omega \), which is close to the lowest values achievable by torsion pendulums, is determined by Eq.~(\ref{simplified_cond}), which requires \( \Omega \) to be as low as possible. The density \( \rho \) corresponds to that of osmium, the heaviest solid material. We propose indicatively a radius \( R = 3\)~cm of each spherical mass, implying \( m = 2.55 \)~kg, as a practical trade-off: the masses must be macroscopic to suppress the relative effect of non-gravitational forces and to approach the condition \( \beta \approx 1 \), while remaining compact enough to ensure compatibility with a cryogenic apparatus. However, we remark that the sensitivity to classical gravity noise is independent of the pendulum mass. The temperature \( T \) is set to the typical value achievable in a continuous dilution refrigerator cryostat. The mechanical quality factor \( Q \) is assumed to be achievable at \( T = 10 \)~mK. The motional detector operates near the quantum limit with \( N \approx 1 \), enabling preparation of the pendulum close to its ground state. We assume the ability to resolve a fraction \( r = 0.01 \) of the thermal noise.
} 
\label{table}
\end{table}

Most torsion pendulums so far have been operated at room temperature, with only very few examples of cryogenics ones, however limited to temperatures above $1$ K and to metallic torsion fibers \cite{Bantel2000, Bantel2014, Fleischer2022}. However, strong suppression of mechanical losses is expected at millikelvin temperature: as a matter of fact, already back to 1977, Braginsky, Thorne and Caves envisioned precisely the possibility of testing a number of (still untested) relativistic effects in laboratory using torsion pendulums at millikelvin temperature \cite{Braginsky1977}. In their visionary work, they estimated that damping times longer than $10^{13}$ s and quality factors $Q \approx 10^{10}$ should be achievable using torsion pendulums comprising fused quartz or single-crystal sapphire fibers at $T < 100$ mK. For amorphous solids, the extrapolations rely on the well-known rapid decrease in the losses of two-level systems with decreasing temperature, typically scaling as $T^3$. For single crystal sapphire oscillators, $Q$ factors close to $10^{10}$ have been measured in mechanical modes at higher frequencies \cite{Braginsky1972}, in par with measurements performed on other crystalline resonators \cite{McGuigan1978,MacCabe2020}. This suggests that intrinsic dissipation of crystalline materials at low temperature can be as low as required by our proposal. 
We also point out that operation at millikelvin temperature will in principle allow suppression of gas damping losses to the desired levels \cite{Braginsky1977}. 

In light of these considerations and the broader relevance to a wide class of precision measurements beyond the specific topic investigated in this article, a systematic study of dissipation in torsion pendulums at millikelvin temperatures is both urgent and timely. Despite the proposal by Braginsky et al. \cite{Braginsky1977}, we are not aware of any experimental attempts to conduct such investigations.

Achieving the required \( Q/T \) factor is a necessary but not sufficient condition for conducting the experiment. Other critical factors include the thermalization of the fiber, the suppression of environmental noise sources, and the implementation of a near-quantum-limited angular detector. To ensure that the thermal noise of the pendulum follows Eq.~(\ref{eq:thermal}), it is essential that the temperature of the dissipative element---the fiber---matches the bath temperature \( T \). This condition is not trivial, as thermalization via conduction in an insulating thin fiber becomes negligible below 1~K. Consequently, the thermalization of the pendulum fiber must rely on the weak conduction channel provided by residual gas, while ensuring that the pressure remains as low as \( 10^{-12} \)~mbar to sufficiently suppress gas damping~\cite{Braginsky1977}. Thus, thermalization of the pendulum is feasible, provided that heat leakage is minimized. This imposes a stringent constraint on the power dissipated by the measurement apparatus.

Suppression of environmental noise is a well-known challenge in any precision measurement involving low-frequency mechanical systems. The proposed experiment will need to be conducted in an extremely quiet environment, such as a deep underground site with minimal seismic and Newtonian noise. If sufficient isolation cannot be achieved on Earth, an alternative would be to consider performing the experiment in space. In fact, the lowest acceleration noise to date of $\sim1.7 \, \mathrm{fm/s}^2/\sqrt{\mathrm{Hz}}$ has been measured in space by the LISA Pathfinder mission \cite{lisa2016,lisa2018}, with only a minor unmodeled 1/f noise component lower than $1 \, \mathrm{fm/s}^2/\sqrt{\mathrm{Hz}}$ at 1 mHz \cite{cesarini2024}. An intermediate option could be a future laboratory on the Moon, which would benefit from the virtual absence of seismic and human-generated noise.

Finally, let us consider the issue of the near-quantum-limited readout. Optomechanical cavities can operate at the standard quantum limit and even beyond in low-frequency macroscopic mechanical resonators, as demonstrated by gravitational wave detectors~\cite{LIGO2023}. However, the optical cavity power required to reach the quantum limit is likely incompatible with the extremely low absorption levels that can be tolerated by an apparatus at millikelvin temperature, thermalized through residual gas. This issue can be circumvented by employing superconducting~\cite{Paik1976} or ferromagnetic~\cite{Vinante2020b} transduction, potentially coupled to a remote quantum-limited SQUID. Microwave-operated SQUID-based tunable resonators can, in principle, achieve quantum-limited performance. While experimental demonstrations in low-frequency mechanical resonators are still lacking, ongoing research is progressing toward this milestone~\cite{Schmidt2024}. A necessary prerequisite is the strong suppression of \( 1/f \) noise, a well-known technical challenge but not a fundamental limitation~\cite{Kumar2016}. In parallel with standard torsion pendulums, other related approaches could be explored and may turn out to be more effective. Levitated systems exhibiting torsion-like behavior can be realized using the Meissner effect~\cite{Ahrens2024}. While current quantum magnetomechanics experiments focus on smaller scales and higher frequencies~\cite{Vinante2020b,Schmidt2024}, levitated systems with masses on the order of kilograms could naturally exhibit libration frequencies in the millihertz range and potentially achieve extremely low dissipation levels, unconstrained by elastic losses in materials. Recent measurements on freely spinning Meissner-levitated magnets have shown very low damping factors $\eta \sim10^{-7}$ Hz \cite{jose2025} at $4.2$ K with foreseeable improvements by orders of magnitude at lower temperature. The latter result suggests a different promising approach, yet to be explored. By spinning and synchronizing a levitated magnetic dumbbell to high frequency, well beyond kHz, one may realize effective torsion pendulums with ultralow dissipation able to evade the issue of low frequency vibrational noise.

\section{Conclusion}
If gravity is classical, then not only it cannot entangle initially separated quantum systems~\cite{bose2017spin,marletto2017gravitationally}, but—this is the first main result of this work—it must also induce diffuse motion in those systems. This fundamental feature paves the way for a new and significantly more accessible class of experiments aimed at testing the  nature of gravity. Instead of requiring the engineering and manipulation of quantum states of massive systems to assess whether their evolution adheres to the unitary dynamics predicted by quantum gravity or deviates from it, one can perform an essentially classical experiment involving macroscopic masses, where their motion is precisely monitored.

\begin{table}[t]
\begin{tabular}{l|c|c|c|c}
\hline
\hline
Requirement & BMV & KM & LPP & Us \\ 
\hline
Extreme isolation $Q/T$ & yes & yes & yes & yes \\

Simultaneous control of two masses & yes & yes & yes & no \\

Tight constraint on size/geometry/alignment & yes & no & no & no \\

Extreme repeatability & yes & no & yes & no \\

Ground state cooling/Quantum Kalman filter & no & yes & yes & no \\

Continuous measurement/power & no & yes & no & no \\

Computation of Quantum Evolution & no & no & yes & no \\
\hline
\hline
\end{tabular}
\caption{\justifying\textbf{Comparison of key operational requirements of different proposals for testing classicality/quantumness of gravity}: the proposals are labeled as BMV (Stern-Gerlach gravitationally induced entanglement according to Bose et al \cite{bose2017spin} and Marletto-Vedral \cite{marletto2017gravitationally}), OM (entanglement generation in continuously monitored optomechanical setup such as in Krisnanda et al \cite{krisnanda2020observable} or in Miki et al \cite{Miki2024}), Lami (test of LOCC via quantum evolution of displaced coherent states without entanglement \cite{lami2024testing}), Us (test of classical gravity via classical noise, this work). The first requirement of extreme environmental isolation is common to all proposals; the present proposal however disposes of the other operational requirements, for the key reason that it does not rely on advanced quantum engineering.}
\label{table2}
\end{table}

 While the first class of quantum experiments demands multiple technological breakthroughs to reach the necessary level of control of quantum coherence, the second class is, though challenging, much more tolerant in terms of technical and operational requirements. This constitutes the second key result of this work. To illustrate this point, we summarize in Table \ref{table2} the requirements of some representative proposals against ours. We consider the Stern-Gerlach gravitationally induced entanglement by BMV \cite{bose2017spin,marletto2017gravitationally}, the gravitationally induced entanglement in a continuously monitored double optomechanical setup \cite{krisnanda2020observable, Miki2024}, and the test of LOCC via quantum evolution of displaced coherent states without entanglement by Lami et al. \cite{lami2024testing}). The first requirement, namely the extreme isolation from environment, i.e. the very high $Q/T$ factor, appears to be similar for all proposals. This is expected, since classical gravity noise is by definition the amount of noise needed to suppress gravitationally-induced entanglement. In the following rows we compare the operational details of the protocol, arguing that our proposal is by far the one featuring the most tolerant requirements: (i) it does not require the ability to simultaneously and independently prepare-evolve-measure or monitor-control two masses; (ii) it does not suffer, as Stern-Gerlach proposals do, from critical size/geometry/alignment requirements, which in turn imply a large  class of potentially uncontrollable decoherence effects (short distance electric and magnetic interactions via impurities, Casimir effects, anisotropies, alignment issues, rototranslational couplings and so on); (iii) being based on a simple reheating rate protocol, it does not require extreme accuracy in state preparation, and is more tolerant to parameter drifts; (iv) it does not require ground-state cooling, although its implementation would optimize the measurement speed; (v) it can be operated in the dark, apart of state preparation, thereby minimizing back-action noise and heating, in contrast with continuous optomechanical proposals which need a constant optical power for high accuracy monitoring \cite{krisnanda2020observable, Miki2024}; (vi) it does not need any computation of quantum evolution following state preparation as in Lami proposal \cite{lami2024testing}, a step which requires an extreme accuracy in the state preparation.

One of the most notable advances in recent research on quantum gravity has been the realization that experimental tests can, in principle, be shifted from the unattainable Planck scale to low-energy, table-top experiments~\cite{carney2019tabletop,bose2025massive}. We have demonstrated a concrete route to making such tests feasible, thereby offering a viable experimental framework to probe the classical vs quantum nature of gravity.

\section*{Acknowledgments}
S.D. acknowledges support from the UKRI through grant EP/X021505/1 and by INFN. A.V. acknowledge support from the QuantERA II Programme (project LEMAQUME) that has received funding from the European Union’s Horizon 2020 research and innovation programme under Grant Agreement No 101017733. A.B. acknowledges financial support from the EIC Pathfinder project QuCoM (GA no. 101046973), the PNRR PE National Quantum Science and Technology Institute (GA no. PE0000023), the University of Trieste and INFN. O.A., G.D.B. and J.L.G.R. acknowledge financial support from the EIC Pathfinder project QuCoM (GA no. 101046973), the University of Trieste and INFN. J.L.G.R. acknowledges financial support from OIST. 

\hfill
\appendix

\section{Approximate faithfulness}\label{app:varepsilon}
In this Appendix we consider the more realistic scenario where the probe provides an accurate but not perfect record of the particle's position via the gravitational interaction, as it actually happens when the noise blurs the interaction.

Assume the conditions of Theorem~2 and the approximate faithfulness conditions~\eqref{eq:faithfulness-eps-narrative} and~\eqref{eq:faithfulness-eps-narrative2} with $\varepsilon \neq 0$. We will prove the analog of theorems 3, 4, 5, 6 of the main text.

{\bf Theorem $\varepsilon$3: $\mathcal M$ is not unitary.}
Assume the classicality condition~\eqref{eq:WeakC2} and the approximate faithfulness
condition~\eqref{eq:faithfulness-eps-narrative} for some injective map $g:X\to Y$ and some
$\varepsilon\in[0,1)$.
Assume moreover that there exist at least two distinct sectors $x\neq x'$.
Then $\mathcal M$ is not unitary.

\smallskip
{\it Proof.} Assume by contradiction that $\mathcal{M} =U(\cdot)U^\dagger $. We will show that for unitary evolutions, because they preserve purity, the only way to satisfy  $\varepsilon$-faithfulness is by satisfying exact faithfulness, i.e. by setting $\varepsilon = 0$. Then Theorem 3 of the main text applies, and thus $\mathcal{M}$ cannot be unitary. 
\\
Consider a probe preparation $\tau_{y_0}=\sum_{i}p_i^{y_0}\ket{\eta_i}\bra{\eta_i}  \in \mathsf{S}_{y_0}$; note that the eigenvectors $\ket{\eta_{i}}$ are also supported on $\Pi_{y_0}$. Then, consider a pure state $\ket{\phi_x}\bra{\phi_x}\in \mathsf{S}_{x}$ and any $\ket{\eta_i}$. By applying approximate faithfulness to the classical state $\ket{\phi_x}\bra{\phi_x}\otimes \ket{\eta_i}\bra{\eta_i}$ we have 
\begin{align}
&\Tr_B[(\mathbb{I}_{B_1}\otimes \Pi_{y_x})\mathcal{M}(\ket{\phi_x}\bra{\phi_x}\otimes\ket{\eta_i}\bra{\eta_i})] =\nonumber\\ &=\Tr_B[(\mathbb{I}_{B_1}\otimes \Pi_{y_x})U(\ket{\phi_x}\bra{\phi_x}\otimes\ket{\eta_i}\bra{\eta_i})U^\dagger]\nonumber\\
&=\big\|(\mathbb I_{B_1}\otimes \Pi_{y_x})\,U(|\phi_x\rangle\otimes|\eta_i\rangle)\big\|^2\nonumber\\
&\ge 1-\varepsilon,
\end{align}
from which it follows that 
\begin{equation}\label{UnitEvoNorm}
    \big\|(\mathbb I_{B_1}\otimes (\mathbb{I}_{B_2}-\Pi_{y_x}))\,U(|\phi_x\rangle\otimes|\eta_i\rangle)\big\|^2\le \varepsilon. 
\end{equation}
Eq.~\eqref{UnitEvoNorm} gives an estimate of the norm of the component of the pure state $\ket{\Psi} = U(\ket{\phi_x}\otimes \ket{\eta_i})$ on sectors other than $\Pi_{y_x}$. If this norm is strictly greater than zero, then the pure state $\ket{\Psi}$ has support on more than one probe sector. For concreteness, suppose that $\ket{\Psi}$ has support on exactly two sectors. This means that $$(\Pi_{y}+\Pi_{y'})\ket{\Psi}= \ket{\Psi},$$
which implies 
\begin{align}
    \ket{\Psi}\bra{\Psi} &= \Pi_y\ket{\Psi}\bra{\Psi}\Pi_y+\Pi_y\ket{\Psi}\bra{\Psi}\Pi_{y'}+\nonumber\\
    &+\Pi_{y'}\ket{\Psi}\bra{\Psi}\Pi_{y}+\Pi_{y'}\ket{\Psi}\bra{\Psi}\Pi_{y'}\,.
\end{align}
From simple inspection, this state is not classical according to~\eqref{eq:WeakC2}: $ \Gamma_Y(\ket{\Psi}\bra{\Psi}) \neq \ket{\Psi}\bra{\Psi}$. 
Yet $\ket{\Psi}$ is the evolved state of a classical pure state $\ket{\phi_x}\otimes \ket{\eta_i}$ and thus must be classical. Therefore, to avoid any contradiction one must have that
\begin{equation}
    \big\|(\mathbb I_{B_1}\otimes (\mathbb{I}_{B_2}-\Pi_{y_x}))\,U(|\phi_x\rangle\otimes|\eta_i\rangle)\big\|^2 = 0.
\end{equation}
But then, for the probe $\tau_{y_0}$ and a generic $\sigma_x = \sum_j p_j^x\ket{\phi^j_x}\bra{\phi^j_x}\in \mathsf{S}_x$ we have
\begin{align}
    &\Tr_B[\widetilde{\mathcal{J}}_{y_x}(\sigma_x)] = \sum_{j}p^x_j\Tr[\tilde{\mathcal{J}}_{y_x}(\ket{\phi^x_j}\bra{\phi^x_j})]\nonumber\\
    &\sum_{j}p^x_j\Tr_B[(\mathbb{I}_{B_1}\otimes \Pi_{y_x})\mathcal{M}(\ket{\phi^x_j}\bra{\phi^x_j}\otimes\tau_{y_0})]\nonumber\\
    &=\sum_{ji}p^x_jp^{y_0}_i\Tr_B[(\mathbb{I}_{B_1}\otimes \Pi_{y_x})\mathcal{M}(\ket{\phi^x_j}\bra{\phi^x_j}\otimes\ket{\eta_i}\bra{\eta_i})]\nonumber\\
    &=\sum_{ji}p^x_jp^{y_0}_i\Tr_B[(\mathbb{I}_{B_1}\otimes \Pi_{y_x})U\ket{\phi^x_j}\bra{\phi^x_j}\otimes\ket{\eta_i}\bra{\eta_i}U^\dagger]\nonumber\\
    &=\sum_{ji}p^x_jp^{y_0}_i\norm{(\mathbb{I}_B\otimes\Pi_{y_x})U\ket{\phi^x_j}\otimes\ket{\eta_i}}^2\nonumber\\
    &=1,
\end{align}
which is nothing but exact faithfulness, as we set out to prove. The conclusion follows from Theorem 3 of the main text. \qed

\smallskip
Under the assumptions of exact faithfulness we were able to fully characterize the supports of the map $\mathcal{M}$ in Theorem 4. In particular we showed that $\mathcal{M}$ depends only on the block diagonal of any input
\begin{equation}
    \mathcal{M}(\hat\rho\otimes \tau_{y_0}) = \mathcal{M}(\Gamma_X\hat\rho\otimes \tau_{y_0}),
\end{equation}
and, furthermore, the output is entirely supported on the set $\{\Pi_x\otimes\Pi_{y_x}\}_{x\in X}$.
\\
Approximate faithfulness, instead, leaves some room for the map to move coherences around, which makes the discussion more delicate. However, even in the case of an inaccurate record, the probe is still capable of distinguishing different configurations of $B_1$, and we are going to show that this still causes the collapse---although not projective.
\\
We firstly address the characterization of the input dependence. By continuity, for small enough $\varepsilon$ one would expect that $\mathcal{M}(\Pi_a X\Pi_b\otimes\tau_{y_0})\sim 0$ for $a\neq b$. In the next theorem we show an example of a sufficient condition under which such intuition can be made precise. 
\\

{\bf Theorem $\varepsilon$4A: Approximate collapse.} Assume that $\forall \hat{\sigma}_x\in \mathsf{S}_x$
\begin{equation}\label{SinglErrors}
\Tr[\widetilde{\mathcal{J}}_y(\hat{\sigma}_x)]\le \varepsilon_y(x)\,,\quad \forall y\neq y_x,
\end{equation}
where the $\varepsilon_y(x)\in[0,1)$ are such that there exists a $\sqrt{\kappa} \in [0,1)$  such that
\begin{equation}\label{convergErr}
    \sum_{y\neq y_x}\sqrt{\varepsilon_y(x)} \le \sqrt{\kappa}.
\end{equation}
Then, given two distinct sectors $\Pi_a,\Pi_b$ where $a\neq b$, for all $X\in \mathcal{T}(\mathcal{H}_{B_1})$ 
\begin{equation}
    \norm{\mathcal{M}(\Pi_aX\Pi_b\otimes \tau_{y_0})}_1\le(2\sqrt{\kappa}+\kappa)\norm{\Pi_aX\Pi_b}_1.
\end{equation}
\\
{\it Remark} Notice that  Eqs.~\eqref{SinglErrors}-\eqref{convergErr} are sufficient conditions for approximate faithfulness. Indeed we have
\begin{align}
   \varepsilon =  \sum_{y\neq y_x}\varepsilon_y(x) \le \qty(\sum_{y\neq y_x}\sqrt{\varepsilon_y(x)} )^2 \le\kappa<1.
\end{align}
\\
{\it Proof.} Defining the effects as in the main text $\widetilde{E}_y \coloneqq \widetilde{\mathcal{J}}_y^\dagger(\mathbb{I}_{B})\le \mathbb{I}_{B_1}$, we have
\begin{align}
    \Tr_{B}[\widetilde{\mathcal{J}}_y(\sigma_x)] &= \Tr_{B_1}[\widetilde{E}_y\sigma_x] \nonumber\\
     &= \Tr_{B_1}[\Pi_x\widetilde{E}_y\Pi_x\sigma_x] \nonumber\\
     &\le \varepsilon_y(x). \nonumber\\
\end{align}
Since the inequality has to be valid $\forall \sigma_x$ we have, by the definition of the operator norm
\begin{equation}
    \norm{\Pi_x\widetilde{E}_y\Pi_x}_\infty \le \varepsilon_y(x)\,,\qquad \forall y\neq y_x.
\end{equation}
We will now compute 
\begin{align}
\norm{\mathcal{M}(\Pi_aX\Pi_b\otimes \tau_{y_0})}_1 &=\norm{\sum_{y}\widetilde{\mathcal{J}}_y(\Pi_aX\Pi_b)}_1\nonumber\\
&\le\sum_{y}\norm{\widetilde{\mathcal{J}}_y(\Pi_aX\Pi_b)}_1,
\end{align}
where we used the triangle inequality. Now each $\widetilde{\mathcal{J}}_y$ is a CP map, therefore by the Stinespring theorem it can be written as
\begin{equation}
    \widetilde{\mathcal{J}}_y(\cdot) = \Tr_{E}[V_y(\cdot)V_y^\dagger],
\end{equation}
on an enlarged Hilbert space $\mathcal{H}_{B}\otimes\mathcal{H}_E$, where, in particular, $\widetilde{E}_y = V_y^\dagger(\mathbb{I}_B\otimes\mathbb{I}_E) V_y$. We then have the following chain 
\begin{align}\label{eq:BoundSum}
\sum_y\norm{\widetilde{\mathcal{J}}_y(\Pi_aX\Pi_b)}_1&\!\!=\!\! \sum_y\norm{\Tr_EV_y(\Pi_aX\Pi_b)V^\dagger_y}_1\nonumber\\
     &\overset{(1)}{\le}\!\! \sum_y\norm{V_y(\Pi_aX\Pi_b)V^\dagger_y}_1\nonumber\\
     &=\!\! \sum_y\norm{V_y\Pi_a(\Pi_aX\Pi_b)\Pi_bV^\dagger_y}_1\nonumber\\
     &\overset{(2)}{\le}\!\! \sum_y\norm{V_y\Pi_a}_\infty\norm{\Pi_aX\Pi_b}_1\norm{\Pi_bV^\dagger_y}_\infty\nonumber\\
      &\!=\!\! \sum_y\big(\norm{V_y\Pi_a}_\infty\norm{V_y\Pi_b}_\infty)\norm{\Pi_aX\Pi_b}_1,\nonumber\\
\end{align}
where $(1)$ follows from the fact that the trace norm is contractive under the action of CPTP maps, such as the partial trace, while (2) is the submultiplicativity of the Schatten norms. Now recall that $\norm{X}_\infty = s_\text{max}$ where $s_\text{max}$ is the largest singular value of $X$. The singular values are the square roots of the eigenvalues of $X^\dagger X$, i.e. $\norm{X}_\infty^2 =\norm{X^\dagger X}_\infty$. Applying it to the terms in the sum~\eqref{eq:BoundSum} we have
\begin{align}
\norm{V_y\Pi_a}_\infty^2 &= \norm{\Pi_aV_y^\dagger V_y\Pi_a}_\infty\nonumber\\
&=\norm{\Pi_a\widetilde{E}_y\Pi_a}_\infty\nonumber\\
&=
    \begin{cases}
        \le 1\ \text{if}\ y= g(a)\\
        \le \varepsilon_y(a)\ \text{otherwise}\,,
    \end{cases}
\end{align}
with a similar bound holding for $\norm{V_y\Pi_b}_\infty$. We are now in a position to bound each term entering the sum. Since the sectors are different $a\neq b$ and $g$ is injective, so $g(a)\neq g(b)$, we have
\begin{align}
&\sum_{y}\norm{V_y\Pi_a}_\infty\norm{V_y\Pi_b}_\infty=\nonumber\\
&=\sum_{y\neq g(a),g(b)} \norm{V_y\Pi_a}_\infty\norm{V_y\Pi_b}_\infty \nonumber\\
&\quad+\norm{V_{g(a)}\Pi_a}_\infty\norm{V_{g(a)}\Pi_b}_\infty+\norm{V_{g(b)}\Pi_a}_\infty\norm{V_{g(b)}\Pi_b}_\infty\nonumber\\
&\le\sum_{y\neq g(a),g(b)} \Big(\sqrt{\varepsilon_y(a)}\sqrt{\varepsilon_y(b)}\Big) + \sqrt{\varepsilon_{g(a)}(b)}+\sqrt{\varepsilon_{g(b)}(a)} \nonumber \\
&\le\Big(\sum_{y\neq g(a)}\sqrt{\varepsilon_y(a)}\Big)\Big(\sum_{y\neq g(b)}\sqrt{\varepsilon_y(b)}\Big) +  2\sqrt{\kappa}\\
&\le \kappa +2\sqrt{\kappa}.
\end{align}
Therefore
\begin{equation}
    \norm{\mathcal{M}(\Pi_aX\Pi_b\otimes \tau_{y_0})}_1\le (\kappa+2\sqrt{\kappa})\norm{\Pi_aX\Pi_b}_1,
\end{equation}
as we set out to prove. \qed

\smallskip
We now move onto addressing the characterization of the output of the map $\mathcal{M}$ under approximate faithfulness. Recall that in the case of exact faithfulness we established in Theorem 4 that the output of the  map is supported precisely on the sectors $\{\Pi_x\otimes\Pi_{y_x}\}_{x\in X}$. It follows that the probability of observing the outcomes $(x',y_x),\ x\neq x'$ is zero:
\begin{equation}
    P(x',y_x|\hat \rho) = \Tr_B[(\Pi_{x'}\otimes \Pi_{y_x})\mathcal{M}(\hat\rho)] = 0,\quad\ \text{for}\ x\neq x'. 
\end{equation}
Whenever the probe does not provide an exact record, this probability will not be exactly zero: let us define the total error probability for a given state $\hat \rho$ associated with the outcome $x$ as
\begin{equation}
    e_{\hat\rho}(x) \coloneqq \sum_{x'\neq x}P(x',y_x|\hat \rho).
\end{equation}
First, we will show that these errors can be bounded for ensembles as those considered in the Gedankenexperiment. 
\\

{\bf Theorem $\varepsilon$4B: $\mathcal{M}$ approximately collapses}\\
Assume the setting of Theorem~2 and the $\varepsilon$--faithfulness conditions
\eqref{eq:faithfulness-eps-narrative}--\eqref{eq:faithfulness-eps-narrative2}
for some $\varepsilon\in[0,1)$ and an injective map $g:X\to Y$.  Fix $x\in X$ and set $y_x:=g(x)$.
Let $\{(r_k,\hat\rho_k)\}_k$ be any ensemble of input states on $B_1$ with average state
\begin{equation}\label{eq:avg_classical_global}
\hat\sigma:=\sum_k r_k\,\hat\rho_k=\sum_{x'\in X} p(x')\,\hat\sigma_{x'},
\end{equation}
for some probability distribution $p(x)$ and localized states $\hat{\sigma}_{x'}\in\mathsf{S}_{x'}$. Then 
\begin{equation}
    \sum_{k}r_k e_{\hat\rho_k}(x) \le \varepsilon.
\end{equation}
\medskip
{\it Proof.} The theorem follows from a direct computation. Indeed, we have
\begin{align}
    \sum_{k}r_ke_{\hat\rho_k}(x) &=\sum_{k} r_k \sum_{x' \neq x}P(x',y_x|\hat\rho_k) \nonumber\\
&=\sum_{k,y}r_k\Tr_B[((\mathbb{I}_{B_1}-\Pi_x)\otimes\Pi_{y_x})\widetilde{\mathcal{J}}_{y}(\hat\rho_k)]\nonumber\\
    &=\Tr_B[((\mathbb{I}_{B_1}-\Pi_x)\otimes\mathbb{I}_{B_2})\widetilde{\mathcal{J}}_{y_x}\Big(\sum_{k}r_k\hat\rho_k\Big)]\nonumber\\
    &=\Tr_B[((\mathbb{I}_{B_1}-\Pi_x)\otimes\mathbb{I}_{B_2})\widetilde{\mathcal{J}}_{y_x}\Big(\sum_{x'}p(x')\hat\sigma_{x'}\Big)]\nonumber\\
    &=\sum_{x'}p(x')\Tr_B[((\mathbb{I}_{B_1}-\Pi_x)\otimes\mathbb{I}_{B_2})\widetilde{\mathcal{J}}_{y_x}(\hat\sigma_{x'})]\nonumber.
\end{align}
Each term in the sum can be bounded by applying the faithfulness conditions. Indeed the term $x'=x$ is bounded by the condition~\eqref{eq:faithfulness-eps-narrative2}
\begin{equation}
    \Tr_B[((\mathbb{I}_{B_1}-\Pi_x)\otimes\mathbb{I}_{B_2})\widetilde{\mathcal{J}}_{y_x}(\hat{\sigma}_{x})]\le \varepsilon\,,
\end{equation}
while for $x'\neq x$ we have
\begin{align}
\Tr_B[((\mathbb{I}_{B_1}-\Pi_x)\otimes\mathbb{I}_{B_2})\widetilde{\mathcal{J}}_{y_x}(\hat{\sigma}_{x'})]&\le\Tr_B[\mathbb{I}_B\widetilde{\mathcal{J}}_{y_x}(\hat{\sigma}_{x'})]\nonumber\\
&\le \sum_{y\neq y_{x'}}\Tr_B[\widetilde{\mathcal{J}}_{y}(\hat{\sigma}_{x'})]\nonumber\\
&\le \varepsilon\,,
\end{align}
by the condition~\eqref{eq:faithfulness-eps-narrative}. Therefore
\begin{equation}
    \sum_{k}r_ke_{\hat\rho_k}(x) \le \sum_{x'}p(x') \varepsilon = \varepsilon.
\end{equation}
\\
\\
{\it Remark} We have shown that the average error probability for ensemble preparations which average to classical states can be bounded by $\varepsilon$. Since we are working with positive random variables, we can use Markov's inequality to estimate the weight of the outliers: let $\delta >0$, then
\begin{equation}
    \sum_{k:\ e_{\hat\rho_k}>\delta}r_k \le \frac{1}{\delta}\sum_{k}r_ke_{\hat\rho_k}(x)\le\frac{\varepsilon}{\delta}.
\end{equation}
Thus, for a given $\delta>0$, the outliers for which the $e_{\hat\rho_k}> \delta$ make up but a fraction $\varepsilon/\delta$ of the ensemble. Thus, for accurate enough probes, these outliers are exceptional.
\\
Now we show that the results of Theorem~$\varepsilon$4B can be used to show that an arbitrary input state must approximately collapse. Given two bounded operators $X$ and $Y$, let us define 
\begin{align}\label{def_rel_entr}
    d_{\text{max}}(X||Y ) &=\text{min}\{\lambda>0:\ X\le \lambda Y\}.
\end{align}
Notice that the max-relative entropy $D_\text{max}(X||Y)$ is readily computed from $d_\text{max}(X||Y)$  as $D_\text{max}(X||Y) = \text{log}_2 \,d_\text{max}(X||Y)$ ~\cite{datta2009min,Wilde2017}. We then have the following theorem.
\\

{\bf Theorem $\varepsilon$4C: collapse of arbitrary input} Assume the setting of Theorem~2 and the $\varepsilon$--faithfulness conditions
\eqref{eq:faithfulness-eps-narrative}--\eqref{eq:faithfulness-eps-narrative2}
for some $\varepsilon\in[0,1)$ and an injective map $g:X\to Y$. Let $\hat \rho\in \mathcal{T}(\mathcal{H}_{B_1})$ be an arbitrary non-classical input state. Let $\hat{\sigma} = \Gamma_X\hat\rho\neq \hat\rho$ be the dephased part of $\hat{\rho}$. Then
\begin{equation}\label{Eq_thr_4c}
    e_{\hat\rho}(x) \le \varepsilon \, d_\text{max}(\hat\rho||\hat\sigma).
\end{equation}
\medskip
{\it Proof.} Consider the set $\{r>0:  r\hat \rho\le \hat\sigma \}$. Notice that for $r\ge 1$ the inequality trivially cannot be satisfied, so $r\in[0,1)$. Then for all such values of $r$ we know that $\hat\sigma - r\hat\rho\ge 0$; therefore we can define
\begin{equation}
    \hat \tau \coloneqq \frac{\hat\sigma -r \hat\rho}{1-r},
\end{equation}
which is a {\it bona fide} quantum state, since it is positive and its trace is one.  Consider then the two elements ensemble 
\begin{equation}
   \{(r_k,\hat\rho_k)\}= \{(r,\hat\rho),(1-r,\hat\tau)\}.
\end{equation}
It is clear that $\hat{\sigma}= \sum_{k}r_k\hat\rho_k$. Therefore
\begin{align}
    \sum_{k}r_k\hat\rho_k &= \sum_{x}\Pi_x\hat\rho\Pi_x= \sum_{x}\Tr[\Pi_x\hat\rho]\frac{\Pi_x\hat\rho}{\Tr[\Pi_x\hat\rho\Pi_x]}\nonumber\\
    &\coloneqq \sum_{x}p(x)\hat{\sigma}_x.
\end{align}
We can therefore apply the results of Theorem $\varepsilon$4B and we will have
\begin{equation}
    r e_{\hat\rho}(x) \le \sum_{k}r_ke_{\hat\rho_k}\le \varepsilon\ \implies\ e_{\hat\rho}(x)\le \frac{\varepsilon}{r}.
\end{equation}
The strongest bound is set by taking as $r$ the max$\{r>0: r\hat \rho\le \hat\sigma\}$, which, from the definition~\eqref{def_rel_entr}, is exactly equal to $d_\text{max}(\hat\rho||\hat\sigma)^{-1}$. Therefore we have
\begin{equation}
    e_{\hat\rho}(x) \le  \varepsilon d_\text{max}(\hat\rho||\hat\sigma),
\end{equation}
as we wanted to prove. \qed
\\
\\
{\it Remark.} Given a state $\hat\rho$ for which $p(y_x|\hat\rho)\neq 0$, we have that Eq.~\eqref{Eq_thr_4c} for the conditioned state $\hat\varrho_{|y_x} = \tilde{\mathcal{J}}_{y_x}(\hat\rho)/p(y_x|\hat\rho)$ reads, explicitly
\begin{equation}
    \Tr_B[((\mathbb{I}_{B_1}-\Pi_x)\otimes \mathbb{I}_{B_2})\hat\varrho_{|y_x}]\le \frac{\varepsilon 2^{D_\text{max}(\hat\rho||\hat\sigma)}}{p(y_x|\hat\rho)}.
\end{equation}
We see that the state conditioned on the outcome $y_x$ is approximately collapsed onto $\Pi_x\otimes\Pi_{y_x}$ with the error being controlled by the faithfulness of the probe and the max-relative entropy between $\hat\rho$ and its dephased counterpart.
\\

{\it Remark.} Note that for $\hat\sigma = \Gamma_X\hat\rho$ we have that supp$\hat\rho\subset$ supp$\hat\sigma$; therefore we have the standard result~\cite{datta2009min}:
\begin{equation}
    d_\text{max}(\hat\rho||\hat\sigma) = \norm{\hat\sigma^{-1/2}\hat\rho\hat\sigma^{-1/2}}_\infty\,.
\end{equation}
This allows the explicit evaluation of the bound~\eqref{Eq_thr_4c} in Theorem $\varepsilon$4C. Clearly, the bound becomes nontrivial whenever 
$\big\|\hat\sigma^{-1/2}\hat\rho\,\hat\sigma^{-1/2}\big\|_\infty$
is $O(1)$. To better understand the size of the bound in the present context, consider rank--$1$ sectors $\Pi_j=|j\rangle\!\langle j|$, $j=1,\dots,n$, and the pure state uniformly delocalized across $n$
sectors,
\begin{equation}
|\psi\rangle=\frac{1}{\sqrt n}\sum_{j=1}^n |j\rangle.
\end{equation}
Its dephased part is the incoherent mixture $
\hat\sigma=\Gamma_X(\ket{\psi}\bra{\psi})=\frac{1}{n}\sum_{j=1}^n |j\rangle\!\langle j|$.
Thus, physically, $\hat\rho$ exhibits full interference across $n$ classical alternatives, whereas
$\hat\sigma$ assigns probability $1/n$ to each of them. Then, a direct computation gives \begin{equation}
\big\|\hat\sigma^{-1/2}\hat\rho\,\hat\sigma^{-1/2}\big\|_\infty=\norm{\sum_{i,j=1}^{n}\ket{i}\bra{j}}_\infty=n.
\end{equation}
As a rule of thumb, for a state with
coherent support over an effective number $n_{\rm coh}$ of classical sectors, one typically expects
\begin{equation}
\big\|\hat\sigma^{-1/2}\hat\rho\,\hat\sigma^{-1/2}\big\|_\infty\sim n_{\rm coh},
\end{equation}
while it reduces to $1$ when $\hat\rho$ is already block--diagonal, i.e.\ when
$\hat\rho=\hat\sigma$. 

Having proven that allowing for an imperfect record the map still approximately collapses the state, we specialize to $\mathcal{H}_{B_1}= \mathcal{H}_{B_2} = L^2(\mathbb{R})$ and move onto showing that this must produce diffusion in the motion of quantum systems. We will assume the same interval coarse graining of the main text: fixing $\ell>0$ we have
\begin{equation}\label{eq:interval_projectors}
I_n:=[n\ell,(n+1)\ell),\qquad 
\Pi_n:=\int_{I_n}\!|x\rangle\!\langle x|\,dx,
\qquad n\in\mathbb Z .
\end{equation}

{\bf Theorem $\varepsilon$5:  Change in total momentum statistics}
Assume $\mathcal{H}_{B_1} = \mathcal{H}_{B_2} = L^2(\mathbb{R})$ and, further, the setting of Theorem $\varepsilon$4A; we then have that
\begin{equation}
\norm{\mathcal{M}(\Pi_aX\Pi_b\otimes\tau_{y_0})}_1\le (\kappa +2\sqrt{\kappa})\norm{\Pi_a X\Pi_b}_1\,.
\end{equation}
Assume that $2(\kappa+2\sqrt{\kappa})<1$; then the total $\hat P_\text{tot} = \hat p_1+\hat p_2$ cannot be conserved.
\\
\\
{\it Proof.} As in the main text, let us define the total momentum characteristic function of a state $\hat\rho$:
\begin{equation}
\Xi_{\hat\rho}(d) = \Tr_B[\hat \rho e^{-\frac{i}{\hbar}d\hat P_\text{tot}}]\,.
\end{equation}
To show that the total momentum $\hat P_\text{tot}$ cannot be conserved we will show that there exist two states whose difference of characteristic functions changes after applying $\mathcal{M}$. Let $\Pi_n$ and $\Pi_{n+1}$ be the projectors corresponding to two adjacent cells and let $\ket{\psi_n}$ and $\ket{\psi_{n+1}}$ denote any two normalized states belonging to such sectors. Defining $X_{a,b} \coloneqq \ket{\psi_a}\bra{\psi_b}$, we consider the states of the form
\begin{equation}
    \hat{\rho}_{\pm} = \frac{1}{2}\Big(X_{n,n}+X_{n+1,n+1}\pm X_{n+1,n}\pm X_{n,n+1}\Big)\otimes\tau_{y_0}.
\end{equation} 
It follows that
\begin{align}
    &\Xi_{\hat \rho_+}(d)\!-\!\Xi_{\hat \rho_-}(d)\! = \!\Tr_{B_1}[(X_{n+1,n}\!+\!X_{n,n+1})e^{-\frac{i}{\hbar}d\hat p_1}]\chi_{\tau_{y_0}}(d)\nonumber\\
    &=\Big(\bra{\psi_n}e^{-\frac{i}{\hbar}d\hat p_1}\ket{\psi_{n+1}}+\bra{\psi_{n+1}}e^{-\frac{i}{\hbar}d\hat p_1}\ket{\psi_{n}}\Big)\chi_{\tau_{y_0}}(d)\nonumber\\
    &=\bra{\psi_{n+1}}e^{-\frac{i}{\hbar}d\hat p_1}\ket{\psi_{n}}\chi_{\tau_{y_0}}(d),\nonumber\\
\end{align}
where we used the fact that $\bra{\psi_{n}}e^{-\frac{i}{\hbar}d\hat p_1}\ket{\psi_{n+1}} = 0$ for such states supported on adjacent cells. Now, $|\chi_{\tau_{y_0}}(0)| = 1$ and, by definition, $|\chi_{\tau_{y_0}}(d)| \le 1\ \forall\ d$. Therefore, by continuity, for any $\epsilon>0$ there always exists a $\delta_\epsilon$ such that for all $|d|\le \delta_\epsilon$ $|\chi_{\tau_{y_0}}(d)| > 1-\epsilon$. In particular, let $\epsilon = 1-2(\kappa +2\sqrt{\kappa})$ and $0<d^*<\delta_\epsilon$. 
\\
Now choose a state $\ket{\psi_{n+1}}$ sufficiently close to the boundary of the sectors corresponding to $\Pi_n$, $\Pi_{n+1}$ such that its translated counterpart $e^{\frac{i}{\hbar}d^*\hat p_1}\ket{\psi_{n+1}}$ lies entirely on the sector $\Pi_n$. Choosing then $\ket{\psi_{n}}=e^{\frac{i}{\hbar}d^*\hat p_1}\ket{\psi_{n+1}}$ we have
\begin{align}\label{chard^*}
    |\Xi_{\hat \rho_+}(d^*)-\Xi_{\hat \rho_-}(d^*) |&=|\bra{\psi_{n+1}}e^{-\frac{i}{\hbar}d^*\hat p_1}\ket{\psi_{n}}\chi_{\tau_{y_0}}(d^*)|\nonumber\\
    &=|\chi_{\tau_{y_0}}(d^*)|\nonumber\\
    &>2(\kappa +2\sqrt{\kappa}). 
\end{align}
Now let us consider what happens to the difference of characteristic functions as the map $\mathcal{M}$ is applied.
\begin{align}\label{charM}
&|\Xi_{\mathcal{M}(\hat\rho_+)}(d)-\Xi_{\mathcal{M}{(\hat \rho_-)}}(d)| = \nonumber\\
&=\Big|\Tr_B[\mathcal{M}((X_{n,n+1}+X_{n+1,n})\otimes \tau_{y_0})e^{-\frac{i}{\hbar}d(\hat p_1+\hat p_2)}]\Big|\nonumber\\
&\le\norm{\mathcal{M}((X_{n,n+1}+X_{n+1,n})\otimes\tau_{y_0})}_1\norm{e^{-\frac{i}{\hbar}d(\hat p_1+\hat p_2)}}_\infty\nonumber\\
&\le\norm{\mathcal{M}(X_{n,n+1}\otimes\tau_{y_0})}_1+\norm{\mathcal{M}(X_{n+1,n}\otimes\tau_{y_0})}_1\nonumber\\
&\le(\kappa + 2\sqrt{\kappa})(\norm{X_{n,n+1}}_1+\norm{X_{n+1,n}}_1)\nonumber\\
&=(\kappa + 2\sqrt{\kappa})(\norm{\ket{\psi_{n}}\bra{\psi_{n+1}}}_1+\norm{\ket{\psi_{n+1}}\bra{\psi_{n}}}_1)\nonumber\\
&=2(\kappa + 2\sqrt{\kappa}).
\end{align}
Comparing, for $d=d^*$, Eq. ~\eqref{chard^*} with Eq.~\eqref{charM} we see that  the difference of characteristic functions has changed. Therefore the total momentum statistics must have changed. \qed
\\
\\
As in the main text,  we now return to the center--of--mass ($C$) and relative ($r$) degrees of freedom. Thus let
\begin{equation}
M:=m_1+m_2,
\quad \hat X_C:=\frac{m_1\hat x_1+m_2\hat x_2}{M}, \quad \hat r:=\hat x_1-\hat x_2,
\end{equation}
and
\begin{equation}
\hat P_C:=\hat p_1+\hat p_2,\qquad \hat p_r:=\frac{m_2\hat p_1-m_1\hat p_2}{M}.
\end{equation}
Through the standard unitary change of variables $\mathcal{U}_{Cr}= U_{Cr}\cdot U_{Cr}^\dagger$ one has
\begin{equation}
\mathcal H_{B_1}\otimes\mathcal H_{B_2}\simeq \mathcal H_C\otimes\mathcal H_r,
\end{equation}
and
\begin{equation}
\hat P_{\mathrm{tot}}:=\hat p_1+\hat p_2=U_{Cr}^\dagger(\hat P_C\otimes \mathbb I_r)U_{Cr}.
\end{equation}

As before, we assume that in these variables the dynamics factorizes:
\begin{equation}\label{eq:M_factorized_C_r_eps}
\mathcal M=\mathcal{U}_{Cr}^{-1}\circ\qty(\mathcal M_C\otimes\mathcal M_r)\circ\mathcal{U}_{Cr},
\end{equation}
with $\mathcal M_C$ and $\mathcal M_r$ CPTP.
Therefore the statistics of the total momentum are entirely governed by the center--of--mass channel
$\mathcal{M}_C$. By Theorem~$\varepsilon$5 we know that the action of $\mathcal M_C$ cannot be a trivial identity. As in the main text we are going to assume that the center of mass dynamics must induce collapse. \\
Considering, as in the main text, an interval coarse graining $\{\Pi_\alpha^C\}$ of the center of mass degrees of freedom, we will assume that a result analogous result to Theorem~$\varepsilon$4A holds for $\mathcal{M}_C$. Concretely we assume the existence of a $\kappa_{C}\in[0,1)$ such that $\forall\ X\in \mathcal{T}(\mathcal{H}_{C})$ we have
\begin{equation}
    \norm{\mathcal{M}_C(\Pi^C_\alpha X\Pi^C_\beta)}_1 \le \kappa_C\norm{\Pi^C_\alpha X\Pi^C_\beta}_1.
\end{equation}
\\
{\bf Theorem $\varepsilon$6: Center of mass diffusion}\\
Assume that $\norm{\mathcal{M}_C(\Pi_\alpha^C X\Pi_\beta^C)}_1 \le \kappa_{C}\norm{\Pi_\alpha^CX\Pi_\beta^C}_1$ for $\alpha\neq \beta $ with $2\kappa_C <1$ and that $\mathcal{M}_C$ is a Galilei covariant channel. Then
\begin{equation}
    \text{Var}_{\mathcal{M}_C(\hat\rho)}(\hat P_C) =\text{Var}_{\hat\rho}(\hat P_C) + \Delta,
\end{equation}
where $\Delta >0$.
\\
\\
{\it Proof.} 
As in the main text we have that if  $\mathcal{M}_C$ is a Galilei covariant channel then 
\begin{equation}
    \text{Var}_{\mathcal{M}_C(\hat\rho)}(\hat P_C) =\text{Var}_{\hat\rho}(\hat P_C) + \text{Var}_{\nu_t}(p),
\end{equation}
where $\nu_t(p) = \int dq\mu_t(p,q)$ and $\Delta = \text{Var}_{\nu_t}(p) $. Let us again assume that $\text{Var}_{\nu_t}(p) = 0$ so that $\nu_t(p) = \delta(p-\bar{p}_t)$ [cf. Eq.~\eqref{barp}] and show that this leads to a contradiction.
The characteristic function of the momentum changes as in the $\kappa_{C} = 0$ case [cf. Eq.~\eqref{chafunc}], that is
\begin{align}
&\chi_{\mathcal{M}_C(\hat\rho)}(d)=\chi_{\hat \rho}(d)e^{- \frac{i}{\hbar}d \bar{p}_t}\,.
\end{align}
Therefore, given any two states $\hat{\rho}_{1,2}$ we have
\begin{align}
|\chi_{\mathcal{M}_C(\hat\rho_1)}(d)-\chi_{\mathcal{M}_C(\hat\rho_2)}(d)| = |\chi_{\hat\rho_1}(d)-\chi_{\hat\rho_2}(d)|.
\end{align}
Consider then a state $\ket{\psi_\alpha^C}$ localized in a center of mass sector $\Pi_\alpha^C$. Construct then $\ket{\Psi_{\pm}} = \frac{1}{\sqrt{2}}(\ket{\psi_\alpha^C} \pm e^{\frac{i}{\hbar}d^*\hat P_C}\ket{\psi_\alpha^C})$ and $\hat\rho_{1,2} = \hat \rho_{\pm} = \ket{\Psi_{\pm}}\bra{\Psi_{\pm}}$ with $d^*> \ell$. Then, for $d=d^*$
\begin{equation}\label{diffChi}
    |\chi_{\hat{\rho}_+}(d^*)-\chi_{\hat{\rho}_-}(d^*)| = 1.
\end{equation}
However since the states $\ket{\psi_\alpha^C}$ and $e^{\frac{i}{\hbar}d \hat P_C}\ket{\psi_\alpha^C}$ are supported on different sectors whenever $d>\ell$ we have that
\begin{align}
&|\chi_{\mathcal{M}_C(\hat\rho_+)}(d)-\chi_{\mathcal{M}_C(\hat\rho_-)}(d)| = \nonumber\\
&=\!|\!\Tr_B[\mathcal{M}_C(e^{\frac{i}{\hbar}d^*\hat P_C}\!\ket{\psi_\alpha^C}\!\bra{\psi_\alpha^C}\!+\!\ket{\psi_\alpha^C}\!\bra{\psi_\alpha^C}\!e^{-\frac{i}{\hbar}d^*\hat P_C})e^{\frac{i}{\hbar}d\hat P_C}]|\nonumber\\
&\le \! \! \norm{\mathcal{M}_C(e^{\frac{i}{\hbar}d^*\hat P_C}\!\ket{\psi_\alpha^C}\!\bra{\psi_\alpha^C}\!+\!\ket{\psi_\alpha^C}\!\bra{\psi_\alpha^C}\!e^{-\frac{i}{\hbar}d^*\hat P_C})}_1\!\norm{e^{\frac{i}{\hbar}d\hat P_C}}_\infty\nonumber\\
&\le \!\! \norm{\mathcal{M}_C(e^{\frac{i}{\hbar}d^*\hat P_C}\ket{\psi_\alpha^C}\!\bra{\psi_\alpha^C})}_1\!\!\!\!+\!\norm{\mathcal{M}_C(\ket{\psi_\alpha^C}\!\bra{\psi_\alpha^C}\!e^{-\frac{i}{\hbar}d^*\hat P_C})}_1\nonumber\\
&\le 2\kappa_{C} <1\,,
\end{align}
which for $d=d^*$ leads to a contradiction with~\eqref{diffChi}.\qed 

\section{Linearized Hamiltonian}\label{App:HamiltonianNewton}
Although familiar in the literature \cite{kafri2014classical}, for the sake of completeness we report the step-by-step derivation of the Hamiltonian used in Eq. (\ref{ME_main}) of the main text. We consider the setup in Fig. \ref{fig2}, where two particles
are harmonically trapped  and  interact through the Newtonian potential $V_N(\hat{r}_{1}-\hat{r}_{2})$. We study the general case where the frequencies $\omega_1,\omega_2$ and the masses $m_1,m_2$ of the oscillators are different:
\begin{equation}
\begin{aligned}
\hat{H}\!&=\frac{\hat{p}_{1}^{2}}{2m_1}\!+\!\frac{\hat{p}_{2}^{2}}{2m_2}\!+\!\frac{1}{2}m_1\omega_1^{2}\left(\hat{r}_{1}\!+\!\frac{d}{2}\right)^{2}\!+\!\frac{1}{2}m_2\omega_2^{2}\left(\hat{r}_{2}\!-\!\frac{d}{2}\right)^{2}\\
&+V_N(\hat{r}_{1}-\hat{r}_{2}).
\end{aligned}\label{meth_Horig}
\end{equation}
We start by introducing the deviations from the equilibrium position
of the oscillators $\hat{q}_{1}=\hat{r}_{1}+\frac{d}{2}$ and $\hat{q}_{2}=\hat{r}_{2}-\frac{d}{2}$
and then we expand the Newtonian potential by assuming that these deviations are
much smaller than the distance $d$ between the two oscillators:
\begin{align} \label{NewtPotApp}
&V_N(\hat{r}_{1}-\hat{r}_{2})=-\frac{Gm_1m_2}{|\hat{r}_{1}-\hat{r}_{2}|}=-\frac{Gm_1m_2}{|d-(\hat{q}_{1}-\hat{q}_{2})|}\nonumber\\
&\simeq-Gm_1m_2\left(\frac{1}{d}+\frac{1}{d^{2}}(\hat{q}_{1}-\hat{q}_{2})+\frac{1}{d^{3}}(\hat{q}_{1}-\hat{q}_{2})^{2}\right).
\end{align}
By replacing the above expression in Eq. (\ref{meth_Horig}) we get:
\begin{equation} \label{eq:jyhgsdf}
\begin{aligned}
\hat{H}=&\frac{\hat{p}_{1}^{2}}{2m_{1}}+\frac{\hat{p}_{2}^{2}}{2m_{2}}+\frac{1}{2}m_{1}\Omega_{1}^{2}\hat{q}_{1}^{2}+\frac{1}{2}m_{2}\Omega_{2}^{2}\hat{q}_{2}^{2}\\
&+K\hat{q}_{1}\hat{q}_{2}-K\frac{d}{2}(\hat{q}_{1}-\hat{q}_{2})-K\frac{d^{2}}{2},
\end{aligned}
\end{equation}
where we introduced:
\begin{equation}
\Omega_{1}^{2}:=\omega_{1}^{2}-\frac{2Gm_{2}}{d^{3}};\;\;\;\Omega_{2}^{2}:=\omega_{2}^{2}-\frac{2Gm_{1}}{d^{3}};\;\;\;K:=\frac{2Gm_{1}m_{2}}{d^{3}}.
\end{equation}
The  linear term $\hat{q}_{1}-\hat{q}_{2}$ can be absorbed by introducing the change of variables: $\hat{x}_{1}=\hat{q}_{1}-a_{1}$ and $\hat{x}_{2}=\hat{q}_{2}+a_{2}$, with $a_1$ and $a_2$ constants to be determined. Requiring that the terms linear in $\hat{x}_{1}$ and $\hat{x}_{2}$ vanish, implies the following two conditions on $a_1$ and $a_2$:
\begin{align}    &m_{1}\Omega_{1}^{2}a_{1}-Ka_{2}-K\frac{d}{2}=0,\\
&m_{2}\Omega_{2}^{2}a_{2}-Ka_{1}-K\frac{d}{2}=0.
\end{align}
By solving for $a_1$ and $a_2$, the change of variables reads:
\begin{align}
\hat{x}_{1}&=\hat{q}_{1}-\frac{1}{m_{1}\omega_{1}^{2}\left(\frac{1}{K}-\frac{1}{m_{1}\omega_{1}^{2}}-\frac{1}{m_{2}\omega_{2}^{2}}\right)}\frac{d}{2}\\
&=\hat{r}_{1}+\left(1-\frac{1}{m_{1}\omega_{1}^{2}\left(\frac{1}{K}-\frac{1}{m_{1}\omega_{1}^{2}}-\frac{1}{m_{2}\omega_{2}^{2}}\right)}\right)\frac{d}{2};\nonumber\\
\hat{x}_{2}&=\hat{q}_{2}+\frac{1}{m_{2}\omega_{2}^{2}\left(\frac{1}{K}-\frac{1}{m_{1}\omega_{1}^{2}}-\frac{1}{m_{2}\omega_{2}^{2}}\right)}\frac{d}{2}\\
&=\hat{r}_{2}-\left(1-\frac{1}{m_{2}\omega_{2}^{2}\left(\frac{1}{K}-\frac{1}{m_{1}\omega_{1}^{2}}-\frac{1}{m_{2}\omega_{2}^{2}}\right)}\right)\frac{d}{2}\nonumber.
\end{align}
Inserting this result into Eq.~\eqref{eq:jyhgsdf},  and  neglecting constant terms, the Hamiltonian becomes:
\begin{equation}\label{777}
\hat{H}=\frac{\hat{p}_{1}^{2}}{2m_{1}}+\frac{\hat{p}_{2}^{2}}{2m_{2}}+\frac{1}{2}m_{1}\Omega_{1}^{2}\hat{x}_{1}^{2}+\frac{1}{2}m_{2}\Omega_{2}^{2}\hat{x}_{2}^{2}+K\hat{x}_{1}\hat{x}_{2}.
\end{equation}

When $\omega_1=\omega_2=\omega$ and $m_1=m_2=m$, as in the case considered in the main text, Eq. (\ref{777}) reduces to
\begin{equation} \label{HB10}
\hat{H}=\frac{\hat{p}_{1}^{2}}{2m}+\frac{\hat{p}_{2}^{2}}{2m}+\frac{1}{2}m\Omega^{2}\hat{x}_{1}^{2}+\frac{1}{2}m\Omega^{2}\hat{x}_{2}^{2}+K\hat{x}_{1}\hat{x}_{2},
\end{equation}
with 
\begin{equation}
\Omega^{2}:=\omega^{2}-\frac{K}{m}\;\;\;\;\;\textrm{and}\;\;\;\;\;K:=\frac{2Gm^{2}}{d^{3}},
\end{equation}
which is the Hamiltonian in Eq. (\ref{ME_main}) of the main text when setting:
\begin{equation}
\hat{H}_0=\frac{\hat{p}_{1}^{2}}{2m}+\frac{\hat{p}_{2}^{2}}{2m}+\frac{1}{2}m\Omega^{2}\hat{x}_{1}^{2}+\frac{1}{2}m\Omega^{2}\hat{x}_{2}^{2}.
\end{equation}

\section{Compatibility with Newton}\label{APP_NewtLimit}
In this Appendix we show that in the Gaussian regime compatibility with the Newtonian limit forces the master equation to take the form of Eq.~\eqref{ME_main} of the main text. In particular, the Kossakowski matrix $\gamma$ must be real and $\hat H^G = \hat V_N$, aptly expanded to second order. To see why this is the case, consider the Taylor expanded Newtonian potential [cf. Eq.~\eqref{NewtPotApp}]
\begin{equation} \label{VNApp}
    \hat V_N \approx -\frac{K}{2}(\hat x_1-\hat x_2)^2\,.
\end{equation}
where we neglected an irrelevant constant term and the linear contributions, as in the main text. In this context, compatibility with Newtonian predictions means that the averages positions and momenta should follow Newtonian trajectories. In the Gaussian regime the equations of motion for the averages read 
\begin{equation}\label{app_average}
\frac{\dd }{\dd t}\langle \hat c \rangle = JH_{V_N}\langle\hat c\rangle, 
\end{equation} 
where the Hamiltonian matrix $H_{V_N}$ determined by the potential $\hat V_N$ in Eq.~\eqref{VNApp} reads
\begin{equation}
    H_{V_N} = \begin{bmatrix}
        -K&K&0&0\\
        K&-K&0&0\\
        0&0&0&0\\
        0&0&0&0
    \end{bmatrix}.
\end{equation}
Let us then consider the general Gaussian dissipator of Eq.~\eqref{ME_main_1} of the main text
\begin{align}\label{app_me}
    \frac{\dd }{\dd t}\hat\rho(t) =& -\frac{i}{\hbar}\comm{\hat H_0 +\frac{1}{2}\sum_{ij}\hat c_i(H_q^G)^{ij}\hat c_j}{\hat \rho(t)}\nonumber\\
    &+\sum_{ij=1}^4\gamma_{ij}\Big[\hat c_i\hat\rho(t)\hat c_j -\frac{1}{2}\acomm{\hat c_j\hat c_i}{\hat \rho(t)}\Big],
\end{align}
and inquire what choices of $H_q^G$ and $\gamma$ can reproduce the same equation of motion~\eqref{app_average}, neglecting $\hat H_0$ which just collects the free evolution and the non gravitational couplings. The equations of motion given by~\eqref{app_me} then read
\begin{equation}\label{app_average_2}
    \frac{\dd }{\dd t}\langle \hat c\rangle = J(H_q^G-\Im \gamma )\langle c\rangle.
\end{equation}

For Eq.~\eqref{app_average} and Eq.~\eqref{app_average_2} to describe the same dynamics we must have $H_{V_N}=H^G_q-\Im \gamma$. However, since $H_{V_N}$ is symmetric and $\Im \gamma$ is structurally antisymmetric, the only choice is $\Im\gamma =0$. Then, absorbing all the non-coupling contributions into $\hat H_0$, the master equation~\eqref{app_me} reduces to 
\begin{equation}\label{app_me_fin}
    \frac{\dd }{\dd t}\hat\rho(t) \!=\! -\frac{i}{\hbar}\comm{\hat H_0 + K\hat x_1\hat x_2}{\hat \rho(t)} -\frac{1}{2}\sum_{ij=1}^4\gamma_{ij}\comm{\hat c_i}{\comm{\hat c_j}{\hat\rho(t)}},
\end{equation}
recovering Eq.~\eqref{ME_main} of the main text. 

Finally, we stress that the above structure is specific to the Gaussian regime. Outside it, it is possible to construct Lindblad dynamics that reproduces the Newtonian force at the level of the equations of motion for the averages without an explicit Hamiltonian term $\hat V_N$~\cite{piccione2025hybrid,carney2023strongly}. Regardless, we have shown that whenever such models are smooth enough to admit a Taylor expansion to a Gaussian regime, the previous argument applies and thus they would fall into the class of equations considered~\eqref{app_me_fin}. For such models, therefore, the results discussed in the the main text  still apply. 

\section{A note on the KTM model and N-particle generalizations}\label{APP_D}
In the main text we considered a master equation for two point particles. The canonical variables appearing in~\eqref{ME_main} are intended as the center of mass variables of two macroscopic masses.   The formulation is agnostic as to how the master equation~\eqref{ME_main}, is related to the microscopic dynamics.

In this respect, it is relevant to put Eq.~\eqref{ME_main} in connection with the seminal proposal of a model of classical gravity by Kafri, Taylor and Milburn~\cite{kafri2014classical} (KTM). There, the authors posit a two particle master equation using a measurement and feedback scheme to describe a classical (in the LOCC sense) gravitational interaction. As in the present case, in order not to generate entanglement they need some minimal amount of decoherence~\cite{kafri2013noise}.
Later~\cite{altamirano2018gravity} it has been shown that if the two-particle KTM equation is generalized to $N$ systems in a pairwise fashion, then this minimal amount is already sufficiently large to be ruled out by interferometric experiments with atomic fountains, if the gravitational pull of the Earth is taken into account. Since the present proposal does not make this extra assumption, it cannot be falsified with the same arguments. 

In this regard, it is import to point out that the dynamics here considered, while similar in spirit, differs from that of KTM in some key aspects, such as the presence of commutators that mix the variables of the two systems. These terms have important implications when the dynamics is extended to a composite body, regardless of the pairwise assumption. Let us consider, for the sake of the argument, a simplified version where only the position operators appear in the irreversible contribution to the master equation
\begin{equation}
    \frac{\dd}{\dd t} \hat\rho(t) = -\frac{1}{2}\sum_{i,j=1}^N\gamma_{ij}\comm{\hat x_i}{\comm{\hat x_j}{\hat\rho_t}}\,,
\end{equation}
and introducing the center of mass coordinate $\hat R = \frac{1}{M}\sum m_i \hat x_i$, where $M$ is the total mass of the body and $m_i$ are the masses of the constituents; the evolution of the center of mass state $\hat\rho_\text{CM} =\Tr_\text{rel}[\hat \rho(t)]$ will be
\begin{equation}
    \frac{\dd}{\dd t} \hat\rho_\text{CM}(t) = -\frac{1}{2}\Big(\sum_{i,j=1}^N\gamma_{ij}\Big)\comm{\hat R}{\comm{\hat R}{\hat\rho_\text{CM}(t)}}\,.
\end{equation}
In  general, $\gamma_\text{CM} = \sum_{ij=1}^N\gamma_{ij}$ can take any value in the interval $[0,2\Tr[\gamma]]$ if no further assumptions are made, since the off-diagonal elements of the matrix are allowed to be negative. In the KTM model such terms are not present and, thus, only positive entries appear in the sum; this need not be true in general~\cite{gaona2021gravitational}.

In Ref.~\cite{altamirano2018gravity} the authors also claim to falsify the KTM proposal for just two point-like masses. The claim is, however, based on an arbitrary assumption that the decoherence coefficients of two highly asymmetric masses are the same. This assumption could and should be questioned: in fact, it is consistent to require that the ``total" decoherence is the minimum needed to avoid entanglement, and at the same time that it is distributed in a highly asymmetric fashion between the two systems. Although using either informational-theoretic or gravitational arguments, educated guesses can be made about the relative strength of the decoherence on the two systems, these guesses ultimately amount to a further restriction the generality of the arguments (just as the pairwise assumption does).

Therefore, the simplest avenue is to remain agnostic and consider the setup which requires the fewest assumptions: that of two equal masses.

\section{Dissipation and non-Markovianity}\label{APP:DissNonM}
In the main text, it was proven that, if gravity does not entangle quantum systems, then we can set lower bounds on the coefficients describing the noise. The result was obtained, consistently with our assumptions, for a purely diffusive and Markovian master equation. In this Appendix, we demonstrate that the results are robust under dissipative and transient non-Markovian extensions of the master equation~\eqref{ME_main} used in the main text. In both of the following paragraphs, we work with dimensionless variables, as defined in Section~\ref{secIV}, and assume equal masses as well as $\Omega_1 = \Omega_2$.

\subsection{Dissipation}
In the linearized limit, neglecting linear terms in the Hamiltonian which would not play any role, the most general Markovian master equation reads
\begin{align}\label{me_diss}
    \frac{\dd }{\dd t} \hat \rho(t) = &-\frac{i}{\hbar}\comm{\bar H_0 +\frac{K}{m\Omega}\bar x_1\bar x_2}{\hat\rho(t)}\nonumber\\
    &+\sum_{ij=1}^{4}\bar \gamma_{ij}\Big[\bar c_i\hat\rho(t)\bar c_j-\frac{1}{2}\acomm{\bar c_j\bar c_i}{\hat\rho(t)}\Big]\,,
\end{align}
where now the Kossakowski matrix is allowed to be complex, albeit Hermitian $\bar\gamma = \bar\gamma^\dagger$ and positive $\bar\gamma \ge 0$. The presence of the imaginary components of the Kossakowski matrix affects the evolutions of the average positions and momenta: these are dissipative deviations from Newton’s $1/r^2$ force law. These introduce decays whose rates are governed by the real parts of the eigenvalues of $J(H-\Im\bar\gamma)$, where $ H$ is given by Eq.~\eqref{Hambar} of the main text. Since these modifications appear at the level of first moments, they would be present also for macroscopic objects in the semiclassical Newtonian limit. Thus their presence is, in principle, incompatible with our assumption 4. Nonetheless, we can still investigate the impact that their presence would have on the conclusions drawn in the main text.

Let us then consider a $\Im\bar\gamma$ which couples the position and momenta of the two systems, as distinctive of dissipative master equations~\cite{caldeira1983path,vacchini2002non} . Since we are assuming the systems to be identical $\Im\bar\gamma$ takes the form:
\begin{equation}\label{diss_gamma}
    \Im\bar\gamma = \begin{bmatrix}
        0&0&-\bar\eta&-\bar\nu\\
        0&0&-\bar\nu&-\bar\eta\\
        \bar\eta&\bar\nu&0&0\\
        \bar\nu&\bar\eta&0&0\,
    \end{bmatrix}.
\end{equation}
In Eq.~\eqref{diss_gamma} $\bar\eta$ can be identified as the coefficient of local dissipation, while $\bar \nu$ encodes potential non-local effects which mix the position of one party with the momentum of the other. This structure encompasses the dissipative extensions of the KTM and TD models~\cite{di2021gravity}. Notice that for the dynamics to be genuinely dissipative, the drift matrix $J(H - \Im\bar\gamma)$ must have eigenvalues with strictly negative real parts, which are given by $ {-\bar\eta\pm \bar\nu}$. Therefore we have the constraint $\bar\eta>0$ and $\bar\eta>|\bar\nu|$. 

We can recalculate the separability bound under the more general evolution~\eqref{me_diss} with the matrix~\eqref{diss_gamma}. The equation of motion for the covariance matrix reads \begin{align}
    \frac{\dd }{\dd t}V(t) =& (J H-J\Im\bar\gamma)V(t)+V(t)(J H-J\Im\bar\gamma)^T
    \nonumber \\
    & -J\Re\bar\gamma J\,.
\end{align}
Following the derivation in the main text,  we take as initial state the ground state, for which $V_0 = \mathbb{I}/2$, and the vector $z_0 = (1,i,i,1)$. Given that $z_0^\dagger(\mathbb{I}_4+i\Lambda J\Lambda)z_0 = 0$ must hold, we obtain that PPT implies $z_0^\dagger\frac{dV(t)}{dt}\Big|_{t=0}z_0\ge 0$, explicitly:
\begin{equation}
    0\le  z_0^\dagger\Big(\frac{1}{2}\comm{J}{ H}-\frac{1}{2}\acomm{J}{\Im\bar\gamma}-J\Re\bar\gamma J\Big)z_0  \,.
\end{equation}
Computing this average and using the same majorization we used in the main text, we obtain
\begin{equation}\label{APP_diss:Newbound}
    \Tr[\bar\gamma] -2\bar\eta\ge \frac{K}{m\Omega}.
\end{equation}
Now, since the coefficient $\bar\eta> 0$ due to the constraint imposed by the request of decay, we see that the bound \eqref{APP_diss:Newbound} improves with the presence of local dissipation in the two systems, in the sense that a larger noise contribution  $\Tr[\bar\gamma]$ is required to maintain separability, compare to Eq. (\ref{MET_EQ:bound}) in the main text with $\Omega_1 = \Omega_2$. Hence, the limit $\bar\eta\rightarrow 0$ of the main text is, in principle, a conservative scenario. Moreover, since $\bar\eta> |\bar\nu|$, absence of local gravitational dissipation implies the absence of the non-local one as well.

The reason why dissipation, counterintuitively, leads to the necessity of more heating is the following. While it is true that dissipation contributes to the sudden death of pre-existing entanglement, in the present problem entanglement generation is being continuously driven by the gravitational Hamiltonian. Dissipation drives the system towards ``colder" states –- the ground states, in absence of noise -–, which are more susceptible to the entangling Hamiltonian: to compensate for this, more heat has to enter the system.

\subsection{Non-Markovianity}
The non-relativistic dynamics that couples  two masses gravitationally is expected to be Markovian.~\cite{christodoulou2023locally} The reason is that, whenever the masses are slow with respect to the speed of light, the dynamics is effectively described by the instantaneous Newtonian potential; therefore it is expected that, in this regime, also the diffusive contributions are instantaneous, i.e. Markovian~\cite{tilloy2016sourcing,layton2023weak}. 

However, it is conceivable that, before settling to the Markovian regime, a non-Markovian transient is present, characterized  by a short timescale $t_{in}$. This could affect the heating induced by the gravitational noise: specifically, if  before vanishing, this transient decreases $\Tr[\bar\gamma]$, then the Markovian  calculation performed in the main text would give an overestimation of the strength of the noise, undermining the conclusion that Eq.~\eqref{MET_EQ:bound} serves to bound the gravitationally induced noise. We  show that this is not the case.

Under the hypotheses of theorem~7 of Section~\ref{tre}, and for small displacements, the non-Markovian dynamics for the two masses is given by the following time local master equation~\cite{chruscinski2010non,Breuer2002}
\begin{align}\label{non_markov_me}
    \frac{\dd }{\dd t}\hat\rho(t) =& -\frac{i}{\hbar}\comm{\bar H_0 +\frac{K}{m\Omega}\bar x_1\bar x_2}{\hat \rho(t)}\nonumber\\
    &+\sum_{ij}\bar\gamma_{ij}(t)\big[\bar c_i\hat \rho(t)\bar c_j-\frac{1}{2}\acomm{\bar c_j\bar c_i}{\hat \rho(t)}\big]\,,
\end{align}
where now all the components of the matrix $\bar\gamma$ are allowed to depend on time. While that matrix $\bar\gamma(t)$ need not be positive semi-definite,  a necessary condition for the evolution to be CP $\bar\gamma(t)|_{t=0}\ge 0$.

We  consider the worst-case-scenario where, over the timescale $t_{in}$, the matrix $\bar\gamma(t)$ determines the minimum possible heating of the systems, which is compatible with the request that the separability of the initial state is maintained over time. This will constrain the rate of change of $\bar\gamma(t)$. In particular, by considering a generic squeezed state as initial state, we are able to show that on timescales $t_{in}$, which are short with respect to the inverse of the frequencies $\Omega$ and $\sqrt{K/m}$, such transients do not  affect the bound~\eqref{MET_EQ:bound} appreciably.

Let us then consider again the PPT condition used in the main text:
\begin{equation}
P(t) := V(t) +\frac{i}{2}\Lambda J \Lambda\, \geq 0.
\end{equation}
Let us differentiate $P(t)$ in time
\begin{align}\label{EqP_appE}
    &\frac{\dd }{\dd t}P(t) = \dot V(t) = AV+VA^T -J\bar\gamma(t) J\nonumber\\
    &=AP(t)+P(t)A^T -\frac{i}{2}(A\Lambda J \Lambda +\Lambda J \Lambda A^T)-J\bar\gamma(t) J\nonumber\\
    &\coloneqq AP(t)+P(t)A^T +D_K +D_\gamma(t)\,,
\end{align}
where 
\begin{equation}
    A=JH =
\begin{bmatrix}
0 & 0 & \Omega & 0\\
0 & 0 & 0 & \Omega\\
-\Omega & -\dfrac{K}{m\Omega} & 0 & 0\\
-\dfrac{K}{m\Omega} & -\Omega & 0 & 0
\end{bmatrix},
\end{equation}
and we separated the contribution $D_K$ given by the free dynamics from the diffusive part $D_\gamma(t)$.

The solution of Eq. (\ref{EqP_appE}) is: 
\begin{align}
    P(t) = e^{At}P_0e^{A^Tt} +\int_0^t ds e^{A(t-s)}(D_K+D_\gamma(s))e^{A^T(t-s)}.
\end{align}
Since
\begin{align}
    \int_0^t ds& e^{A(t-s)}D_Ke^{A^T(t-s)} =\nonumber\\
    &=-\frac{i}{2}\int_0^t ds e^{A(t-s)}(A\Lambda J\Lambda +\Lambda J\Lambda A^T )e^{A^T(t-s)}\nonumber\\
    &=\frac{i}{2}\int_0^t ds \frac{d}{ds}\Big(e^{A(t-s)}\Lambda J\Lambda e^{A^T(t-s)}\Big)\nonumber\\
    &=\frac{i}{2}\Lambda J\Lambda -e^{A t}\frac{i}{2}\Lambda J \Lambda e^{A^T t}\,,
\end{align}
by rearranging the terms, we obtain
\begin{equation}
    P(t) = \Big(e^{A t}V_0e^{A^T t}+\frac{i}{2}\Lambda J\Lambda\Big) +\int_0^t \!\!ds e^{A(t-s)}D_\gamma(s)e^{A^T(t-s)},
\end{equation}
where we recognize that the first term in round brackets is  $P^K(t)$, the PPT matrix for the evolution {\it without} the extra diffusive elements, that is, for the standard Newtonian evolution. We thus have the following matrix inequality which must be satisfied if the evolution has to preserve the separability condition ($P(t) \geq 0$):
\begin{equation}\label{PPT_int}
    \int_0^tdse^{A(t-s)}(-J\bar\gamma(s) J)e^{A^T(t-s)} \ge -P^{K} (t)\,.
\end{equation}

We  evaluate this matrix inequality on the vector $z_t = e^{-A^T t}z_0$: 
\begin{align}\label{PPT_NM_av}
z_0^\dagger\int_0^t dse^{-As}&(-J\bar\gamma(s)J)e^{-A^Ts}z_0 \ge\nonumber\\
&-z_0^\dagger\Big(V_0+\frac{i}{2}e^{-A t}\Lambda  J\Lambda e^{-A^T t}\Big)z_0\,.
\end{align}
We now Taylor expand ~\eqref{PPT_NM_av} to second order in time and impose the inequality to be saturated at each order: this is the condition for the minimal noise required to be compatible with the separability of the evolution along the vector $z_t$. Since the matrix elements of $A$ depend on $\Omega$ and $\sqrt{K/m}$, as anticiapted the expansion is valid for times shorter than the inverse of these frequencies. 

Let us denote by $f_t$ the left hand side of~\eqref{PPT_NM_av}; its Taylor expansion reads
\begin{align}
    f_t =& 0 + tz_0^\dagger(-J\bar\gamma|_0J)z_0+\nonumber\\
    &+\frac{t^2}{2}z_0^\dagger (AJ\bar\gamma|_0J+J\bar\gamma|_0JA^T -J\dot{\bar\gamma}|_0J)z_0 +...
\end{align}
Expanding the right hand side, instead, gives
\begin{align}
    V_0 &+\frac{i}{2}e^{-At}\Lambda J \Lambda e^{-A^T t} =\nonumber\\
     &=V_0 +\frac{i}{2}\Lambda J\Lambda+\nonumber\\
     &-\frac{i}{2}t\Big(A\Lambda J \Lambda+\Lambda J \Lambda A^T\Big)\nonumber\\
     &+t^2\frac{i}{2}\Big(\frac{A^2}{2}\Lambda J\Lambda +\Lambda J\Lambda\frac{(A^T)^2}{2} +A\Lambda J \Lambda A^T\Big).
\end{align}
We can now match the various orders in time. The 0-th order terms give:
\begin{equation}\label{PPT_t0}
0 = - z_0^\dagger(V_0 + \frac{i}{2}\Lambda J\Lambda)z_0,
\end{equation}
meaning that $z_0$ has to be taken in the kernel of $V_0 +\frac{i}{2}\Lambda J\Lambda$, as done in the main text. The first order terms give
\begin{equation}
z_0^\dagger(-J\bar\gamma|_0J)z_0 = \frac{i}{2}z_0^\dagger(A\Lambda J\Lambda+\Lambda J\Lambda A^T)z_0\,,
\end{equation}
which determines a constraint on the initial values of the matrix $\bar{\gamma}$. At second order, we have:
\begin{align}\label{PPT_t2}
&z_0^\dagger\Big(AJ\bar\gamma|_0J+J\bar\gamma|_0JA^T-J\dot{\bar\gamma}|_0 J\Big)z_0=\nonumber\\
&\quad\frac{1}{i}z_0^\dagger(\frac{A^2}{2}\Lambda J\Lambda +\Lambda J\Lambda\frac{(A^T)^2}{2} +A\Lambda J \Lambda A^T)z_0\,,
\end{align} 
which  constrains the time derivative of  $\bar\gamma$.

In the main text,  $V_0$ was chosen to be the ground state. Here, it is convenient to consider a separable state corresponding to both parties being initially in a squeezed state with squeezing parameter $r\in \mathbb{R}$., i.e. $V_0 = \text{diag}[e^{r},e^{r},{e^{-r},e^{-r}}]/2$.

Taking $z_0 = (e^{-r},\ ie^{-r},\ i,\ 1)^T$, the 0-th order condition~\eqref{PPT_t0} is satisfied automatically . The first order condition, instead, gives 
\begin{align}\label{app:bound_r}
\Big[\bar\gamma_{11} -2e^{-r}\bar\gamma_{14}+e^{-2r}\bar\gamma_{33}\Big]\Big|_0\!= \frac{K}{m\Omega}\,,
\end{align}
where note that, given the symmetry assumptions that we have adopted, we have used $\bar\gamma_{11} = \bar\gamma_{22}$, $\bar\gamma_{33} = \bar\gamma_{44}$ as well as $\bar
\gamma_{14} = \bar\gamma_{23}$. Remembering that $\bar{\gamma}(t)$ does not depend on the state of the system, Eq.~\eqref{app:bound_r} holds true for any value of $r$, that is, for any squeezed state, if
\begin{equation}\label{Bound_nm_1}
\bar\gamma_{11}|_0= \frac{K}{m\Omega}\,,
\end{equation} 
as well as
\begin{equation}
    \begin{cases}
    \bar \gamma_{33}|_0 = 0\,,\label{mom-gamma-0}\\
    \bar \gamma_{14}|_0 = 0\,.
    \end{cases}
\end{equation}
Since $\bar\gamma_{33}|_0= \bar\gamma_{44}|_=0=0$ we have that  $\bar\gamma|_0\ge 0$, implies $\bar\gamma_{34}|_0=0$. Then
\begin{equation}
\bar\gamma = \begin{bmatrix}
\bar\gamma_x & \bar\gamma_{xp}\\
\bar\gamma_{xp}^T & 0
\end{bmatrix}\,,
\end{equation}
For this matrix to be positive semi-definite we must also have $\bar\gamma_{xp}|_0=0$. Overall, therefore, we have $\bar\gamma\big|_0 = \gamma_x|_0\oplus 0$.

We can now evaluate the condition which is set on the derivatives of $\bar\gamma$ by Eq.~\eqref{PPT_t2}, which, given the constraints here above, becomes

\begin{align}
\frac{\dd }{\dd t}&\Big[\bar\gamma_{11}-2e^{-r}\bar\gamma_{14}+e^{-2r}\bar\gamma_{33}\Big]\Big|_0= 2e^{-r}\Omega\bar\gamma_{12}|_0.
\end{align}
Again, for the equality to hold for any value of $r$, the following conditions must hold
\begin{equation}\label{constraint-derivatives}
    \frac{\dd}{\dd t}\bar\gamma_{11}|_0 = 0,\qquad
    \frac{\dd}{\dd t}\bar\gamma_{33}|_0 = 0,\qquad
    \frac{\dd}{\dd t}\bar\gamma_{14}|_0 = -\Omega\bar\gamma_{12}|_0 \,. 
\end{equation}
Let us now consider the implications of Eqs.~\eqref{constraint-derivatives} and~\eqref{Bound_nm_1} for the total heating rate of the system, to be compared with the Markovian estimates of the main text. 

The heating rate due to the non-Markovian master equation~\eqref{non_markov_me} is given by
\begin{align}
    \Gamma_{G}(t) &\coloneqq \frac{1}{\Omega}\frac{\dd}{\dd t}\langle\bar H_0 \rangle=\frac{1}{2\Omega}\sum_{ij}\gamma_{ij}(t)\langle\comm{\comm{\bar H_0}{\bar c_i}}{\bar c_j}\rangle\nonumber\\
    &=(\bar\gamma_{11}(t) + \bar\gamma_{33}(t)).
\end{align}
Using the constraints of Eq.~\eqref{Bound_nm_1} and Eq.~\eqref{constraint-derivatives}  we have that
\begin{align}
    \Gamma_G(t) &= \Gamma_G(t=0) + \mathcal{O}(t^2)\nonumber\\
    &= \frac{K}{m\Omega} + \mathcal{O}(t^2)\,.
\end{align}
$\Gamma_G(t=0)$ is the heating rate in the Markovian setting, see Eq.~\eqref{simplified_cond2} of the main text, up to geometry-dependent numerical prefactors.  Therefore, assuming that $\bar\gamma(t)$ is such that the trajectory of minimal heating for $z_t$ occurs, over short timescales $t_{in}$ with respect to $\text{min}[\Omega^{-1},m\Omega/K]$, the bound Eq.~\eqref{MET_EQ:bound} of the main text cannot  decrease appreciably.

\section{Spectral density for symmetric setup.}\label{app:Spectrum}
In the main text, we assumed that classical gravity remains universal when coupled to quantum matter, implying that the coefficients \( \gamma_{ij} \) depend only on the masses and relative distances. While this assumption is reasonable, in principle it leaves open the loophole of a potential dependence on $\gamma = \gamma(\Omega)$. We will show that this loophole can be in principle closed by considering a completely symmetric setup, and with similar experimental requirements as those considered in the main text. If we choose to work with two harmonic oscillators with equal masses $m$ and renormalized frequencies $\Omega$, the symmetry under exchange guarantees us that $\gamma_{11} = \gamma_{22}$ and $\gamma_{33}= \gamma_{44}$, but also $\gamma_{13} = \gamma_{24}$ and $\gamma_{14} = \gamma_{23}$. Then, the separability bound~\eqref{MAIN_EQ:BoundOmegas} of the main text simplifies to 
\begin{equation}\label{APP_E:bound}
    \gamma_{11} + m^2\Omega^2\gamma_{33} \ge \frac{G m^2}{\hbar d^3}\,.
\end{equation}
where we note that, differently from Eq.~\eqref{MAIN_EQ:bound}, here $\Omega$ is specifically the renormalized frequency of the oscillators. 

In this setup, to calculate the DNS it is necessary to properly take into account the dynamics of both systems; nonetheless, we still require the monitoring of just one of the two. We recall that the Hamiltonian, for a symmetric configuration, is:
\begin{equation} \label{eq:gh}
\hat{H}=\frac{\hat{p}_1^2}{2m}+\frac{\hat{p}_2^2}{2m}+\frac{1}{2}m \Omega^2 \hat{x}_1^2+\frac{1}{2}m \Omega^2 \hat{x}_2^2+K \hat{x}_1 \hat{x}_2,
\end{equation}
where
\begin{equation}
\Omega^{2}:=\omega^{2}-\frac{2Gm}{d^{3}}\ \quad \text{and}\ \quad  K:=\frac{2Gm^2}{d^{3}}.
\end{equation}
The dynamics described by the master equation \eqref{ME_main} with the Hamiltonian as in Eq.~\eqref{eq:gh} is statistically equivalent to the following set of Heisenberg-Langevin equations for the position and momentum operators:
\begin{equation}\label{APP_E:LangeviSyst}
\begin{cases}
\frac{\dd \hat{x}_1}{\dd t}=\frac{\hat{p}_1}{m} + \hbar w_{3}(t), \\
\frac{\dd \hat{x}_2}{\dd t}=\frac{\hat{p}_2}{m} + \hbar w_{4}(t),  \\
\frac{\dd \hat{p}_1}{\dd t}=- m \Omega^2 \hat{x}_1 -K \hat{x}_2  - \hbar w_{1}(t)-\eta \hat{p}_1 +\hat\xi_{1}(t), \\
\frac{\dd \hat{p}_2}{\dd t}=-m \Omega^2 \hat{x}_2 -K \hat{x}_1  - \hbar w_2(t)-\eta \hat{p}_2 + \hat\xi_{2}(t),
\end{cases}
\end{equation}
where $w_i(t)$ are correlated white noises: $\mathbb{E}[w_i(t)w_j(t')]~=~ \gamma_{ij}\delta(t-t')$. We have further assumed that both masses interact with a common environment, modeled as a thermal bath, which causes the systems to dissipate energy (controlled by the constant $\eta$) and adds additional noise:
\begin{align}
\mathbb{E}[\langle\hat\xi_{i}(t)\hat\xi_{j}(t')\rangle]=&\delta_{ij}\frac{\hbar\eta m}{2\pi}\times\nonumber\\
&\int \dd \omega \, \omega e^{-i \omega (t-t')}\left[1+\coth\left(\frac{\hbar \omega}{2 k_\text{B} T} \right) \right].
\end{align}
The DNS is, via the Wiener-Kinchine theorem, the Fourier transform of the autocorrelation of the corresponding stochastic quantity $\mathbb{E}[\langle\acomm{\hat{x}_1(t_1)}{\hat{x}_1(t_2)}\rangle]$. A more practical definition of the DNS, equivalent to the one used in the main text, can be given in Fourier space~\cite{paternostro2006reconstructing}:
\begin{equation}
S_{x_1 x_1}(\omega)=\frac{1}{2\pi}\int \dd \omega' e^{-i (\omega+\omega')t}\mathbb{E}[\hat{x}_1(\omega)\hat{x}_1(\omega')].
\end{equation}

To evaluate the DNS we solve the system~\eqref{APP_E:LangeviSyst} in Fourier space, finding:
\begin{equation}
\left( \begin{array}{c}\hat{x}_1(\omega) \\ \hat{x}_2(\omega) \end{array} \right)=A_x(\omega)B_x(\omega),   
\end{equation}
where
\begin{equation}
A_x(\omega)=\chi(\omega)\left(\begin{array}{cc}
 m (\Omega^2-\omega^2-i \eta \omega)    & -K \\
   -K  & m (\Omega^2-\omega^2-i \eta \omega) 
\end{array} \right),
\end{equation}
with
\begin{equation}
\chi(\omega)=\left(-K^2 + m^2(\Omega^2-\omega^2 - i \eta \omega)^2\right)^{-1},
\end{equation}
and
\begin{equation}
B_x(\omega)=\left(\begin{array}{c}
- \hbar \widetilde{w}_1(\omega)+(\eta-i\omega)m \hbar \widetilde{w}_3(\omega)-\widetilde{\xi}_1(\omega)   \\
-\hbar \widetilde{w}_2(\omega)+(\eta-i \omega)m \hbar \widetilde{w}_4(\omega)-\widetilde{\xi}_2(\omega) 
\end{array} \right).  
\end{equation}
In particular, we have:
\begin{equation}
\hat{x}_1(\omega)=(A_x)_{11}(\omega)(B_x)_{1}(\omega)+(A_x)_{12}(\omega)(B_x)_{2}(\omega)\,.    
\end{equation}
Noting that $\mathbb{E}[w_i(t)] =\mathbb{E}[\hat \xi(t)] =0$ and that the gravitational and thermal noises are uncorrelated ($\mathbb{E}[w_i(t)\hat \xi(t)] = 0$), a long but otherwise straightforward calculation leads to:
\begin{widetext}
\begin{eqnarray}\label{APP_E:SpectraDiffMass}
S_{x_1 x_1}(\omega)& = &|(A_x)_{11}(\omega)|^2\hbar^2 \left(\gamma_{11}+(\eta^2+\omega^2)m^2 \gamma_{33}-2 \eta m \gamma_{13}+\frac{\eta m}{ \hbar} \omega \left[1+\coth \left(\frac{\hbar \omega}{2k_\text{B} T} \right) \right]\right) \nonumber \\
&+&|(A_x)_{12}(\omega)|^2 \hbar^2 \left(\gamma_{22}+(\eta^2+\omega^2)m^2 \gamma_{44}- 2\eta m \gamma_{24}+ \frac{\eta m}{\hbar} \omega \left[1+\coth \left(\frac{\hbar \omega}{2 k_\text{B} T} \right) \right] \right) \nonumber \\
&+& 2 \hbar^2 \mathfrak{Re} \Big[(A_x)_{11}(\omega)(A_x)_{12}^*(\omega)\left(\gamma_{12}-(\eta-i \omega)m \gamma_{23}-(\eta+i \omega)m \gamma_{14}+(\eta^2+ \omega^2) m^2 \gamma_{34} \right) \Big].
\end{eqnarray}
\end{widetext}
We see why the symmetric setup is particularly convenient: since gravity should not distinguish between two identical systems, we can assume that all the coefficients $\gamma_{ij}$ referring to one mass must be the same as those for the other. This means, as emphasized previously, that $\gamma_{11} = \gamma_{22}$, but also, for example, $\gamma_{13} = \gamma_{24}$. Thus  the first two lines of Eq. \eqref{APP_E:SpectraDiffMass} simply add. Further evaluating the spectrum on the renormalized frequency $\omega=\Omega$, the last line of~\eqref{APP_E:SpectraDiffMass} vanishes, and thus we have:
\begin{widetext}
\begin{equation}\label{APP_E:SpectrumRes}
         S_{x_1x_1}(\Omega) =\frac{\hbar^2}{K^2+m^2\eta^2\Omega^2} \Big[\gamma_{11}+m^2 \Omega^2 \gamma_{33} +\frac{\eta m}{\hbar} \Omega\Big(1+\coth\Big(\frac{\hbar\Omega}{2k_\text{B}T}\Big)\Big)+m^2 \eta^2 \gamma_{33}-2m\eta\gamma_{13} \Big].
\end{equation}
\end{widetext}
 We see that the spectrum has the same peak as that in Eq.~\eqref{main_eq:Spectrum} of the main text. Furthermore, on $\omega = \Omega$ we can use the bound~\eqref{APP_E:bound} and arrive at similar experimental requirements as those discussed in Section~\ref{secV}.

\bibliography{ref}

\begin{thebibliography}{100}

\bibitem{dewitt2011role}
C{\'e}cile~M DeWitt and Dean Rickles.
\newblock {\em The role of gravitation in physics: report from the 1957 Chapel
  \newline Hill Conference}.
\newblock Edition Open Access, 2011.

\bibitem{Schellekens2013}
A.~N. Schellekens.
\newblock Life at the interface of particle physics and string theory.
\newblock {\em Reviews of Modern Physics}, 85:1491, 2013.

\bibitem{Rovelli2008}
Carlo Rovelli.
\newblock Loop quantum gravity.
\newblock {\em Living Reviews in Relativity}, 11:5, 2008.

\bibitem{Dyson2013}
Freeman Dyson.
\newblock Is a graviton detectable?
\newblock {\em International Journal of Modern Physics A}, 28:1330041, 2013.

\bibitem{Rothman2006}
Tony Rothman and Stephen Boughn.
\newblock Can gravitons be detected?
\newblock {\em Foundations of Physics}, 36:1801, 2006.

\bibitem{Carney2024}
Daniel Carney, Valerie Domcke, and Nicholas~L. Rodd.
\newblock Graviton detection and the quantization of gravity.
\newblock {\em Physical Review D}, 109:044009, 2024.

\bibitem{Palessandro2024}
Andrea Palessandro.
\newblock Graviton-photon oscillations as a probe of quantum gravity.
\newblock {\em Classical and Quantum Gravity}, 41:215011, 2024.

\bibitem{Parikh2020}
Maulik Parikh, Franck Wilczek, and George Zahariade.
\newblock The noise of gravitons.
\newblock {\em International Journal of Modern Physics D}, 29:2042001, 2020.

\bibitem{Kanno2021}
Sugumi Kanno, Jiro Soda, and Junsei Tokuda.
\newblock Noise and decoherence induced by gravitons.
\newblock {\em Physical Review D}, 103:044017, 2021.

\bibitem{Tobar2024}
Germain Tobar, Sreenath~K. Manikandan, Thomas Beitel, and Igor Pikovski.
\newblock Detecting single gravitons with quantum sensing.
\newblock {\em Nature Communications}, 15:7229, 2024.

\bibitem{belenchia2018quantum}
Alessio Belenchia, Robert~M Wald, Flaminia Giacomini, Esteban Castro-Ruiz,
  {\v{C}}aslav Brukner, and Markus Aspelmeyer.
\newblock Quantum superposition of massive objects and the quantization of
  gravity.
\newblock {\em Physical Review D}, 98(12):126009, 2018.

\bibitem{rydving2021gedanken}
Erik Rydving, Erik Aurell, and Igor Pikovski.
\newblock Do gedanken experiments compel quantization of gravity?
\newblock {\em Physical Review D}, 104(8):086024, 2021.

\bibitem{Carney2022}
Daniel Carney.
\newblock Newton, entanglement, and the graviton.
\newblock {\em Physical Review D}, 105:024029, 2022.

\bibitem{danielson2022gravitationally}
Daine~L Danielson, Gautam Satishchandran, and Robert~M Wald.
\newblock Gravitationally mediated entanglement: Newtonian field versus
  gravitons.
\newblock {\em Physical Review D}, 105(8):086001, 2022.

\bibitem{Fragkos2022inf}
Vasileios Fragkos, Michael Kopp, and Igor Pikovski.
\newblock On inference of quantization from gravitationally induced
  entanglement.
\newblock {\em AVS Quantum Science}, 4:045601, 2022.

\bibitem{karolyhazy1966gravitation}
Frederick Karolyhazy.
\newblock Gravitation and quantum mechanics of macroscopic objects.
\newblock {\em Il Nuovo Cimento A (1965-1970)}, 42(2):390--402, 1966.

\bibitem{diosi1987universal}
Lajos Diosi.
\newblock A universal master equation for the gravitational violation of
  quantum mechanics.
\newblock {\em Physics letters A}, 120(8):377--381, 1987.

\bibitem{penrose2014gravitization}
Roger Penrose.
\newblock On the gravitization of quantum mechanics 1: Quantum state reduction.
\newblock {\em Foundations of Physics}, 44:557--575, 2014.

\bibitem{penrose1996gravity}
Roger Penrose.
\newblock On gravity's role in quantum state reduction.
\newblock {\em General relativity and gravitation}, 28:581--600, 1996.

\bibitem{gasbarri2017gravity}
Giulio Gasbarri, Marko Toro{\v{s}}, Sandro Donadi, and Angelo Bassi.
\newblock Gravity induced wave function collapse.
\newblock {\em Physical Review D}, 96(10):104013, 2017.

\bibitem{diosi1984gravitation}
L.~Di{\'o}si.
\newblock Gravitation and quantum-mechanical localization of macro-objects.
\newblock {\em Physics Letters A}, 105(4-5):199--202, 1984.

\bibitem{tilloy2018ghirardi}
Antoine Tilloy.
\newblock Ghirardi-rimini-weber model with massive flashes.
\newblock {\em Physical Review D}, 97(2):021502, 2018.

\bibitem{oppenheim2023postquantum}
Jonathan Oppenheim.
\newblock A postquantum theory of classical gravity?
\newblock {\em Physical Review X}, 13(4):041040, 2023.

\bibitem{layton2023weak}
Isaac Layton, Jonathan Oppenheim, Andrea Russo, and Zachary Weller-Davies.
\newblock The weak field limit of quantum matter back-reacting on classical
  spacetime.
\newblock {\em Journal of High Energy Physics}, 2023(8):1--43, 2023.

\bibitem{donadi2022seven}
Sandro Donadi and Angelo Bassi.
\newblock Seven nonstandard models coupling quantum matter and gravity.
\newblock {\em AVS Quantum Science}, 4(2), 2022.

\bibitem{bose2017spin}
Sougato Bose, Anupam Mazumdar, Gavin~W Morley, Hendrik Ulbricht, Marko
  Toro{\v{s}}, Mauro Paternostro, Andrew~A Geraci, Peter~F Barker, MS~Kim, and
  Gerard Milburn.
\newblock Spin entanglement witness for quantum gravity.
\newblock {\em Physical review letters}, 119(24):240401, 2017.

\bibitem{marletto2017gravitationally}
Chiara Marletto and Vlatko Vedral.
\newblock Gravitationally induced entanglement between two massive particles is
  sufficient evidence of quantum effects in gravity.
\newblock {\em Physical review letters}, 119(24):240402, 2017.

\bibitem{krisnanda2020observable}
Tanjung Krisnanda, Guo~Yao Tham, Mauro Paternostro, and Tomasz Paterek.
\newblock Observable quantum entanglement due to gravity.
\newblock {\em npj Quantum Information}, 6(1):12, 2020.

\bibitem{christodoulou2019possibility}
Marios Christodoulou and Carlo Rovelli.
\newblock On the possibility of laboratory evidence for quantum superposition
  of geometries.
\newblock {\em Physics Letters B}, 792:64--68, 2019.

\bibitem{Guff2022}
Thomas Guff, Nicolas Boulle, and Igor Pikovski.
\newblock Optimal fidelity witnesses for gravitational entanglement.
\newblock {\em Physical Review A}, 105:022444, 2022.

\bibitem{Tilly2021}
Jules Tilly, Ryan~J. Marshman, Anupam Mazumdar, and Sougato Bose.
\newblock Qudits for witnessing quantum-gravity-induced entanglement of masses
  under decoherence.
\newblock {\em Physical Review A}, 104:052416, 2021.

\bibitem{Schut2022}
Martine Schut, Jules Tilly, Ryan~J. Marshman, Sougato Bose, and Anupam
  Mazumdar.
\newblock Improving resilience of quantum-gravity-induced entanglement of
  masses to decoherence using three superpositions.
\newblock {\em Physical Review A}, 105:032411, 2022.

\bibitem{Li2023}
Pan Li, Yi~Ling, and Zhangping Yu.
\newblock Generation rate of quantum gravity induced entanglement with multiple
  massive particles.
\newblock {\em Physical Review D}, 107:064054, 2023.

\bibitem{VandeKamp2020}
Thomas~W. van~de Kamp, Ryan~J. Marshman, Sougato Bose, and Anupam Mazumdar.
\newblock Quantum gravity witness via entanglement of masses: Casimir
  screening.
\newblock {\em Physical Review A}, 102:062807, 2020.

\bibitem{Marshman2020}
Ryan~J. Marshman, Anupam Mazumdar, and Sougato Bose.
\newblock Locality and entanglement in table-top testing of the quantum nature
  of linearized gravity.
\newblock {\em Physical Review A}, 101:052110, 2020.

\bibitem{Schut2023}
Martine Schut, Alexey Grinin, Andrew Dana, Sougato Bose, Andrew Geraci, and
  Anupam Mazumdar.
\newblock Relaxation of experimental parameters in a quantum-gravity-induced
  entanglement of masses protocol using electromagnetic screening.
\newblock {\em Physical Review Research}, 5:043170, 2023.

\bibitem{Gunnik2023}
Fabian Gunnik, Anupam Mazumdar, Martine Schut, and Marko Toroš.
\newblock Gravitational decoherence by the apparatus in the
  quantum-gravity-induced entanglement of masses.
\newblock {\em Classical and Quantum Gravity}, 40:235006, 2023.

\bibitem{Christodoulou2023}
Marios Christodoulou, Andrea~Di Biagio, Richard Howl, and Carlo Rovelli.
\newblock Gravity entanglement, quantum reference systems, degrees of freedom.
\newblock {\em Classical and Quantum Gravity}, 40:047001, 2023.

\bibitem{Großardt2020}
André Großardt.
\newblock Acceleration noise constraints on gravity-induced entanglement.
\newblock {\em Physical Review A}, 102:040202(R), 2020.

\bibitem{Kent2021}
Adrian Kent and Damián Pitalúa-García.
\newblock Testing the nonclassicality of spacetime: What can we learn from
  bell-bose \textit{et al}.-marletto-vedral experiments?
\newblock {\em Physical Review D}, 104:126030, 2021.

\bibitem{Chevalier2020}
Hadrien Chevalier, A.~J. Paige, and M.~S. Kim.
\newblock Witnessing the nonclassical nature of gravity in the presence of
  unknown interactions.
\newblock {\em Physical Review A}, 102:022428, 2020.

\bibitem{Matsumara2020}
Akira Matsumara and Kazuhiro Yamamoto.
\newblock Gravity-induced entanglement in optomechanical systems.
\newblock {\em Physical Review D}, 102:106021, 2020.

\bibitem{Krisnanda2023}
Tanjung Krisnanda, Tomasz Paterek, Mauro Paternostro, and Timothy C.~H. Liew.
\newblock Quantum neuromorphic approach to efficient sensing of gravity-induced
  entanglement.
\newblock {\em Physical Review D}, 107:086014, 2023.

\bibitem{Miki2024}
Daisuke Miki, Akira Matsumara, and Kazuhiro Yamamoto.
\newblock Feasible generation of gravity-induced entanglement by using
  optomechanical systems.
\newblock {\em Physical Review D}, 110:024057, 2024.

\bibitem{Higgins2024}
Gerard Higgins, Andrea~Di Biagio, and Marios Christodoulou.
\newblock Truly relativistic gravity mediated entanglement protocol using
  superpositions of rotational energies.
\newblock {\em Physical Review D}, 110:L101901, 2024.

\bibitem{Miki2021}
Daisuke Miki, Akira Matsumara, and Kazuhiro Yamamoto.
\newblock Entanglement and decoherence of massive particles due to gravity.
\newblock {\em Physical Review D}, 103:026017, 2021.

\bibitem{Feng2022}
Tianfeng Feng and Vlatko Vedral.
\newblock Amplification of gravitationally induced entanglement.
\newblock {\em Physical Review D}, 106:066013, 2022.

\bibitem{Carney2021}
Daniel Carney, Holger M\"uller, and Jacob~M. Taylor.
\newblock Using an atom interferometer to infer gravitational entanglement
  generation.
\newblock {\em Physical Review X Quantum}, 2:030330, 2021.

\bibitem{hosten2022constraints}
Onur Hosten.
\newblock Constraints on probing quantum coherence to infer gravitational
  entanglement.
\newblock {\em Physical Review Research}, 4(1):013023, 2022.

\bibitem{ma2022limits}
Yue Ma, Thomas Guff, Gavin~W Morley, Igor Pikovski, and MS~Kim.
\newblock Limits on inference of gravitational entanglement.
\newblock {\em Physical Review Research}, 4(1):013024, 2022.

\bibitem{streltsov2022significance}
Kirill Streltsov, Julen~Simon Pedernales, and Martin~Bodo Plenio.
\newblock On the significance of interferometric revivals for the fundamental
  description of gravity.
\newblock {\em Universe}, 8(2):58, 2022.

\bibitem{carney2022erratum}
D~Carney, H~M{\"u}ller, and JM~Taylor.
\newblock Erratum: Using an atom interferometer to infer gravitational
  entanglement generation [prx quantum 2, 030330 (2021)].
\newblock {\em PRX Quantum}, 3(1):010902, 2022.

\bibitem{Marletto2021}
Chiara Marletto and Vlatko Vedral.
\newblock Sagnac interferometer and the quantum nature of gravity.
\newblock {\em Journal of Physics Communications}, 5:051001, 2021.

\bibitem{Nguyen2020}
H.~Chau Nguyen and Fabian Bernards.
\newblock Entanglement dynamics of two mesoscopic objects with gravitational
  interaction.
\newblock {\em The European Physical Journal D}, 74:69, 2020.

\bibitem{Yant2023}
Jackson Yant and Miles Blencowe.
\newblock Gravitationally induced entanglement in a harmonic trap.
\newblock {\em Physical Review D}, 107:106018, 2023.

\bibitem{Ghosal2024}
Pratik Ghosal, Arkaprabha Ghosal, and Somshubrho Bandyopadhyay.
\newblock Distribution of quantum gravity induced entanglement in many-body
  systems.
\newblock {\em Journal of Physics A: Mathematical and Theoretical}, 57:445302,
  2024.

\bibitem{Cui2023}
Dianzhen Cui and X.~X. Yi.
\newblock Exponentially enhanced gravitationally induced entanglement between
  quantum systems with a two-phonon drive.
\newblock {\em Physical Review A}, 108:023502, 2023.

\bibitem{Bose2022}
Sougato Bose, Anupam Mazumdar, Martine Schut, and Marko~Marko Toroš.
\newblock Mechanism for the quantum natured gravitons to entangle masses.
\newblock {\em Physical Review D}, 105:106028, 2022.

\bibitem{Toroš2021}
Marko Toroš, Thomas~W. van~de Kamp, Ryan~J. Marshman, M.~S. Kim, Anupam
  Mazumdar, and Sougato Bose.
\newblock Relative acceleration noise mitigation for nanocrystal matter-wave
  interferometry: Applications to entangling masses via quantum gravity.
\newblock {\em Physical Review Research}, 3:023178, 2021.

\bibitem{Zhang2024a}
Chi Zhang and Fu-Wen Shu.
\newblock Gravity-induced entanglement between two massive microscopic
  particles in curved spacetime: I. the schwarzschild background.
\newblock {\em The European Physical Journal C}, 84:256, 2024.

\bibitem{Zhang2024b}
Chi Zhang and Fu-Wen Shu.
\newblock Gravity-induced entanglement between two massive microscopic
  particles in curved spacetime: Ii. friedmann–lemaître–robertson–walker
  universe.
\newblock {\em The European Physical Journal C}, 84:500, 2024.

\bibitem{martin2023gravity}
Eduardo Mart{\'\i}n-Mart{\'\i}nez and T~Rick Perche.
\newblock What gravity mediated entanglement can really tell us about quantum
  gravity.
\newblock {\em Physical Review D}, 108(10):L101702, 2023.

\bibitem{Ma2022}
Yue Ma, Thomas Guff, Gavin~W. Morley, Igor Pikovski, and M.~S. Kim.
\newblock Limits on inference of gravitational entanglement.
\newblock {\em Physical Review Research}, 4:013024, 2022.

\bibitem{Gollapudi2025}
Praveer~K. Gollapudi, M.~Kemal D\"oner, and André Großardt.
\newblock State swapping via semiclassical gravity.
\newblock {\em Physical Review A}, 111:012208, 2025.

\bibitem{Husain2022}
Viqar Husain, Irfan Javed, and Suprit Singh.
\newblock Dynamics and entanglement in quantum and quantum-classical systems:
  Lessons for gravity.
\newblock {\em Physical Review Letters}, 129:111302, 2022.

\bibitem{Hanif2024}
Farhan Hanif, Debarshi Das, Jonathan Halliwell, Dipankar Home, Anupam Mazumdar,
  Hendrik Ulbricht, and Sougato Bose.
\newblock Testing whether gravity acts as a quantum entity when measured.
\newblock {\em Physical Review Letters}, 133:180201, 2024.

\bibitem{horodecki2009quantum}
Ryszard Horodecki, Pawe{\l} Horodecki, Micha{\l} Horodecki, and Karol
  Horodecki.
\newblock Quantum entanglement.
\newblock {\em Reviews of modern physics}, 81(2):865--942, 2009.

\bibitem{rijavec2021decoherence}
Simone Rijavec, Matteo Carlesso, Angelo Bassi, Vlatko Vedral, and Chiara
  Marletto.
\newblock Decoherence effects in non-classicality tests of gravity.
\newblock {\em New Journal of Physics}, 23(4):043040, 2021.

\bibitem{aspelmeyer2022zeh}
Markus Aspelmeyer.
\newblock When zeh meets feynman: How to avoid the appearance of a classical
  world in gravity experiments.
\newblock In {\em From Quantum to Classical: Essays in Honour of H.-Dieter
  Zeh}, pages 85--95. Springer, 2022.

\bibitem{lami2024testing}
Ludovico Lami, Julen~S. Pedernales, and Martin~B. Plenio.
\newblock Testing the quantumness of gravity without entanglement.
\newblock {\em Phys. Rev. X}, 14:021022, May 2024.

\bibitem{Datta2021}
Animesh Datta and Haixing Miao.
\newblock Signatures of the quantum nature of gravity in the differential
  motion of two masses.
\newblock {\em Quantum Science and Technology}, 6:045014, 2021.

\bibitem{Margalit2021}
Yair Margalit, Or~Dobkowski, Zhifan Zhou, Omer Amit, Yonathan Japha, Samuel
  Moukouri, Daniel Rohrlich, Anupam Mazumdar, Sougato Bose, Carsten Henkel, and
  Ron Folman.
\newblock Realization of a complete stern-gerlach interferometer: Toward a test
  of quantum gravity.
\newblock {\em Science Advances}, 7:2879, 2021.

\bibitem{kryhin2025distinguishable}
Serhii Kryhin and Vivishek Sudhir.
\newblock Distinguishable consequence of classical gravity on quantum matter.
\newblock {\em Physical Review Letters}, 134(6):061501, 2025.

\bibitem{howl2021non}
Richard Howl, Vlatko Vedral, Devang Naik, Marios Christodoulou, Carlo Rovelli,
  and Aditya Iyer.
\newblock Non-gaussianity as a signature of a quantum theory of gravity.
\newblock {\em PRX Quantum}, 2(1):010325, 2021.

\bibitem{fein2019quantum}
Yaakov~Y Fein, Philipp Geyer, Patrick Zwick, Filip Kia{\l}ka, Sebastian
  Pedalino, Marcel Mayor, Stefan Gerlich, and Markus Arndt.
\newblock Quantum superposition of molecules beyond 25 kda.
\newblock {\em Nature Physics}, 15(12):1242--1245, 2019.

\bibitem{Ma2021}
X.~Ma, J.~J. Viennot, S.~Kotler, J.~D. Teufel, and K.~W. Lehnert.
\newblock Non-classical energy squeezing of a macroscopic mechanical
  oscillator.
\newblock {\em Nature Physics}, 17:322, 2021.

\bibitem{Bild2023}
Marius Bild, Matteo Fadel, Yu~Yang, Uwe von Lüpke, Phillip Martin, Alessandro
  Bruno, and Yiwen Chu.
\newblock Schrödinger cat states of a 16-microgram mechanical oscillator.
\newblock {\em Science}, 380:274, 2023.

\bibitem{Bahrami2014}
Mohammad Bahrami, Mauro Paternostro, Angelo Bassi, and Hendrik Ulbricht.
\newblock Proposal for a noninterferometric test of collapse models in
  optomechanical systems.
\newblock {\em Physical Review Letters}, 112(21):210404, 2014.

\bibitem{Diosi2015}
Lajos Di\'osi.
\newblock Testing spontaneous wave-function collapse models on classical
  mechanical oscillators.
\newblock {\em Physical Review Letters}, 114:050403, 2015.

\bibitem{Vinante2017}
A.~Vinante, R.~Mezzena, P.~Falferi, M.~Carlesso, and A.~Bassi.
\newblock Improved noninterferometric test of collapse models using ultracold
  cantilevers.
\newblock {\em Physical Review Letters}, 119:110401, 2017.

\bibitem{bilardello2016bounds}
Marco Bilardello, Sandro Donadi, Andrea Vinante, and Angelo Bassi.
\newblock Bounds on collapse models from cold-atom experiments.
\newblock {\em Physica A: Statistical Mechanics and its Applications},
  462:764--782, 2016.

\bibitem{schrinski2017collapse}
Bj{\"o}rn Schrinski, Benjamin~A Stickler, and Klaus Hornberger.
\newblock Collapse-induced orientational localization of rigid rotors.
\newblock {\em Journal of the Optical Society of America B}, 34(6):C1--C7,
  2017.

\bibitem{altamura2024noninterferometric}
Davide Giordano~Ario Altamura, Matteo Carlesso, Sandro Donadi, and Angelo
  Bassi.
\newblock Noninterferometric rotational test of the continuous spontaneous
  localization model: Enhancement of the collapse noise through shape
  optimization.
\newblock {\em Physical Review A}, 109(6):062212, 2024.

\bibitem{donadi2021novel}
Sandro Donadi, Kristian Piscicchia, Raffaele Del~Grande, Catalina Curceanu,
  Matthias Laubenstein, and Angelo Bassi.
\newblock Novel csl bounds from the noise-induced radiation emission from
  atoms.
\newblock {\em The European Physical Journal C}, 81:1--10, 2021.

\bibitem{donadi2021underground}
Sandro Donadi, Kristian Piscicchia, Catalina Curceanu, Lajos Di{\'o}si,
  Matthias Laubenstein, and Angelo Bassi.
\newblock Underground test of gravity-related wave function collapse.
\newblock {\em Nature Physics}, 17(1):74--78, 2021.

\bibitem{arnquist2022search}
IJ~Arnquist, FT~Avignone~III, AS~Barabash, CJ~Barton, KH~Bhimani, E~Blalock,
  B~Bos, M~Busch, M~Buuck, TS~Caldwell, et~al.
\newblock Search for spontaneous radiation from wave function collapse in the
  majorana demonstrator.
\newblock {\em Physical Review Letters}, 129(8):080401, 2022.

\bibitem{carlesso2022present}
Matteo Carlesso, Sandro Donadi, Luca Ferialdi, Mauro Paternostro, Hendrik
  Ulbricht, and Angelo Bassi.
\newblock Present status and future challenges of non-interferometric tests of
  collapse models.
\newblock {\em Nature Physics}, 18(3):243--250, 2022.

\bibitem{donadi2023collapse}
Sandro Donadi, Luca Ferialdi, and Angelo Bassi.
\newblock Collapse dynamics are diffusive.
\newblock {\em Physical Review Letters}, 130(23):230202, 2023.

\bibitem{gisin1989stochastic}
Nicolas Gisin.
\newblock Stochastic quantum dynamics and relativity.
\newblock {\em Helv. Phys. Acta}, 62(4):363--371, 1989.

\bibitem{polchinski1991weinberg}
Joseph Polchinski.
\newblock Weinberg’s nonlinear quantum mechanics and the
  einstein-podolsky-rosen paradox.
\newblock {\em Physical Review Letters}, 66(4):397, 1991.

\bibitem{weinberg1989testing}
Steven Weinberg.
\newblock Testing quantum mechanics.
\newblock {\em Annals of Physics}, 194(2):336--386, 1989.

\bibitem{weinberg1989precision}
Steven Weinberg.
\newblock Precision tests of quantum mechanics.
\newblock {\em Physical Review Letters}, 62(5):485, 1989.

\bibitem{kafri2014classical}
D~Kafri, JM~Taylor, and GJ~Milburn.
\newblock A classical channel model for gravitational decoherence.
\newblock {\em New Journal of Physics}, 16(6):065020, 2014.

\bibitem{tilloy2016sourcing}
Antoine Tilloy and Lajos Di{\'o}si.
\newblock Sourcing semiclassical gravity from spontaneously localized quantum
  matter.
\newblock {\em Physical Review D}, 93(2):024026, 2016.

\bibitem{oppenheim2023gravitationally}
Jonathan Oppenheim, Carlo Sparaciari, Barbara {\v{S}}oda, and Zachary
  Weller-Davies.
\newblock Gravitationally induced decoherence vs space-time diffusion: testing
  the quantum nature of gravity.
\newblock {\em Nature Communications}, 14(1):7910, 2023.

\bibitem{diosi1989models}
Lajos Di{\'o}si.
\newblock Models for universal reduction of macroscopic quantum fluctuations.
\newblock {\em Physical Review A}, 40(3):1165, 1989.

\bibitem{galley2023any}
Thomas~D Galley, Flaminia Giacomini, and John~H Selby.
\newblock Any consistent coupling between classical gravity and quantum matter
  is fundamentally irreversible.
\newblock {\em Quantum}, 7:1142, 2023.

\bibitem{bahrami2014schrodinger}
Bahrami Mohammad, Gro{\ss}ardt Andr{\'e}, Donadi Sandro, and Bassi Angelo.
\newblock The schr{\"o}dinger--newton equation and its foundations.
\newblock {\em New Journal of Physics}, 16(11):115007, 2014.

\bibitem{piccione2025hybrid}
Nicol{\`o} Piccione and Angelo Bassi.
\newblock Hybrid classical-quantum newtonian gravity with stable vacuum.
\newblock {\em Classical and Quantum Gravity}, 42(22):225002, 2025.

\bibitem{Simon2001}
Christoph Simon, Vladimír Bužek, and Nicolas Gisin.
\newblock No-signaling condition and quantum dynamics.
\newblock {\em Physical Review Letters}, 87:170405, 2001.

\bibitem{hughston1993complete}
Lane~P Hughston, Richard Jozsa, and William~K Wootters.
\newblock A complete classification of quantum ensembles having a given density
  matrix.
\newblock {\em Physics Letters A}, 183(1):14--18, 1993.

\bibitem{Heinosaari2012}
Teiko Heinosaari and M\'ario Ziman.
\newblock {\em The Mathematical Language of Quantum Theory. From Uncertainty to
  Entanglement}.
\newblock Cambridge University Press, 2012.

\bibitem{Watrous2018}
John Watrous.
\newblock {\em The Theory of Quantum Information}.
\newblock Cambridge University Press, 2018.

\bibitem{aziz2025classical}
Joseph Aziz and Richard Howl.
\newblock Classical theories of gravity produce entanglement.
\newblock {\em Nature}, 646(8086):813--817, 2025.

\bibitem{freedman1972experimental}
Stuart~J Freedman and John~F Clauser.
\newblock Experimental test of local hidden-variable theories.
\newblock {\em Physical review letters}, 28(14):938, 1972.

\bibitem{aspect1982experimental}
Alain Aspect, Jean Dalibard, and G{\'e}rard Roger.
\newblock Experimental test of bell's inequalities using time-varying
  analyzers.
\newblock {\em Physical review letters}, 49(25):1804, 1982.

\bibitem{weihs1998violation}
Gregor Weihs, Thomas Jennewein, Christoph Simon, Harald Weinfurter, and Anton
  Zeilinger.
\newblock Violation of bell's inequality under strict einstein locality
  conditions.
\newblock {\em Physical Review Letters}, 81(23):5039, 1998.

\bibitem{tittel1998violation}
Wolfgang Tittel, J{\"u}rgen Brendel, Hugo Zbinden, and Nicolas Gisin.
\newblock Violation of bell inequalities by photons more than 10 km apart.
\newblock {\em Physical review letters}, 81(17):3563, 1998.

\bibitem{giustina2013bell}
Marissa Giustina, Alexandra Mech, Sven Ramelow, Bernhard Wittmann, Johannes
  Kofler, J{\"o}rn Beyer, Adriana Lita, Brice Calkins, Thomas Gerrits, Sae~Woo
  Nam, et~al.
\newblock Bell violation using entangled photons without the fair-sampling
  assumption.
\newblock {\em Nature}, 497(7448):227--230, 2013.

\bibitem{shalm2015strong}
Lynden~K Shalm, Evan Meyer-Scott, Bradley~G Christensen, Peter Bierhorst,
  Michael~A Wayne, Martin~J Stevens, Thomas Gerrits, Scott Glancy, Deny~R
  Hamel, Michael~S Allman, et~al.
\newblock Strong loophole-free test of local realism.
\newblock {\em Physical review letters}, 115(25):250402, 2015.

\bibitem{di2021gravity}
Giovanni Di~Bartolomeo, Matteo Carlesso, and Angelo Bassi.
\newblock Gravity as a classical channel and its dissipative generalization.
\newblock {\em Physical Review D}, 104(10):104027, 2021.

\bibitem{Wilde2017}
Mark~M. Wilde.
\newblock {\em Quantum Information Theory}.
\newblock Cambridge University Press, 2017.

\bibitem{Trillo2025}
David Trillo and Miguel Navascués.
\newblock Diósi-penrose model of classical gravity predicts gravitationally
  induced entanglement.
\newblock {\em Physical Review D}, 111:L121101, 2025.

\bibitem{Angeli2025}
Oliviero Angeli and Matteo Carlesso.
\newblock Entanglement in markovian hybrid classical-quantum theories of
  gravity.
\newblock {\em Physical Review D}, 112(2):024047, 2025.

\bibitem{diosi1995quantum}
Lajos Diosi.
\newblock Quantum dynamics with two planck constants and the semiclassical
  limit.
\newblock {\em arXiv preprint quant-ph/9503023}, 1995.

\bibitem{Diosi2000}
Lajos Diósi, Nicolas Gisin, and Walter~T. Strunz.
\newblock Quantum approach to coupling classical and quantum dynamics.
\newblock {\em Physical Review A}, 61:022108, 2000.

\bibitem{diosi2023hybrid}
Lajos Di{\'o}si.
\newblock Hybrid completely positive markovian quantum-classical dynamics.
\newblock {\em Physical Review A}, 107(6):062206, 2023.

\bibitem{Ghirardi1986}
Gian~Carlo Ghirardi, Alberto Rimini, and Tullio Weber.
\newblock Unified dynamics for microscopic and macroscopic systems.
\newblock {\em Physical Review D}, 34:470, 1986.

\bibitem{Pearle1989}
Philip Pearle.
\newblock Combining stochastic dynamical state-vector reduction with
  spontaneous localization.
\newblock {\em Physical Review A}, 39:2277, 1989.

\bibitem{Ghirardi1990}
Gian~Carlo Ghirardi, Philip Pearle, and Alberto Rimini.
\newblock Markov processes in hilbert space and continuous spontaneous
  localization of systems of identical particles.
\newblock {\em Physical Review A}, 42:78, 1990.

\bibitem{Note1}
While this holds exactly for unitary evolution under the Newtonian potential;
  it is expected to hold approximately for the classical gravity models
  considered here. For example, in the KTM~\cite {kafri2014classical} and the
  Diósi-Penrose models, it holds exactly for two equal masses, but small
  deviations are present for different masses. Consider the linearized
  Diósi-Penrose with two masses at distances $d\gg R_0$ with $R_0$ the
  smearing parameter~\cite {gaona2021gravitational}. Then center of mass motion
  and relative coordinates are coupled in the dissipator through a coefficient
  $\gamma _{Rr}\propto \protect \frac {G\mu \Delta m}{\hbar R_0^3}$ where $\mu
  $ is the reduced mass and $\Delta m$ the difference.

\bibitem{FondaGhirardi1970}
Luciano Fonda and Gian~Carlo Ghirardi.
\newblock {\em Symmetry Principles in Quantum Physics}, volume~1 of {\em
  Theoretical Physics}.
\newblock M. Dekker, New York, 1970.

\bibitem{Note2}
As a further justification, in~\cite {Angeli2025} a toy model describing
  collapse only in the relative coordinate via a GKLS master equation was
  investigated and it was found that it produces entanglement between
  well-separated systems, which is incompatible with the classicality
  condition~\protect \eqref {eq:WeakC2}.

\bibitem{Haapasalo2019}
Erkka Haapasalo.
\newblock Compatibility of covariant quantum channels with emphasis on weyl
  symmetry.
\newblock {\em Annales Henri Poincar{\'e}}, 20:3163--3195, 2019.

\bibitem{holevo2005additivity}
AS~Holevo.
\newblock Additivity conjecture and covariant channels.
\newblock {\em International Journal of Quantum Information}, 3(01):41--47,
  2005.

\bibitem{bassi2005collapse}
Angelo Bassi.
\newblock Collapse models: analysis of the free particle dynamics.
\newblock {\em Journal of Physics A: Mathematical and General},
  38(14):3173--3192, 2005.

\bibitem{Note3}
Entanglement has been discussed also withing the framework of generalized
  probabilistic theories \cite {chiribella2015entanglement,galley2022no}; it is
  a interesting question to assess how our framework carries over to such
  situations~\cite {galley2023any}.

\bibitem{gorini1976completely}
Vittorio Gorini, Andrzej Kossakowski, and Ennackal Chandy~George Sudarshan.
\newblock Completely positive dynamical semigroups of n-level systems.
\newblock {\em Journal of Mathematical Physics}, 17(5):821--825, 1976.

\bibitem{lindblad1976generators}
Goran Lindblad.
\newblock On the generators of quantum dynamical semigroups.
\newblock {\em Communications in mathematical physics}, 48:119--130, 1976.

\bibitem{bahrami2014role}
Bahrami Mohammad, Smirne Andrea, and Bassi Angelo.
\newblock Role of gravity in the collapse of a wave function: A probe into the
  di{\'o}si-penrose model.
\newblock {\em Physical Review A}, 90(6):062105, 2014.

\bibitem{di2023linear}
Giovanni Di~Bartolomeo, Matteo Carlesso, Kristian Piscicchia, Catalina
  Curceanu, Maaneli Derakhshani, and Lajos Di{\'o}si.
\newblock Linear-friction many-body equation for dissipative spontaneous
  wave-function collapse.
\newblock {\em Physical Review A}, 108(1):012202, 2023.

\bibitem{gaona2021gravitational}
Jos{\'e}~Luis Gaona-Reyes, Matteo Carlesso, and Angelo Bassi.
\newblock Gravitational interaction through a feedback mechanism.
\newblock {\em Physical Review D}, 103(5):056011, 2021.

\bibitem{peres1996separability}
Asher Peres.
\newblock Separability criterion for density matrices.
\newblock {\em Physical Review Letters}, 77(8):1413, 1996.

\bibitem{ferraro2005gaussian}
Alessandro Ferraro, Stefano Olivares, and Matteo~GA Paris.
\newblock Gaussian states in continuous variable quantum information.
\newblock {\em arXiv preprint quant-ph/0503237}, 2005.

\bibitem{simon2000peres}
Rajiah Simon.
\newblock Peres-horodecki separability criterion for continuous variable
  systems.
\newblock {\em Physical Review Letters}, 84(12):2726, 2000.

\bibitem{Note4}
In Appendix~\ref {app:Spectrum} we discuss the case where the coefficients
  $\gamma _{ij}$ do depend on the trapping frequencies, showing that it is
  still possible identify a suitable experimental setup testing the
  gravitational bound.

\bibitem{kafri2013noise}
Dvir Kafri and JM~Taylor.
\newblock A noise inequality for classical forces.
\newblock {\em arXiv preprint arXiv:1311.4558}, 2013.

\bibitem{amelino1999gravity}
Giovanni Amelino-Camelia.
\newblock Gravity-wave interferometers as quantum-gravity detectors.
\newblock {\em Nature}, 398(6724):216--218, 1999.

\bibitem{amelino2001phenomenological}
Giovanni Amelino-Camelia.
\newblock A phenomenological description of space-time noise in quantum
  gravity.
\newblock {\em Nature}, 410(6832):1065--1067, 2001.

\bibitem{petruzziello2021quantum}
Luciano Petruzziello and Fabrizio Illuminati.
\newblock Quantum gravitational decoherence from fluctuating minimal length and
  deformation parameter at the planck scale.
\newblock {\em Nature Communications}, 12(1):4449, 2021.

\bibitem{verlinde2021observational}
Erik~P Verlinde and Kathryn~M Zurek.
\newblock Observational signatures of quantum gravity in interferometers.
\newblock {\em Physics Letters B}, 822:136663, 2021.

\bibitem{zurek2022vacuum}
Kathryn~M Zurek.
\newblock On vacuum fluctuations in quantum gravity and interferometer arm
  fluctuations.
\newblock {\em Physics Letters B}, 826:136910, 2022.

\bibitem{arzano2023fundamental}
Michele Arzano, Vittorio D’Esposito, and Giulia Gubitosi.
\newblock Fundamental decoherence from quantum spacetime.
\newblock {\em Communications Physics}, 6(1):242, 2023.

\bibitem{sharmila2025signatures}
B~Sharmila, Sander~M Vermeulen, and Animesh Datta.
\newblock Signatures of correlation of spacetime fluctuations in laser
  interferometers.
\newblock {\em Nature Communications}, 2025.

\bibitem{paternostro2006reconstructing}
Mauro Paternostro, Sylvain Gigan, Myung~Shik Kim, Florian Blaser, HR~B{\"o}hm,
  and Markus Aspelmeyer.
\newblock Reconstructing the dynamics of a movable mirror in a detuned optical
  cavity.
\newblock {\em New Journal of Physics}, 8(6):107, 2006.

\bibitem{Note5}
Cf. equation (5) of~\cite {adler2007collapse} with the choice $\xi = i$.

\bibitem{vinante2020}
A.~Vinante, M.~Carlesso, A.~Bassi, A.~Chiasera, S.~Varas, P.~Falferi,
  B.~Margesin, R.~Mezzena, and H.~Ulbricht.
\newblock Narrowing the parameter space of collapse models with ultracold
  layered force sensors.
\newblock {\em Physical Review Letters}, 125(10), sep 2020.

\bibitem{vinantepontin2019}
A.~Vinante, A.~Pontin, M.~Rashid, P.~Barker, M.~Toros, and H.~Ulbricht.
\newblock Testing collapse models with levitated nanoparticles: Detection
  challenge.
\newblock {\em Physical review A}, 100:012119, 2019.

\bibitem{Vinante2016}
A.~Vinante, M.~Bahrami, A.~Bassi, O.~Usenko, G.~Wijts, and T.~H. Oosterkamp.
\newblock Upper bounds on spontaneous wave-function collapse models using
  millikelvin-cooled nanocantilevers.
\newblock {\em Physical Review Letters}, 116:090402, 2016.

\bibitem{Braginsky1972}
V.B. Braginsky and V.I. Panov.
\newblock Verification of the equivalence of inertial and gravitational mass.
\newblock {\em Journal of Experimental and Theoretical Physics}, 34:463, 1972.

\bibitem{cavalleri2009}
A.~Cavalleri, G.~Ciani, R.~Dolesi, A.~Heptonstall, M.~Hueller, D.~Nicolodi,
  S.~Rowan, D.~Tombolato, S.~Vitale, P.~J. Wass, and W.~J. Weber.
\newblock Increased brownian force noise from molecular impacts in a
  constrained volume.
\newblock {\em Physical Review Letters}, 103:140601, 2009.

\bibitem{Bantel2000}
M.K. Bantel and R.D. Newman.
\newblock High precision measurement of torsion fiber internal friction at
  cryogenic temperatures.
\newblock {\em Journal of Alloys and Compounds}, 310:233, 2000.

\bibitem{Bantel2014}
Riley Newman, Michael Bantel, Eric Berg, and William Cross.
\newblock A measurement of g with a cryogenic torsion pendulum.
\newblock {\em Philosophical Transactions of the Royal Society A},
  372:20140025, 2014.

\bibitem{Fleischer2022}
S.M. Fleischer, M.P. Ross, K.~Venkateswara, C.A. Hagedorn, E.A. Shaw,
  E.~Swanson, B.R. Heckel, and J.H. Gundlach.
\newblock A cryogenic torsion balance using a liquid-cryogen free, ultra-low
  vibration cryostat.
\newblock {\em Review of Scientific Instruments}, 93:064505, 2022.

\bibitem{Braginsky1977}
V.B. Braginsky, C.M. Caves, and K.S. Thorne.
\newblock Laboratory experiments to test relativistic gravity.
\newblock {\em Physical Review D}, 15:2047, 1977.

\bibitem{McGuigan1978}
D.F. McGuigan, C.C. Lam, R.Q. Gram, A.W. Hoffman, D.H. Douglass, and H.~W.
  Gutche.
\newblock Measurements of the mechanical {Q} of single-crystal silicon at low
  temperatures.
\newblock {\em Journal of Low Temperature Physics}, 30:621, 1978.

\bibitem{MacCabe2020}
Gregory~S MacCabe, Hengjiang Ren, Jie Luo, Justin~D Cohen, Hengyun Zhou, Alp
  Sipahigil, Mohammad Mirhosseini, and Oskar Painter.
\newblock Nano-acoustic resonator with ultralong phonon lifetime.
\newblock {\em Science}, 370:840, 2020.

\bibitem{lisa2016}
Michele Armano, Heather Audley, Gerard Auger, Jonathon~T Baird, Massimo Bassan,
  Pierre Binetruy, Michael Born, Daniele Bortoluzzi, Nico Brandt, Maria Caleno,
  et~al.
\newblock Sub-femto-g free fall for space-based gravitational wave
  observatories: {LISA} pathfinder results.
\newblock {\em Physical review letters}, 116(23):231101, 2016.

\bibitem{lisa2018}
M.~Armano and et~al.
\newblock Beyond the required {LISA} free-fall performance: New {LISA}
  pathfinder results down to 20 {$\mu$⁢Hz}.
\newblock {\em Physical review letters}, 120:061101, 2018.

\bibitem{cesarini2024}
A.~Cesarini et~al. {(LISA Pathfinder Collaboration}).
\newblock In-depth analysis of {LISA} pathfinder performance results: Time
  evolution, noise projection, physical models, and implications for {LISA}.
\newblock {\em Physical Review D}, 110:042004, 2024.

\bibitem{LIGO2023}
D.~Ganapathy and et~al.
\newblock Broadband quantum enhancement of the ligo detectors with
  frequency-dependent squeezing.
\newblock {\em Physical Review X}, 47:041021, 2023.

\bibitem{Paik1976}
Ho~Jung Paik.
\newblock Superconducting tunable‐diaphragm transducer for sensitive
  acceleration measurements.
\newblock {\em Journal of Applied Physics}, 47:1168, 1976.

\bibitem{Vinante2020b}
A.~Vinante, P.~Falferi, G.~Gasbarri, A.~Setter, C.~Timberlake, and H.~Ulbricht.
\newblock Ultralow mechanical damping with meissner-levitated ferromagnetic
  microparticles.
\newblock {\em Physical Review Applied}, 13:064027, 2020.

\bibitem{Schmidt2024}
Philip Schmidt, Remi Claessen, Gerard Higgins, Joachim Hofer, Jannek~J. Hansen,
  Peter Asenbaum, Martin Zemlicka, Kevin Uhl, and Reinhold Kleiner.
\newblock Remote sensing of a levitated superconductor with a flux-tunable
  microwave cavity.
\newblock {\em Physical Review Applied}, 22:014078, 2024.

\bibitem{Kumar2016}
P.~Kumar, S.~Sendelbach, M.A. Beck, J.W. Freeland, Zhe Wang, Hui Wang, Clare~C.
  Yu, R.Q. Wu, D.P. Pappas, and R.~McDermott.
\newblock Origin and reduction of 1/f magnetic flux noise in superconducting
  devices.
\newblock {\em Physical Review Applied}, 6:041001, 2016.

\bibitem{Ahrens2024}
Felix Ahrens, Wei Ji, Dmitry Budker, Chris Timberlake, Hendrik Ulbricht, and
  Andrea Vinante.
\newblock Levitated ferromagnetic magnetometer with energy resolution well
  below {$\hbar$}.
\newblock {\em arXiv:2401.03774}, 2024.

\bibitem{jose2025}
Joel~K Jose, Andrea Marchese, Marion Cromb, Hendrik Ulbricht, Andrejs Cebers,
  Ping~Koy Lam, Tao Wang, and Andrea Vinante.
\newblock Cryogenic pressure sensing with an ultrafast meissner-levitated
  microrotor.
\newblock {\em arXiv:2509.24964}, 2025.

\bibitem{carney2019tabletop}
Daniel Carney, Philip~CE Stamp, and Jacob~M Taylor.
\newblock Tabletop experiments for quantum gravity: a user’s manual.
\newblock {\em Classical and Quantum Gravity}, 36(3):034001, 2019.

\bibitem{bose2025massive}
Sougato Bose, Ivette Fuentes, Andrew~A Geraci, Saba~Mehsar Khan, Sofia
  Qvarfort, Markus Rademacher, Muddassar Rashid, Marko Toro{\v{s}}, Hendrik
  Ulbricht, and Clara~C Wanjura.
\newblock Massive quantum systems as interfaces of quantum mechanics and
  gravity.
\newblock {\em Reviews of Modern Physics}, 97(1):015003, 2025.

\bibitem{datta2009min}
Nilanjana Datta.
\newblock Min-and max-relative entropies and a new entanglement monotone.
\newblock {\em IEEE Transactions on Information Theory}, 55(6):2816--2826,
  2009.

\bibitem{carney2023strongly}
Daniel Carney and Jacob~M Taylor.
\newblock Strongly incoherent gravity.
\newblock {\em arXiv preprint arXiv:2301.08378}, 2023.

\bibitem{altamirano2018gravity}
Natacha Altamirano, Paulina Corona-Ugalde, Robert~B Mann, and Magdalena Zych.
\newblock Gravity is not a pairwise local classical channel.
\newblock {\em Classical and Quantum Gravity}, 35(14):145005, 2018.

\bibitem{caldeira1983path}
Amir~O Caldeira and Anthony~J Leggett.
\newblock Path integral approach to quantum brownian motion.
\newblock {\em Physica A: Statistical mechanics and its Applications},
  121(3):587--616, 1983.

\bibitem{vacchini2002non}
Bassano Vacchini.
\newblock Non-abelian linear boltzmann equation and quantum correction to
  kramers and smoluchowski equation.
\newblock {\em Physical Review E}, 66(2):027107, 2002.

\bibitem{christodoulou2023locally}
Marios Christodoulou, Andrea Di~Biagio, Markus Aspelmeyer, {\v{C}}aslav
  Brukner, Carlo Rovelli, and Richard Howl.
\newblock Locally mediated entanglement in linearized quantum gravity.
\newblock {\em Physical Review Letters}, 130(10):100202, 2023.

\bibitem{chruscinski2010non}
Dariusz Chru{\'s}ci{\'n}ski and Andrzej Kossakowski.
\newblock Non-markovian quantum dynamics: local versus nonlocal.
\newblock {\em Physical review letters}, 104(7):070406, 2010.

\bibitem{Breuer2002}
Heinz-Peter Breuer and Francesco Petruccione.
\newblock {\em The Theory of Open Quantum Systems}.
\newblock Oxford University Press, 2002.

\bibitem{chiribella2015entanglement}
Giulio Chiribella and Carlo~Maria Scandolo.
\newblock Entanglement and thermodynamics in general probabilistic theories.
\newblock {\em New Journal of Physics}, 17(10):103027, 2015.

\bibitem{galley2022no}
Thomas~D Galley, Flaminia Giacomini, and John~H Selby.
\newblock A no-go theorem on the nature of the gravitational field beyond
  quantum theory.
\newblock {\em Quantum}, 6:779, 2022.

\bibitem{adler2007collapse}
Stephen~L Adler and Angelo Bassi.
\newblock Collapse models with non-white noises.
\newblock {\em Journal of Physics A: Mathematical and Theoretical},
  40(50):15083, 2007.

\end{thebibliography}
\end{document}